\documentclass[11pt,a4paper]{article}
\pdfoutput=1
\usepackage{jcappub}

\usepackage{amsmath}
\usepackage{amssymb}
\usepackage{graphicx}
\usepackage{bbm}

\def\lsp{\tilde{\nu}_1}
\def\mlsp{m_{\tilde{\nu}_1}}
\def\anu{A_{\tilde\nu}}

\def\sn{\sin\theta_{\tilde\nu}}

\title{Mixed sneutrino dark matter in light of the 2011 XENON and LHC results}
\author[a]{B\'eranger Dumont,}
\author[b]{Genevi\`eve B\'elanger,}
\author[c]{Sylvain Fichet,}
\author[a]{Sabine Kraml,}
\author[d]{and Thomas Schwetz}

\affiliation[a]{\,Laboratoire de Physique Subatomique et de Cosmologie, UJF Grenoble 1, CNRS/IN2P3, INPG, 
53 Avenue des Martyrs, F-38026 Grenoble, France}
\affiliation[b]{\,LAPTH, Univ. de Savoie, CNRS, 
B.P.\,110, F-74941 Annecy-le-Vieux, France}
\affiliation[c]{\,International Institute of Physics, UFRN, 
Av.\ Odilon Gomes de Lima, 1722, Capim Macio 59078-400, Natal-RN, Brazil}
\affiliation[d]{\,Max-Planck-Institut f\"ur Kernphysik, 
Saupfercheckweg 1, 69117 Heidelberg, Germany}

\emailAdd{beranger.dumont@lpsc.in2p3.fr}
\emailAdd{belanger@lapp.in2p3.fr}
\emailAdd{sylvain.fichet@lpsc.in2p3.fr}
\emailAdd{sabine.kraml@lpsc.in2p3.fr}
\emailAdd{schwetz@mpi-hd.mpg.de}

\abstract{
In the context of supersymmetric models in which small Dirac neutrino masses are generated by supersymmetry breaking, a mainly right-handed (RH) mixed sneutrino can be an excellent cold dark matter (DM) candidate. We perform a global analysis of the Minimal Supersymmetric Standard Model (MSSM)+RH neutrino parameter space by means of Markov Chain Monte Carlo sampling. We include all relevant constraints from collider and dark matter searches, paying particular attention to nuclear and astrophysical uncertainties. 
Two distinct cases can satisfy all constraints:  heavy sneutrino DM with mass of order 100~GeV, as well as light sneutrino DM with mass of about 3--6~GeV. We discuss the implications for direct and indirect dark matter searches, as well as for SUSY and Higgs searches at the LHC for both, the light and the heavy sneutrino dark matter case. The light sneutrino case is excluded by the 125--126~GeV Higgs signal.}

\begin{document}
\maketitle

%===============================================================================
\section{Introduction}
%===============================================================================

The nature of dark matter (DM)~\cite{Jungman:1995df,Bertone:2004pz,Bertone:2010zz}, 
the nature of the physics stabilizing the electroweak scale~\cite{Giudice:2007qj,Bhattacharyya:2009gw,Grojean:2009fd},  
and the origin of neutrino masses~\cite{Bilenky:1987ty,Mohapatra:2005wg,GonzalezGarcia:2007ib} 
range among the most challenging open problems in particle physics. Huge efforts on both the experimental and theoretical sides are undertaken worldwide to shed light on these questions and eventually unravel a more fundamental theory beyond the current Standard Model (SM) of electroweak and strong interactions. 

Weak-scale supersymmetry (SUSY)~\cite{Martin:1997ns,Drees:2004jm,Baer:2006rs,Binetruy:2006ad} is a fascinating option for such a theory beyond the SM, potentially addressing all three of the above open problems. Indeed SUSY, if realized at the TeV scale, elegantly solves the gauge hierarchy problem and, if R-parity is conserved, provides an excellent particle dark matter candidate: the lightest supersymmetric particle (LSP). Moreover, in a certain class of models, small neutrino masses may naturally arise from F-term SUSY breaking~\cite{ArkaniHamed:2000bq,Borzumati:2000mc}. 
In addition to providing an explanation for neutrino masses, this class of SUSY models offers a particular DM candidate: a mainly right-handed (RH) mixed sneutrino. 

Mixed sneutrinos as thermal DM are indeed a very interesting alternative to the conventional neutralino LSP of the Minimal Supersymmetric Standard Model (MSSM). They have received much attention recently, in part because of their intriguing phenomenology and in part because they provide a possibility for light SUSY DM below 10~GeV.  
Many studies of sneutrino DM have been performed in models with extra singlets, extra gauge groups or 
models with Majorana neutrino masses, see e.g., \cite{Hall:1997ah,Kolb:1999wx,Asaka:2005cn, 
Asaka:2006fs,Lee:2007mt,Arina:2008yh,Arina:2008bb,Cerdeno:2008ep,Deppisch:2008bp,Allahverdi:2009ae,
Cerdeno:2009dv,Demir:2009kc,Allahverdi:2009se,Cerdeno:2009zz,Allahverdi:2009kr,
Kumar:2009sf,MarchRussell:2009aq,Ishiwata:2009gs,Khalil:2011tb,Kang:2011wb,Bandyopadhyay:2011qm,Cerdeno:2011qv,An:2011uq,Belanger:2011rs}. 
In our work, we concentrate instead on the MSSM+RH neutrino model \cite{ArkaniHamed:2000bq,Borzumati:2000mc} with only Dirac masses for neutrinos.  
The phenomenology of this model was investigated in detail in~\cite{ArkaniHamed:2000bq,Belanger:2010cd}. 
Indirect detection signatures were discussed in~\cite{Arina:2007tm,Choi:2012ap}, implications for $\Omega_b/\Omega_{\rm DM}$ in \cite{Hooper:2004dc}, and LHC signatures in~\cite{Thomas:2007bu,Belanger:2011ny}.

The crucial point of this model is that one can have a weak-scale trilinear sneutrino coupling 
$A_{\tilde\nu}$ that is not suppressed by a small Dirac-neutrino Yukawa coupling. It can hence induce a large mixing between left-handed and right-handed sneutrinos even though the Yukawa couplings may be extremely small.  The lightest sneutrino can thus become the LSP and a viable thermal DM candidate.  
Note that the mainly RH sneutrino LSP is not sterile but couples to SM gauge and Higgs 
bosons through the mixing with its LH partner. Sufficient mixing provides efficient 
annihilation so that the sneutrino relic density $\Omega h^2 \simeq 0.11$ as 
extracted from cosmological observations~\cite{Komatsu:2010fb}. 

Direct detection (DD) experiments however pose severe constraints on Dirac or complex 
scalar, i.e.\ not self-conjugated, DM particles because the spin-independent 
elastic scattering cross section receives an important contribution from $Z$ 
exchange, which typically exceeds experimental bounds. 
In the mixed sneutrino model, this cross section is suppressed by the sneutrino 
mixing angle. Therefore, on the one hand a viable sneutrino DM candidate requires 
enough mixing to provide sufficient pair-annihilation, on the other hand the 
mixing should not be too large in order not to exceed the DD limits. 

In \cite{Belanger:2010cd}, some of us explored the parameter space of the Dirac sneutrino DM model where these conditions are satisfied for light sneutrinos with a mass below 10~GeV.
This mass range was motivated by hints of DM signals in DD experiments~\cite{Bernabei:2008yi,Aalseth:2010vx}. 
In the present study, we explore a much wider range of masses, considering both light DM below 10~GeV as well as heavier DM of the order of 100~GeV. 
Moreover, we explore the parameter space by means of Markov Chain Monte Carlo (MCMC) sampling, using Bayesian statistics 
to confront the model predictions with the data. In taking into account the limits from DD experiments 
%(CDMS~\cite{Ahmed:2010wy}, CoGeNT~\cite{Aalseth:2011wp}, XENON100~\cite{Aprile:2011hi}, XENON10~\cite{Angle:2011th}) 
we pay special attention to uncertainties stemming from astrophysical parameters (local DM density and velocity distribution) and to uncertainties in the quark contents of the nucleons (relevant in particular when there is a large Higgs-exchange contribution). Finally, we consider the impact of the LHC results, which push the masses of squarks and gluinos above the TeV scale.
Our results are presented as posterior probability densities of parameters and derived quantities, in particular of the DM mass and direct and indirect detection cross sections.

The paper is organized as follows. 
In Section~\ref{sec:framework}, we briefly recall the main features of the mixed sneutrino model. 
In Section~\ref{sec:analysis}, we then describe in detail the setup of our MCMC analysis. 
The results of this analysis for light sneutrino DM are presented in Section~\ref{sec:light}, 
and for heavy sneutrino DM  in Section~\ref{sec:heavy}. 
Conclusions are given in Section~\ref{sec:conclusions}. 
Appendix~\ref{aTQCD} contains a discussion of the uncertainty in the effective degrees of freedom in the early Universe, 
Appendix~\ref{aTables} contains summary tables for our results, and Appendix~\ref{logprior} illustrates the prior dependence by comparing to log-prior results.

%===============================================================================
\section{Framework}\label{sec:framework}
%===============================================================================

The framework for our study is the model of \cite{ArkaniHamed:2000bq,Borzumati:2000mc} with only 
Dirac masses for neutrinos. In this case, the usual MSSM soft-breaking terms are extended by
\begin{equation}
  \Delta {\cal L}_{\rm soft} = m^2_{\tilde N_i}  |\tilde N_i |^2 +  
                                            A_{\tilde\nu_i} \tilde L_i \tilde N_i H_u + {\rm h.c.} \,,
\end{equation}
where ${m}^2_{\widetilde{N}}$ and $A_{\tilde\nu}$ are weak-scale soft terms, which we assume to 
be flavor-diagonal. Note that the lepton-number violating bilinear term, which appears 
in case of Majorana neutrino masses, is absent. 
Neglecting the tiny Dirac masses, the $2\times2$ sneutrino mass matrix for one generation is 
given by 
\begin{equation}
  m^2_{\tilde\nu} =
  \left( \begin{array}{cc}
   {m}^2_{\widetilde{L}} +\frac{1}{2} m^2_Z \cos 2\beta \quad &  \frac{1}{\sqrt{2}} A_{\tilde\nu}\, v \sin\beta\\
   \frac{1}{\sqrt{2}}    A_{\tilde\nu}\, v \sin\beta&  {m}^2_{\widetilde{N}}
  \end{array}\right) \,.
\label{eq:sneutrino_tree}
\end{equation}
Here ${m}^2_{\widetilde{L}}$ is the SU(2) slepton soft term, $v^2=v_1^2+v_2^2=(246\;{\rm GeV})^2$ 
with $v_{1,2}$ the Higgs vacuum expectation values, and $\tan\beta=v_2/v_1$.  
The main feature of this model is that ${m}^2_{\widetilde{L}}$, ${m}^2_{\widetilde{N}}$
and $A_{\tilde\nu}$ are all of the order of the weak scale, and 
$A_{\tilde\nu}$ does not suffer any suppression from Yukawa couplings. 
In the following, we will always assume $m_{\widetilde N}<m_{\widetilde L}$ so that the lighter mass 
eigenstate, $\tilde\nu_1$, is mostly a $\tilde\nu_R$. 
This is in fact well motivated from renormalization group evolution, 
since for the gauge-singlet $\widetilde{N}$ the running at 1-loop is driven exclusively by the 
$A_{\tilde\nu}$ term:
\begin{equation}
  \frac{{\rm d}m_{{\widetilde N}_i}^2}{{\rm d}t} = \frac{4}{16\pi^2} A_{\tilde{\nu}_i}^2 \,,
\end{equation}
while
\begin{equation}
  \frac{{\rm d}m_{{\widetilde L}_i}^2}{{\rm d}t} = \rm{(MSSM\ terms)} + \frac{2}{16\pi^2}A_{\tilde{\nu}_i}^2 \,.
\end{equation}

\noindent
The renormalization group equation (RGE) for the $A$-term is:
\begin{equation}
\label{eq:runA}
  \frac{{\rm d}A_{\tilde{\nu}_i}}{{\rm d}t} = \frac{2}{16\pi^2}\left(-\frac{3}{2}g_2^2 - \frac{3}{10}g_1^2 + \frac{3}{2}y_t^2 + \frac{1}{2}y^2_{l_i}\right) A_{\tilde{\nu}_i} \,.
\end{equation} 
Here, $g_1$ and $g_2$ are the U(1) and SU(2) gauge couplings, and $y_t$ and $y_{l_i}$ are the top and charged lepton Yukawa couplings.

\noindent
A large $A_{\tilde\nu}$ term in the sneutrino mass matrix will induce a significant 
mixing between the RH and LH states, 
\begin{equation}
\label{eq:mixingA}
  \left(\begin{array}{c}
    \tilde\nu_{1}\\
    \tilde\nu_{2}
  \end{array}\right) = 
  \left(\begin{array}{lr}
    \cos\theta_{\tilde\nu}\, & -\sin\theta_{\tilde\nu}\\
    \sin\theta_{\tilde\nu} & \cos\theta_{\tilde\nu}
  \end{array}\right) 
  \left(\begin{array}{c}
    \tilde\nu_{R}\\
    \tilde\nu_{L}
  \end{array}\right) ,
  \quad
  \sin2\theta_{\tilde\nu} = 
     \frac{\sqrt{2} A_{\tilde\nu} v \sin\beta}{m_{\tilde\nu_2}^2 - m_{\tilde\nu_1}^2}\,,
\end{equation}
and a sizable splitting between the two mass eigenstates $\tilde{\nu}_1$ and $\tilde{\nu}_2$ 
(with $m_{\tilde{\nu}_1}<m_{\tilde{\nu}_2}$). 

One immediate consequence of this mixing is that the mainly RH state, $\tilde{\nu}_1$, is no longer sterile. However, its left-handed couplings are suppressed by $\sin \theta_{\tilde{\nu}}$. This allows the $\tilde{\nu}_1$ to have a large enough pair-annihilation rate to be a viable candidate for thermal dark matter, while at the same time evading the limits from direct dark matter searches~\cite{ArkaniHamed:2000bq,Belanger:2010cd,Thomas:2007bu}. A mainly RH $\tilde{\nu}_1$ as the LSP will also have a significant impact on collider phenomenology, as it alters the particle decay chains as compared to the ``conventional'' MSSM. 
Moreover, it can have a significant impact on Higgs phenomenology: first, a light mixed sneutrino can give a large {\it negative} loop correction to $m_{h^0}$  which is $\propto |A_{\tilde{\nu}}|^4$ \cite{Belanger:2010cd}; second, a large $A_{\tilde{\nu}}$ can lead to dominantly invisible Higgs decays if $m_{\tilde{\nu}_1}<m_{h^0}/2$.

In the following, we will assume that electron and muon sneutrinos are mass-degenerate, 
$m_{\tilde{\nu}_{ie}}=m_{\tilde{\nu}_{i\mu}}$ with $i=1,2$. Moreover, by default we will assume that the tau-sneutrino, $\tilde{\nu}_{1\tau}$ is lighter than the $\tilde{\nu}_{1e}$ and is the LSP. This is motivated by the contribution in the running of the $A$-term coming from the Yukawa coupling,
see eq.~\eqref{eq:runA}. In this case, 
we take  $m_{\tilde\nu_1}$, $m_{\tilde\nu_2}$, $\sin\theta_{\tilde{\nu}}$ and $\tan \beta$ as input parameters 
in the sneutrino sector, from which we compute $m_{\widetilde{L}}$, $m_{\widetilde{N}}$, $A_{\tilde{\nu}}$ 
(all parameters are taken at the electroweak scale). 

%===============================================================================
\section{Analysis}\label{sec:analysis}
%===============================================================================

%-------------------------------------------------------------------------------
\subsection{Method}
%-------------------------------------------------------------------------------

We choose to confront the sneutrino DM model to experimental constraints by means of Bayesian inference. In this kind of analysis, one starts with an a priori probability density function (prior PDF) $p(\theta|\mathcal{M})$ for the parameters $\theta=\{\theta_{1\ldots n}\}$ of the model $\mathcal{M}$, and some experimental information enclosed in a likelihood function $p(d|\theta,\mathcal{M})\equiv \mathcal{L}(\theta)$. The purpose is to combine these two pieces of knowledge, to obtain the so-called posterior PDF, possibly marginalized to some subset of parameters. Splitting the parameter set as $\theta=(\psi,\lambda)$, Bayesian statistics tells us that the  posterior PDF of the parameter subset $\psi$  is
\begin{equation}
p(\psi|\mathcal{M})\propto \int d\lambda\,\, p(\psi,\lambda|\mathcal{M}) \mathcal{L}(\psi,\lambda)\,.
\end{equation}
That means one simply integrates over unwanted parameters to obtain the marginalized posterior PDFs. These unwanted parameters can be model parameters, but can also be nuisance parameters. 

In this work, we evaluate posterior PDFs by means of a Markov Chain Monte Carlo (MCMC) method. The basic idea of a MCMC is setting a random walk in the parameter space such that the density of points tends to reproduce the posterior PDF. Any marginalisation is then reduced to a summation over the points of the Markov chain. We refer to~\cite{Allanach:2005kz,Trotta:2008qt} for details on MCMCs and Bayesian inference. Our MCMC method uses the Metropolis-Hastings algorithm with a symmetric, Gaussian proposal function, basically following the procedure explained in \cite{Belanger:2009ti}. 
We use uniform (linear) priors for all parameters. The impact of logarithmic priors in the sneutrino sector is presented in Appendix~\ref{logprior}. For each of the scenarios which we study,  
we run 8 chains with $10^6$ iterations each, and we check their convergence using the Gelman and Rubin test with multiple chains~\cite{Gelman92}, requiring $\sqrt{\hat{R}} < 1.05$ for each parameter. First iterations are discarded (burn-in), until a point with $\log(\mathcal{L})>-5$ is found. 

The likelihood function ${\mathcal L}$ can be constructed as the product of the likelihoods $\mathcal{L}_{i}$ associated to the $N$ observables $O_{i}$,
\begin{equation}
  {\mathcal L} = \prod_{i=1}^N {\mathcal L}_i \, .
  \label{Ltot}
\end{equation}
Available experimental data fall into two categories: measurements of a central value, and upper/lower limits. In the former case, the central value $O_{\rm exp}$ comes with an uncertainty given at some confidence level CL. It is reasonable to assume that the likelihood function for this kind of measurement is a Gaussian distribution,
\begin{equation}
  {\mathcal L}_i = \mathcal{N}(O- O_{\rm exp}, \Delta O) 
                 = \exp\left( \frac{-(O-O_{\rm exp})^2}{2(\Delta O)^2} \right) \, .
  \label{Linterval}
\end{equation}
Here $\Delta O$ is the uncertainty at $1\sigma$.  
For combining experimental and theoretical uncertainties, we add them in quadrature. 
When $O_{\rm exp}$ is a (one-sided) limit at a given CL, it is less straightforward 
to account for the experimental uncertainty. 
Taking a pragmatic approach, we approximate the likelihood by a smoothed 
step function centered at the 95\% CL limit $O_{\rm exp,\,95\%}$, 
\begin{equation}
  {\mathcal L}_i = \mathbf{F}(O, O_{\rm exp,\,95\%}) 
                 = \frac{1}{1+\exp[\pm(O-O_{\rm exp,\,95\%})/\Delta O]}\,,
  % +: upper bound, -: lower bound
  \label{Llimit}
\end{equation}
with $\Delta O = 1\% \times O_{\rm exp,\,95\%}$. 
The $\pm$ sign in the exponent is chosen depending on whether we are dealing with an upper or lower bound: for an upper bound the plus sign applies, for a lower bound the minus sign. Using a smeared step function rather than a hard cut also helps the MCMC to converge.

Finally, when the $\chi^2$ of the limit is available (this will be the case for the direct detection limits), we compute the likelihood as ${\mathcal L}_i = e^{-\chi_i^2/2}$. 

To carry out the computations, we make use of a number of public tools. 
In particular, we use \texttt{micrOMEGAs 2.6.c}~\cite{Belanger:2008sj,Belanger:2010gh} for the calculation of the relic density and for direct and indirect detection cross sections. This is linked to an appropriately modified~\cite{Belanger:2010cd} version of \texttt{SuSpect 2.4}~\cite{Djouadi:2002ze} for the calculation of the sparticle (and Higgs) spectrum. Decays of the Higgs bosons are computed using a modified version of \texttt{HDECAY 4.40}~\cite{Djouadi:1997yw}, and 
Higgs mass limits are evaluated with \texttt{HiggsBounds 3.6.1beta}~\cite{Bechtle:2008jh,Bechtle:2011sb}. Regarding the computation of the direct detection limits, we  make use of a private code described in section~\ref{sec:directdetection}.

%-------------------------------------------------------------------------------
\subsection{Parameters of the model}
%-------------------------------------------------------------------------------

We parametrize the model with 12 parameters as follows. 
The sneutrino sector is fixed by three parameters per generation (the two mass eigenvalues $m_{\tilde{\nu}_{1}}$, $m_{\tilde{\nu}_{2}}$ and the mixing angle $\sin\theta_{\tilde{\nu}}$, or the soft breaking parameters $m_{\widetilde L}$,  $m_{\widetilde N}$, $\anu$) plus $\tan\beta$.  
Assuming degeneracy between electron and muon sneutrinos, this gives seven parameters to scan over.  
The soft term for the LH sneutrino, $m_{\widetilde L}$, also defines the mass of the LH charged slepton (of each generation); the remaining free parameter in the slepton sector is $m_{\widetilde R}$, the soft mass of the RH charged slepton, which we fix by  
$m_{\widetilde R}=m_{\widetilde L}$ for simplicity. 

The chargino--neutralino sector is described by the gaugino mass parameters $M_1$, $M_2$ and the higgsino mass parameter $\mu$. Moreover, we need the gluino soft mass $M_3$. Motivated by gauge coupling unification, we assume [approximate] GUT relations for the gaugino masses,  $M_3=3M_2=6M_1$,\footnote{This assumption  is central when applying the gluino mass limits from LHC searches.}
so we have $M_2$ and $\mu$ as two additional parameters in the scan. 
For stops/sbottoms we assume a common mass parameter $m_{03}\equiv m_{\widetilde Q_3}=m_{\widetilde U_3}=m_{\widetilde D_3}$, which we allow to vary together with $A_t$ (other trilinear couplings are neglected). The masses of the 1st and 2nd generation squarks, on the other hand, are fixed at 2~TeV without loss of generality.
Finally, we need the pseudoscalar Higgs mass $M_A$ to fix the Higgs sector. 
The model parameters and their allowed ranges are summarized in Table~\ref{tab:modelpars}. 

\begin{table}[t]
\begin{center}
\begin{tabular}{|c|c|c|c|}
\hline
$i$     & Parameter     & \multicolumn{2}{c|}{Scan bounds}\\
        & $p_i$       & light sneutrinos & HND sneutrinos \\
\hline\hline
1  & $m_{\tilde{\nu}_{1\tau}}$ & $[1,\,M_Z/2]$ & $[M_Z/2,\,1000]$ \\
\hline
2  & $m_{\tilde{\nu}_{2\tau}}$ & $[m_{\tilde{\nu}_{1\tau}}\!+1,\,3000]$ & $[m_{\tilde{\nu}_{1\tau}}\!+1,\,3000]$ \\
\hline
3  & $\sin\theta_{\tilde{\nu}_{\tau}}$ & $[0,\,1]$ & $[0,\,1]$ \\
\hline
\hline
4  & $m_{\tilde{\nu}_{1e}} = m_{\tilde{\nu}_{1\mu}}$ & $[m_{\tilde{\nu}_{1\tau}}\!+1,\,M_Z/2]$ & $[m_{\tilde{\nu}_{1\tau}}\!+1,\,3000]$ \\
\hline
5  & $m_{\tilde{\nu}_{2e}} = m_{\tilde{\nu}_{2\mu}}$ & $[m_{\tilde{\nu}_{1e}}\!+1,\,3000]$ & $[m_{\tilde{\nu}_{1e}}\!+1,\,3000]$ \\
\hline
6  & $\sin\theta_{\tilde{\nu}_{e}} = \sin\theta_{\tilde{\nu}_{\mu}}$ & $[0,\,1]$ & $[0,\,1]$ \\
\hline
\hline
7  & $\tan \beta$ &  \multicolumn{2}{c|}{$[3,\,65]$} \\
\hline
8  & $\mu$ &  \multicolumn{2}{c|}{$[-3000,\,3000]$} \\
\hline
9  & $M_2 = 2M_1 = M_3/3$ &  \multicolumn{2}{c|}{$[30,\,1000]$} \\
\hline
10 & $m_{\widetilde{Q}_3} = m_{\widetilde{U}_3} = m_{\widetilde{D}_3}$ &  \multicolumn{2}{c|}{$[100,\,3000]$} \\
\hline
11 & $A_t$ &  \multicolumn{2}{c|}{$[-8000,\,8000]$} \\
\hline
12 & $M_A$ &  \multicolumn{2}{c|}{$[30,\,3000]$} \\
\hline
\end{tabular}
\caption{Parameters and scan ranges for the light and the heavy non-democratic (HND) sneutrino cases. All masses and the $A$-term are given in GeV units. In the heavy democratic (HD) case, the same bounds as in the HND case are applied for quantities $i=1\mbox{--}3$ and $7\mbox{--}12$, but entries  $4\mbox{--}6$ are computed from  $m_{\widetilde{N}_e} \in m_{\widetilde{N}_\tau}\!\pm 5\%$, 
$m_{\widetilde{L}_e} \in m_{\widetilde{L}_\tau}\!\pm 5\%$,  and 
$A_{\tilde{\nu}_e} \in A_{\tilde{\nu}_\tau} \pm 5\%$,  with a flat distribution, see text.
\label{tab:modelpars}}
\end{center}
\end{table}

The requirement of having enough sneutrino annihilation to achieve $\Omega h^2 \simeq 0.11$ while having a low enough scattering cross section off protons and neutrons to pass the DD limits, together with the constraints from the $Z$ invisible width, splits the parameter space into two disconnected regions with sneutrinos lighter or heavier than $M_Z/2$ (or more precisely, as we will see, $\mlsp\lesssim 7$~GeV and $\mlsp \gtrsim 50$~GeV). We call this the ``light''  and ``heavy'' cases in the following.
 
In the  ``light'' case, we assume that the $\tau$-sneutrino is the LSP, but the $e/\mu$ sneutrinos are not too different in mass from the $\tau$-sneutrino. More specifically, we assume that $m_{\tilde{\nu}_{1e}}$ lies within $[m_{\tilde{\nu}_{1\tau}}\! + 1$~GeV,~$M_Z/2]$, i.e.\ the tau sneutrino is the LSP and all the three sneutrinos are potentially in the region sensitive to the constraint on the invisible decays of the $Z$ boson. The 1~GeV minimal mass splitting is a quite natural assumption considering the sensitivity of $\mlsp$ to small variations in $\anu$, and suppresses co-annihilation effects (note that the degenerate case was previously studied in~\cite{Belanger:2010cd}).\footnote{We also performed MCMC sampling allowing $m_{\tilde{\nu}_{1e}}>M_Z/2$ up to 3~TeV, keeping only the $\tilde{\nu}_{1\tau}$ light, but the conclusions remain unchanged. So we will present our results only for the case $m_{\tilde{\nu}_{1\tau}}<m_{\tilde{\nu}_{1e}}<M_Z/2$.}

In the ``heavy'' case, we distinguish two different scenarios. First, in analogy to the light case, we assume that the $\tau$-sneutrino is the LSP, with $m_{\tilde{\nu}_{1\tau}} \in [M_Z/2,\,1000~{\rm GeV}]$, and we allow $m_{\tilde{\nu}_{1e}}$ to vary within $[m_{\tilde{\nu}_{1\tau}}\!+1,\,3000]$~GeV. We call this the ``heavy non-democratic'' (HND) case in the following. 
Second, we also consider a ``heavy democratic'' (HD) case, in which $m_{\tilde{\nu}_{1}}$, $m_{\tilde{\nu}_{2}}$ and $\sin\theta_{\tilde{\nu}}$ of the 3rd and the 1st/2nd generation are taken to be close to each other. As before,   
we use $m_{\tilde{\nu}_{1\tau}}$, $m_{\tilde{\nu}_{2\tau}}$ and $\sin\theta_{\tilde{\nu}_\tau}$ as input parameters, from which we compute $m_{\widetilde{N}_\tau}$, $m_{\widetilde{L}_\tau}$ and $A_{\tilde{\nu}_\tau}$.  
For the 1st/2nd generation, we then take 
$m_{\widetilde{N}_e} \in [m_{\widetilde{N}_\tau}\!-5\%,\,m_{\widetilde{N}_\tau}\!+5\%]$, 
$m_{\widetilde{L}_e} \in [m_{\widetilde{L}_\tau}\!-5\%,\,m_{\widetilde{L}_\tau}\!+5\%]$, and 
$A_{\tilde{\nu}_e} \in [A_{\tilde{\nu}_\tau}\!-5\%,\,A_{\tilde{\nu}_\tau}\!+5\%]$ 
with a flat distribution. 
This way either ${\tilde{\nu}_{1\tau}}$ or ${\tilde{\nu}_{1e,\mu}}$ can be the LSP; moreover ${\tilde{\nu}_{1\tau}}$ and ${\tilde{\nu}_{1e,\mu}}$ can be almost degenerate. 
In the latter case, co-annihilations have a sizable effect.\footnote{Note that if the electron/muon/tau sneutrinos are co-LSPs, this has important consequences for 
the relic density~\cite{Belanger:2010cd}. The $e,\mu,\tau$ sneutrino mass hierarchy moreover has important consequences for the LHC phenomenology (more electrons and muons instead of tau leptons from cascade decays), and for the annihilation channels for indirect detection signals. Furthermore, for a very light $\tau$-sneutrino, $m_{\tilde{\nu}_{1\tau}} < m_{\tau} \simeq 1.78$~GeV, annihilation into a pair of tau leptons is kinematically forbidden, while for $\tilde{\nu}_{1e,\mu}$ of the same mass annihilations into electrons or muons would be allowed.}
Nevertheless it turns out that the results for the HND and HD setups are almost the same, so we will take the HND scenario as our standard setup for the heavy case, see Table~\ref{tab:modelpars}, and discuss only what is different in the HD case.

%--------------------------------------------------------------------------------
\subsection{Nuisance parameters}
%--------------------------------------------------------------------------------
\label{nuisanceparams}

Nuisance parameters are experimentally determined quantities which are not of immediate interest to the analysis but which induce a non-negligible uncertainty in the (model) parameters which we want to infer. 
The Bayesian approach allows us to deal easily with nuisance parameters. 
In order to account for experimental uncertainties impacting the results, we choose 10 nuisance parameters, listed in Table~\ref{tab:nuipar}. They fall into three categories: astrophysical parameters (related to dark matter searches), nuclear uncertainties (related to the computation of the DM-nucleon scattering cross section) and Standard Model uncertainties.

In order to compute limits from direct detection experiments, we need to know the properties of the dark matter halo of our galaxy. We assume a Standard Halo Model, taking into account variations of the velocity distribution ($v_0$, $v_{\rm esc}$) and of the local dark matter density ($\rho_{\rm DM}$). To this end, we follow~\cite{McCabe:2010zh} and take the naive weighted average of the quoted values for each parameter (an alternative determination of $\rho_{\rm DM}$ can be found in Ref.~\cite{Catena:2009mf,Catena:2011kv,Bovy:2012tw}). Note that considering $v_0$ and $v_{\rm esc}$ as nuisance parameters is particularly important in the light DM case, because of its sensitivity to the tail of the velocity distribution; indeed a departure from the canonical value $v_0 = 220$~km/s may have a sizable impact on the direct detection limits at low masses. 

Turning to nuclear uncertainties, the Higgs exchange contribution to the elastic scattering cross section depends on the quark contents of the nucleons. The light quark contents can be determined via the ratio of the masses of the light quarks, $m_u/m_d$ and $m_s/m_d$, and the light-quark sigma term $\sigma_{\pi N} = (m_u + m_d) \langle N | \bar{u}u + \bar{d}d | N \rangle/2$.  Moreover, we need the strange quark content of the nucleon,  $\sigma_s = m_s \langle N | \bar{s}s | N \rangle$, which is actually the main source of uncertainty here. We take the latest  results for $\sigma_{\pi N}$ and $\sigma_s$ from lattice QCD~\cite{Thomas:2012tg}. 
We stress that the new direct determinations of $\sigma_s$ lead to a much lower value as
compared to previous estimates based on octet baryon masses and SU(3) symmetry breaking effect.

The Standard Model uncertainties that we include as nuisance parameters in the MCMC sampling are $m_t$, the top pole mass, $m_b(m_b)$, the bottom mass at the scale $m_b$ in the $\overline{\rm MS}$ scheme, and $\alpha_s(M_Z)$, the strong coupling constant at the scale $M_Z$. They impact the derivation of the SUSY and Higgs spectrum. Moreover, the mass of the bottom quark is relevant in the light sneutrino case because if $m_{\tilde{\nu}_{1\tau}} < m_b$, annihilation into $b\bar{b}$ is kinematically forbidden. 

\begin{table}[t]
\begin{center}
\begin{tabular}{|c|c|c|c|}
\hline
$i$     & Nuisance parameter     & Experimental result     & Likelihood function \\
        & $\lambda_i$       & $\Lambda_i$                   &  $\mathcal{L}_i$ \\
\hline\hline
1  & $m_u/m_d$ & $0.553 \pm 0.043$~\cite{Leutwyler:1996qg} & Gaussian \\
\hline
2  & $m_s/m_d$ & $18.9 \pm 0.8$~\cite{Leutwyler:1996qg} & Gaussian \\
\hline
3  & $\sigma_{\pi N}$ & $44 \pm 5$~MeV~\cite{Thomas:2012tg} & Gaussian \\
\hline
4  & $\sigma_s$ & $21 \pm 7$~MeV~\cite{Thomas:2012tg} & Gaussian \\
\hline
5  & $\rho_{\rm DM}$ & $0.3 \pm 0.1$~GeV/cm$^3$~\cite{Weber:2009pt} & Weighted Gaussian average \\
   &                 & $0.43 \pm 0.15$~GeV/cm$^3$~\cite{Salucci:2010qr} & \\
   &                 & $\Rightarrow 0.34 \pm 0.09$~GeV/cm$^3$ & \\
\hline
6  & $v_0$ & $242 \pm 12$~km/s~\cite{Ghez:2008ms} & Weighted Gaussian average \\
   &       & $239 \pm 11$~km/s~\cite{Gillessen:2008qv} & \\
   &       & $221 \pm 18$~km/s~\cite{Koposov:2009hn} & \\
   &       & $225 \pm 29$~km/s~\cite{McMillan:2009yr} & \\
   &       & $\Rightarrow 236 \pm 8$~km/s & \\
\hline
7  & $v_{\rm esc}$ & $550 \pm 35$~km/s~\cite{Smith:2006ym} & Gaussian \\
\hline
8  & $m_t$ & $173.3 \pm 1.1$~GeV~\cite{tevatron:1900yx} & Gaussian \\
\hline
9 & $m_b(m_b)$ & $4.19^{+0.18}_{-0.06}$~GeV~\cite{Nakamura:2010zzi} & Two-sided Gaussian \\
\hline
10 & $\alpha_s(M_Z)$ & $0.1184 \pm 0.0007$~\cite{Nakamura:2010zzi} & Gaussian \\
\hline
\end{tabular}
\caption{\label{tab:nuipar} Nuisance parameters in the scan. The values of the astrophysical parameters are taken from Ref.~\cite{McCabe:2010zh}.}
\end{center}
\end{table}

%--------------------------------------------------------------------------------
\subsection{Experimental constraints entering the likelihood}
%--------------------------------------------------------------------------------
\label{expconst}

We confront our model with the observables listed in Table~\ref{tab:const}. 
Below we comment on the various constraints. 
 
\begin{table}[t]
\begin{center}
\begin{tabular}{|c|c|c|c|}
\hline
$i$     & Observable     & Experimental result     & Likelihood function \\
        & $\mu_i$        & $D_i$                   &  $\mathcal{L}_i$ \\
\hline\hline
1  & $\Omega h^2$ & $0.1123 \pm 0.0035$~\cite{Komatsu:2010fb} & Gaussian \\
    & & (augmented by 10\% theory uncertainty) & \\
\hline
2  & $\sigma_N$ & $\left(m_{\rm DM},\sigma_N\right)$ constraints from & $\mathcal{L}_{2} = e^{-\chi^2_{\rm DD}/2}$ \\
   &                   & XENON10~\cite{Angle:2011th}, XENON100~\cite{Aprile:2011hi}, & \\
   &                   & CDMS~\cite{Ahmed:2010wy} and CoGeNT~\cite{Aalseth:2011wp} & \\
\hline
3  & $\Delta\Gamma_Z$ & $< 2$~MeV (95\% CL)~\cite{CERN-EP-2000-016} & $\mathcal{L}_{3} =\mathbf{F}(\mu_3, 2\rm{\ MeV})$ \\
\hline
4  & Higgs mass & from  & $\mathcal{L}_4 = 1$ if allowed \\
   & limits   & {\tt HiggsBounds 3.6.1beta}~\cite{Bechtle:2008jh,Bechtle:2011sb} & $\mathcal{L}_4 = 10^{-9}$ if not\; \\
\hline
5  & $m_{\tilde{\chi}^{+}_1}$ & $> 100$~GeV~\cite{lepsusywgCH} & $\mathcal{L}_5 = 1$ if allowed \\
   &                          &                                & $\mathcal{L}_5 = 10^{-9}$ if not\; \\
\hline
6  & $m_{\tilde{e}_R} = m_{\tilde{\mu}_R}$ & $> 100$~GeV~\cite{lepsusywgSL} & $\mathcal{L}_6 = 1$ if allowed \\
   &                                       &                                & $\mathcal{L}_6 = 10^{-9}$ if not\; \\
\hline
7  & $m_{\tilde{\tau}_1}$ & $> 85$~GeV~\cite{lepsusywgSL} & $\mathcal{L}_7 = 1$ if allowed \\
   &                      &                               & $\mathcal{L}_7 = 10^{-9}$ if not\; \\
\hline
8  & $m_{\tilde{g}}$ & $> 750, \, 1000$~GeV~\cite{Aad:2011ib,cmsrazor2011} & not included \\
   &                 & or none                                             & (a posteriori cut) \\
\hline
9 & ${\cal B}(b \rightarrow s\gamma)$ & $(3.55 \pm 0.34) \times 10^{-4}$~\cite{Asner:2010qj,Misiak:2006zs} & Gaussian \\
\hline
10 & ${\cal B}(B_s \rightarrow \mu^+\mu^-)$ & $< 1.26 \times 10^{-8}$~(95\% CL)~\cite{cmsblhcbbsmumu,Akeroyd:2011kd} & $\mathbf{F}(\mu_{10}, 1.26 \times 10^{-8})$ \\
\hline
11  & $\Delta a_{\mu}$ & $(26.1 \pm 12.8) \times 10^{-10}$~\cite{Hagiwara:2011af,Bennett:2006fi,Stockinger:2006zn} & Gaussian \\
\hline
\end{tabular}
\caption{\label{tab:const} Experimental constraints used to construct the likelihood. 
Where relevant, experimental and theoretical uncertainties are added in quadrature; 
in particular for $\Omega h^2$ we assume an overall uncertainty of 
$(0.0035^2 + 0.01123^2)^{1/2}=0.0118$.}
\end{center}
\end{table}

%--------------------------------------------------------------------------------
\subsubsection{Relic density of sneutrinos}\label{sec:relic}
%--------------------------------------------------------------------------------

We assume the standard freeze-out picture for computing the sneutrino relic abundance. 
The main annihilation channels for mixed sneutrino dark matter are  
{\it i)}~$\lsp\lsp\to \nu\nu$ ($\lsp^*\lsp^*\to \bar\nu\bar\nu$)
through neutralino t-channel exchange,  
{\it ii)}~$\lsp\lsp^* \to f\bar{f}$ through s-channel $Z$ exchange, and
{\it iii)}~$\lsp\lsp^* \to b\bar{b}$ through s-channel exchange of a light Higgs.   
Moreover, if the $\lsp$ is heavy enough, it can also annihilate into $W^+W^-$ (dominant),
$ZZ$ or $t\bar t$.  Note that for the heavy LSP the annihilation into neutrino pairs is always much suppressed while the annihilation into other channels can be enhanced by the heavy scalar Higgs resonance.

The annihilation into neutrino pairs proceeds mainly through the wino component of the 
t-channel neutralino and is proportional to $\sin^4\theta_{\tilde\nu}$; it is largest for light winos. 
The $Z$ exchange is also proportional to $\sin^4\theta_{\tilde\nu}$. 
The light Higgs exchange, on the other hand, is proportional to $(\anu\sn)^2$.  
The dependence of $\Omega h^2$ on the sneutrino mass and mixing angle has been analyzed in \cite{Belanger:2010cd,Thomas:2007bu}. 

We assume a 10\% theory uncertainty on $\Omega h^2$, mostly to account for unknown higher-order effects.  
In the light DM cases, one also has to worry about the change in the number of effective degrees of freedom in the early Universe,  $g_{\rm eff}$, especially when $m_{\rm DM} \approx 20 \, T_{\rm QCD}$. While we do take into account the change of $g_{\rm eff}$ in the calculation of the relic density, the uncertainty related to it is not accounted for separately. Rather, we assume that it falls within the overall 10\% theory uncertainty.  (We discuss the issue of $g_{\rm eff}$ in more detail in Appendix~\ref{aTQCD}.)

The same annihilation channels will be relevant for indirect DM detection experiments,  looking for gamma-rays ($Fermi$-LAT, H.E.S.S.),   charged particles (positrons, antiprotons; PAMELA, $Fermi$-LAT, AMS) 
or neutrinos (Super-Kamiokande, IceCube, ANTARES),   
that could be produced by annihilation of dark matter, especially in high density regions, see Section~\ref{sec:indirect}.

%--------------------------------------------------------------------------------
\subsubsection{Direct detection limits}\label{sec:directdetection}
%--------------------------------------------------------------------------------

The spin-independent (SI) scattering of $\lsp$ on nucleons occurs
through $Z$ or Higgs exchange.  The $Z$ exchange is again suppressed
by the sneutrino mixing angle, while the Higgs exchange is enhanced by
the $\anu$ term.  A peculiarity of the $Z$-exchange contribution is
that the proton cross section is much smaller than the neutron one,
with the ratio of amplitudes $f_p/f_n=(1-4\sin^2\theta_W)$.  The Higgs
contribution on the other hand, which becomes dominant for large
values of $\anu$, is roughly the same for protons and neutrons.
The total SI cross section on a nucleus $N$ is obtained after averaging over the $\lsp N$
and ${\lsp}^*N$ cross sections, where we assume equal numbers of
sneutrinos and anti-sneutrinos. We note that the interference between
the $Z$ and $h^0$ exchange diagrams has opposite sign for $\lsp N$ and
${\lsp}^*N$, leading to an asymmetry in sneutrino and anti-sneutrino
scattering if both $Z$ and Higgs exchange are important.
All these effects are taken into account when we compute the normalized scattering 
cross section $\sigma_N$:
\begin{equation}
  \sigma_N = \frac{4 \mu_\chi^2}{\pi}\frac{\left( Z f_p+ (A-Z)f_n\right)^2}{A^2} \,,
\end{equation}
where $\mu_\chi$ is the sneutrino--nucleon reduced mass, $Z$ is the atomic number and $A$ the mass number.
This cross section can be directly compared to the experimental limits on $\sigma_p^{\rm SI}$, which are extracted from the observed 
limits on the LSP--nucleus scattering cross section assuming $f_p=f_n$.

We consider the limits coming from various direct detection experiments.  In
particular, we take into account the light dark matter results from
XENON10~\cite{Angle:2011th} and CDMS~\cite{Ahmed:2010wy}, as well as the
latest XENON100~\cite{Aprile:2011hi} and CoGeNT~\cite{Aalseth:2011wp}
results. Thus, we are using the best limits from both low and high mass
regions, with Xenon (XENON10/100) and Germanium (CDMS and CoGeNT) detectors.
We include the data from these experiments using a private code based on
Refs.~\cite{Kopp:2011yr, Schwetz:2011xm, HerreroGarcia:2012fu}, where
further details on the analysis can be found. For XENON100 we adopt the
best-fit light-yield efficiency $L_{\rm eff}$ curve from
\cite{Aprile:2011hi}. Especially for the low DM mass region, the energy
resolution close to the threshold is important. We take into account the
energy resolution due to Poisson fluctuations of the number of single
electrons. The XENON10 analysis is based on the so-called S2 ionization
signal which allows to go to a rather low threshold. In this case we follow
the conservative approach of \cite{Angle:2011th} and impose a sharp cut-off
of the efficiency below the threshold, which excludes the possibility of
upward fluctuations of a signal from below the threshold. Our analysis tries
to approximate as close as possible the one performed in
\cite{Angle:2011th}.  From CDMS we use results from an analysis of Ge data
with a threshold as low as 2~keV~\cite{Ahmed:2010wy}. We use the binned data
from Fig.~1 of \cite{Ahmed:2010wy} and build a $\chi^2$, where we only take
into account bins where the predicted rate is larger than the observed data.
This ensures that only an upper bound is set on the cross section. We
proceed for CoGeNT in a similar way. We ignore the possibility that hints
for an annual modulation in CoGeNT are due to DM (see also
\cite{Ahmed:2012vq}), and use a similar $\chi^2$ method as for CDMS to set
an upper bound on the scattering cross section. The code allows for a
consistent variation of the astrophysical parameters $v_0$, $v_{\rm esc}$
and $\rho_{\rm DM}$ for all considered experiments. 

The information from DD is included in the Bayesian analysis in the
following way. For XENON10 and XENON100 data, we apply the so-called
maximum-gap method \cite{Yellin:2002xd} to calculate an upper bound on the
scattering cross section for a given mass. The probability returned by the
maximum-gap method as a function of the model parameters as well as
astrophysical parameters (appropriately normalized) is considered as the
likelihood function which then is converted into the posterior PDF within
the Bayesian analysis. This is an approximation to a pure Bayesian treatment
with the advantage that it allows us to use the maximum-gap method, which
offers a conservative way to set a limit in the presence of an unknown
background. Since the shape of the expected background distribution is
neither provided for XENON10 nor XENON100, it is not possible to construct a
``true'' likelihood from the data and we stick to the above mentioned
approximation based on the maximum-gap method.\footnote{In
\cite{Bertone:2011nj} XENON100 data has been implemented in a Bayesian study
by constructing a likelihood function from the Poisson distribution based on
the total number of expected signal and background events. We have checked
that such a procedure leads to similar results as our approach based on the
maximum-gap method.} For CDMS and CoGeNT, the likelihood is obtained from
the individual $\chi^2$ functions as $\mathcal{L} \propto \exp(-\chi^2/2)$.
The method to construct the $\chi^2$ described in the previous paragraph
amounts to introducing the unknown background in each bin $i$ as a nuisance
parameter $b_i$ which is allowed to vary by maximizing the likelihood
function under the condition $b_i \ge 0$. Again this is an approximation to
a pure Bayesian approach (in which the posterior PDF would be integrated
over the nuisance parameters), which suffices for our purpose.

%--------------------------------------------------------------------------------
\subsubsection{$Z$ invisible width}
%--------------------------------------------------------------------------------

A light sneutrino with $m_{\tilde\nu}<M_Z/2$ will contribute to the invisible width of the $Z$ boson, well measured at LEP~\cite{CERN-EP-2000-016}, thus putting a constraint on the sneutrino mixing:
\begin{equation}
    \Delta\Gamma_Z=\sum_{i=1}^{N_f} \Gamma_\nu\, \frac{\sin^4\theta_{\tilde{\nu}_i}}{2} 
      \left(1-\left(\frac{2 m_{\tilde{\nu}_i}}{M_Z} \right)^2 \right)^{3/2}< 2~{\rm MeV}
\end{equation}
where $\Gamma_\nu= 166$~MeV is the partial width into one neutrino flavor.
For one light sneutrino with $\mlsp=5$ (20) GeV, this leads only to a mild constraint on the  
mixing angle of $\sn<0.39$ ($0.43$). 
For $m_{\tilde{\nu}_{1\tau}} = 4$~GeV, $m_{\tilde{\nu}_{1e}} = m_{\tilde{\nu}_{1\mu}} = 5$~GeV and assuming a common mixing angle, this constraint becomes stricter: $\sin \theta_{\tilde{\nu}} < 0.3$.

On the other hand, a minimum amount of mixing is needed for light $\lsp$'s to achieve large enough annihilation cross section. In \cite{Belanger:2010cd} we found $\sn\gtrsim 0.12$ for LSP masses above the $b$-threshold, where annihilation into $b\bar{b}$ through $Z$ or $h^0$ can contribute significantly, 
and $\sn\gtrsim 0.25$ for $m_{\lsp}<m_b$. Therefore, for light sneutrinos, the mixing angle should be not far from the limit imposed by the $Z$ invisible width. Such a large mixing is however in conflict with DD limits unless $m_{\lsp}\lesssim 7$~GeV. For sneutrino LSPs with masses of, roughly, 7--40~GeV, the DD limits constrain $\sn$ to be smaller than about 0.05--0.07, which makes it impossible to achieve low enough $\Omega h^2$. For heavier masses, one needs $m_{\lsp}$ near the Higgs pole or above the $W^+W^-$ threshold to satisfy both DD and relic density constraints. 
This was also discussed in~\cite{Thomas:2007bu}. As mentioned, this splits our parameter space into two distinct regions where the Markov Chains converge, one with $m_{\lsp}\lesssim 7$~GeV and one with $m_{\lsp} > M_Z/2$ (more precisely, $m_{\lsp} \gtrsim 50$~GeV).

%--------------------------------------------------------------------------------
\subsubsection{Higgs and SUSY mass limits}
%--------------------------------------------------------------------------------

In the MCMC sampling, we impose chargino and charged slepton mass limits~\cite{lepsusywgCH,lepsusywgSL} from LEP as listed in Table~\ref{tab:const}.  We here choose conservative values because the LEP analyses in principle assumed a neutralino LSP, and hence the parametrization of the LEP limits in terms of e.g.\ the chargino--neutralino mass difference as implemented in {\tt micrOMEGAs} does not apply. 
To evaluate Higgs mass constraints based on LEP, Tevatron and LHC data, we use {\tt HiggsBounds\,3.6.1beta}. (The latest CMS limit on $A/H \rightarrow \tau\tau$~\cite{cmsAtautau} is also included via \texttt{HiggsBounds}.)
Here note that for large sneutrino mixing, which as detailed above is necessary for light $m_{\lsp}$, the light Higgs mass receives an important negative correction from the sneutrino loop, which is proportional to $|\anu|^4/(m_{\widetilde{\nu}_2}^2 - m_{\widetilde{\nu}_1}^2)^2$. Thus the lower limit on $m_{h^0}$ also somewhat constrains the sneutrino sector. In order to take into account the theoretical uncertainty in $m_{h^0}$, we smear the Higgs mass computed with {\tt SuSpect}  by 
a Gaussian with a width of 1.5~GeV before feeding it to {\tt HiggsBounds}.
In the light sneutrino case, the Higgs decays into sneutrinos are always kinematically allowed, and they are enhanced as $A_{\tilde{\nu}}$; as a result the $h^0$ decays almost completely invisibly in this case. (In the heavy sneutrino case, only a small fraction of the points have $m_{\lsp}<m_{h^0}/2$.) The Higgs decays into sneutrinos are properly taken into account in our {\tt HiggsBounds} interface.

An important point of our study is how SUSY mass limits from the 2011 LHC searches affect the sneutrino DM scenarios. Here note that squarks and gluinos undergo the same cascade decays into charginos and neutralinos as in the conventional MSSM. Since we assume gaugino mass unification, the gluino and squark mass limits derived in the CMSSM limits from jets$+E_T^{\rm miss}$ searches apply to good approximation. We have checked several $\lsp$ LSP benchmark points and found $m_{\tilde g}\gtrsim 750$~GeV for $m_{\tilde q}\sim 2$~TeV based on a fast simulation of the ATLAS 0-lepton analysis for 1~fb$^{-1}$~\cite{Aad:2011ib}. This is in very good agreement with the corresponding gluino mass limit in the CMSSM for large $m_0$. For 5~fb$^{-1}$ of data, this limit should improve to $m_{\tilde g}\gtrsim 1$~TeV. 

A word of caution is in order however. For $m_{\tilde q}\gg m_{\tilde g}$ we expect $\tilde g\to q\bar{q}\tilde\chi^0_{1,2}$ and $\tilde g\to q\bar{q}'\tilde\chi^\pm_{1}$ as in the MSSM with a neutralino LSP. In our model, the $\tilde{\chi}^0_{1,2}$ decay  further into the $\lsp$ LSP; if this decay is direct, $\tilde{\chi}^0_{1,2}\to \nu\lsp$, it is completely invisible. Indeed, the invisible $\tilde{\chi}^0_{1,2}$ decays often have close to 100\% branching ratio. We do not expect however that this has a large effect on the exclusion limits. The situation is different for chargino decays. In most cases, the $\tilde\chi^\pm_{1}$ decays dominantly into a sneutrino and a charged lepton ($e$, $\mu$ or $\tau$, depending on the sneutrino flavor). This can lead to a much larger rate of single lepton or dilepton events. As a consequence, we expect the limits from 0-lepton  jets$+E_T^{\rm miss}$ searches to weaken, while  single lepton or dilepton $+E_T^{\rm miss}$ searches should become more effective than in the CMSSM. Overall, assuming gaugino mass unification, the gluino mass limit should remain comparable to the limit derived in the CMSSM.

A detailed analysis of the SUSY mass limits in the sneutrino DM model is left for a separate work. In the present paper, we are interested in the effect of the LHC pushing the gluino mass limit to $m_{\tilde g}\gtrsim 750$~GeV or $m_{\tilde g}\gtrsim 1000$~GeV, see above. In order to illustrate this effect without having to run the MCMC several times (which would have been too CPU intensive), we apply the gluino mass constraint a posteriori. As we will see, it is only relevant for the light sneutrino case.

%--------------------------------------------------------------------------------
\subsubsection{Low-energy observables}
%--------------------------------------------------------------------------------

Further important constraints on the model come from flavor physics and from the muon anomalous magnetic moment.
Regarding flavor physics constraints, we use the HFAG average value of ${\cal B}(b \rightarrow s\gamma)=(3.55\pm0.24\pm0.09)\times 10^{-4}$~\cite{Asner:2010qj} with a theoretical uncertainty of $0.23\times 10^{-4}$~\cite{Misiak:2006zs}. Moreover, we use the combined LHCb and CMS limit on ${\cal B}(B_s \rightarrow \mu^+\mu^-)$~\cite{cmsblhcbbsmumu}, augmented by a 20\% theory uncertainty (mainly due to $f_{B_s}$) as suggested in~\cite{Akeroyd:2011kd}. After completion of the MCMC runs, a new limit of ${\cal B}(B_s \rightarrow \mu^+\mu^-) < 4.5 \times 10^{-9}$ (95\% CL)~\cite{Aaij:2012ac} became available. We impose this new limit a posteriori, again assuming 20\% theory uncertainty, but the effect of this on the posterior distributions is marginal.\footnote{Effectively, we impose ${\cal B}(B_s \rightarrow \mu^+\mu^-) < 5.4 \times 10^{-9}$ as a hard cut, but we have checked that this makes no difference as compared to reweighing the likelihood according to eq.~(\ref{Llimit}).} 

Regarding the supersymmetric contribution to the anomalous magnetic moment of the muon, $\Delta a^{\rm SUSY}_\mu$, we implement the 1-loop calculation taking into account the mixing between RH and LH $\tilde{\nu}_\mu$.  Then we compare this value to $\Delta a_{\mu} = a^{\rm exp}_\mu - a^{\rm SM}_\mu$, where for $a^{\rm exp}_\mu$ we take the experimental value reported by the E821 experiment~\cite{Bennett:2006fi}, and for $a^{\rm SM}_\mu$ we take the result of Ref.~\cite{Hagiwara:2011af} (note however the slightly lower $a^{\rm SM}_{\mu}$ reported in \cite{Davier:2010nc}). 
Guided by \cite{Stockinger:2006zn} and because of our ignorance of the 2-loop effects involving mixed sneutrinos, we assume a conservative theoretical uncertainty of $10 \times 10^{-10}$.  
This brings us to $\Delta a_{\mu}^{\rm SUSY}=(26.1 \pm 12.8) \times 10^{-10}$ in Table~\ref{tab:const}.

%--------------------------------------------------------------------------------
\subsubsection{Indirect detection of photons and antiprotons}\label{sec:indirect}
%--------------------------------------------------------------------------------

Dwarf Spheroidal galaxies (dSphs) in the Milky Way provide a good probe of DM through the observation of gamma-rays. Although the photon signal is weaker than from the Galactic center, the signal-to-noise ratio is more favorable since dSphs are DM dominated and the background from astrophysical sources is small.  From measurements of the gamma-rays from ten different dSphs~\cite{Ackermann:2011wa}, the {\it Fermi}-LAT collaboration has extracted
an upper limit on the  DM  annihilation cross section in three different channels: $W^+W^-$, $b\bar{b}$, and $\tau^+\tau^-$. 
For this one assumes a NFW~\cite{Navarro:1996gj} dark matter profile. For DM lighter than 40 GeV, both the $b\bar{b}$ and $\tau^+\tau^-$ channels have the sensitivity to probe the canonical DM annihilation cross section, $\sigma v> 3\times 10^{-26}~\rm{cm}^3 /{\rm s}$. We will not use these constraints in the fit but rather  compare our predictions for the annihilation cross section in different channels with the limit provided by {\it Fermi}-LAT. We will see in the next section that this measurement constrains  sneutrino DM in only a few scenarios for three reasons. 
First, for light sneutrinos we have a sizable $\lsp$ ($\lsp^*$) pair annihilation into $\nu\nu$ ($\bar{\nu}\bar{\nu}$), which clearly cannot lead to a photon signal. Second, {\it Fermi}-LAT has not published results for DM particles lighter than 5 GeV, where the bulk of our  light DM sample that survives direct detection constraints lies. Third, {\it Fermi}-LAT's sensitivity is still one order of magnitude above the canonical cross section for DM masses at the electroweak scale or above. 

Annihilation of DM  in the Milky Way will also, after hadronisation of the decay products of SM particles, lead to antiprotons. 
This antiproton flux has been measured by PAMELA~\cite{Adriani:2010rc} and fits rather well the astrophysics background~\cite{Maurin:2006hy}.  There is however a large uncertainty in  the background at low energies (below a few GeV) due to solar modulation effects that are not well known. Furthermore the antiprotons---as well as any other charged particle---propagate 
 through the Galactic halo  and  their energy spectrum at the Earth differs from the one produced at the source.
The propagation model  introduces additional model dependence in the prediction of the antiproton flux from DM annihilation.
As for photons above, we will not use the antiproton flux  as a  constraint in the fit, but compare our predictions  for different propagation model parameters with the measurements of PAMELA.  We will see that the largest flux, and the largest deviation from the background, are observed at low energies when the sneutrino DM has a mass of a few GeV, thus leading to an excess of events for some values of the propagation parameters.

Finally, a comment is in order regarding annihilation into neutrinos. Indeed, neutrino telescopes (Super-Kamiokande, IceCube, ANTARES) may probe sneutrino DM annihilation into neutrinos, e.g.\ from the Galactic Center or from accretion in the Sun. The neutrino flux from annihilation of DM captured by the Sun is determined by the cross section for sneutrino scattering on nucleons discussed in \cite{Belanger:2010cd} and Section~\ref{sec:directdetection}. 
We do not include a possible neutrino signal in this analysis but leave it for a future study. 

%================================================================================
\section{Results}
%================================================================================

Let us now present the results of this analysis. 
As mentioned, for each of the three scenarios which we study,  we run 8 Markov chains with $10^6$ iterations each. 
The distributions of the points in these chains map the likelihood of the parameter space. We hence present our results 
in terms of posterior probability distributions shown in the form of histograms (1-dimensional distributions) with 100 bins and of contour graphs (2-dimensional distributions) with $100 \times 100$ bins. Results based on alternative (logarithmic) priors in the sneutrino sector can be found in Appendix~\ref{logprior}.

%--------------------------------------------------------------------------------
\subsection{Light sneutrino DM with mass below 10 GeV}\label{sec:light}
%--------------------------------------------------------------------------------

\begin{figure}[!ht] 
   \centering
   \includegraphics[width=4.96cm]{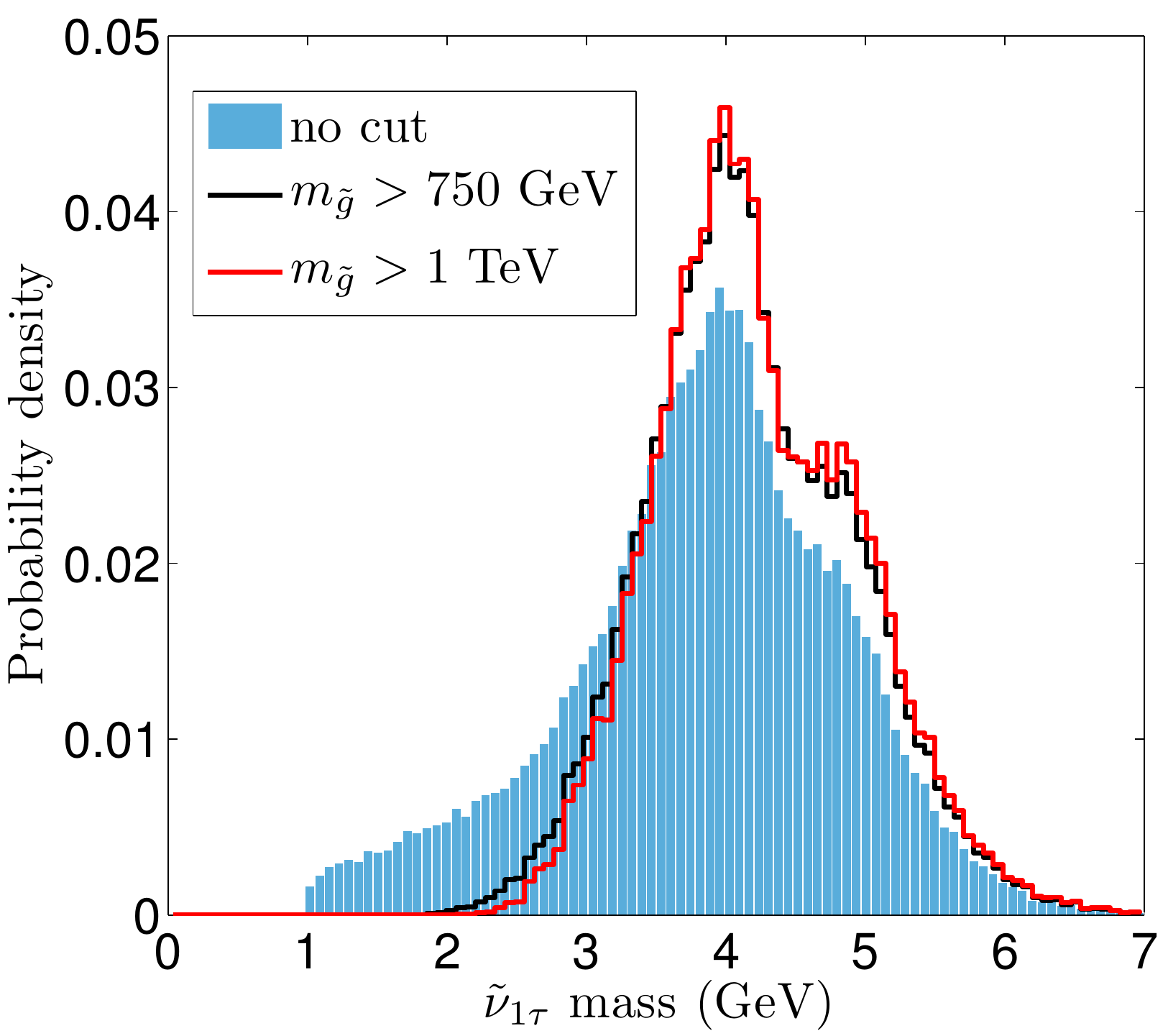}
   \includegraphics[width=4.96cm]{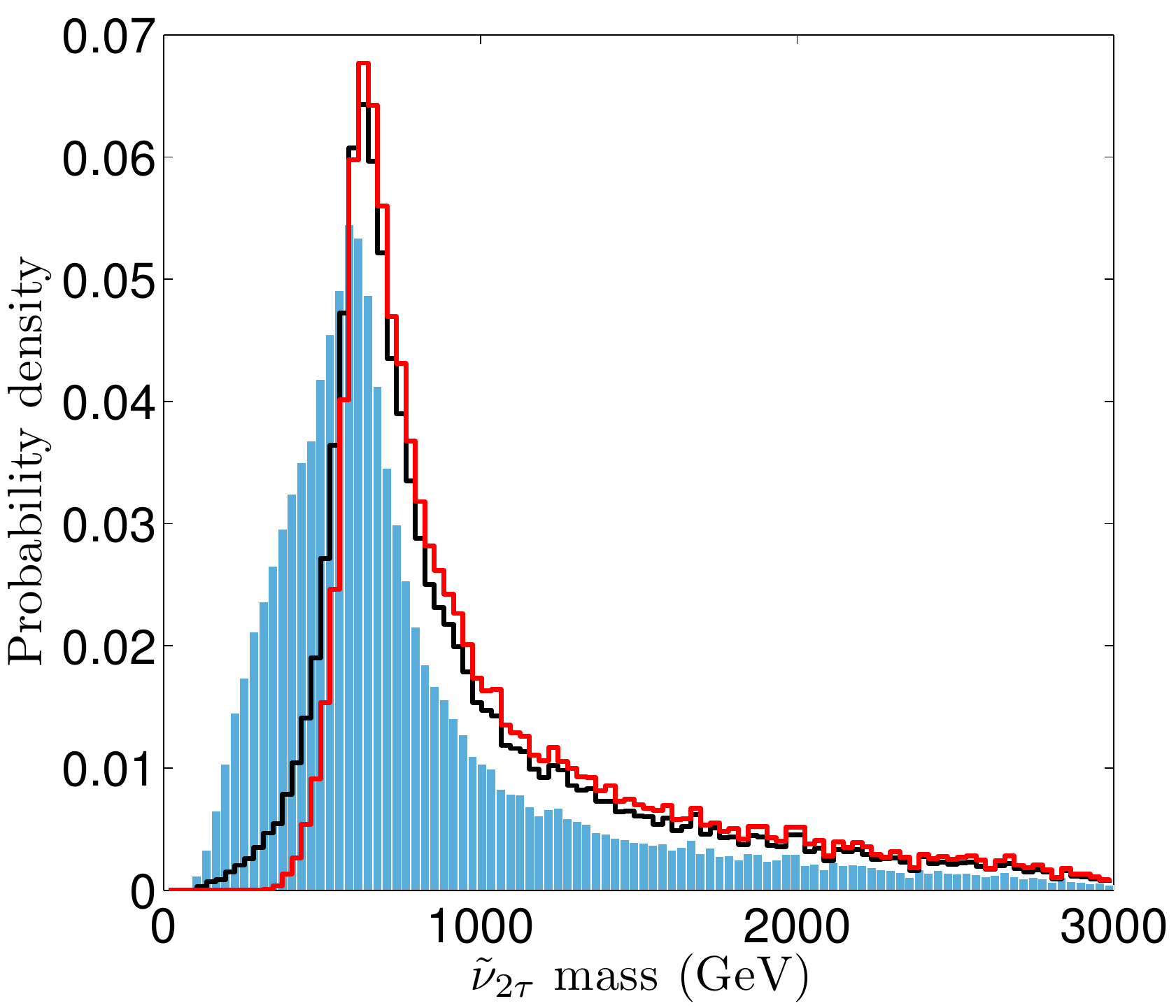}
   \includegraphics[width=4.96cm]{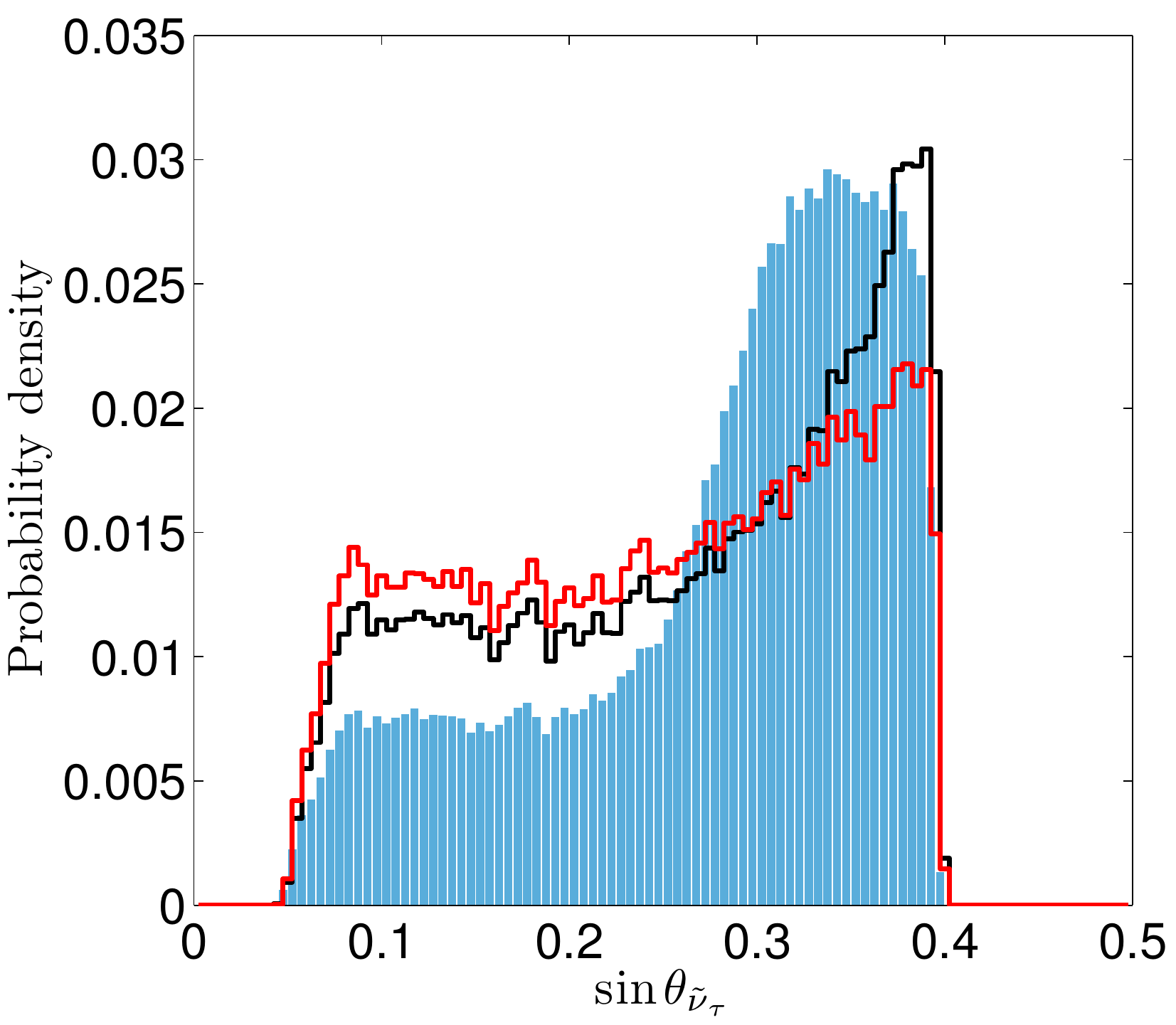}
   \includegraphics[width=4.96cm]{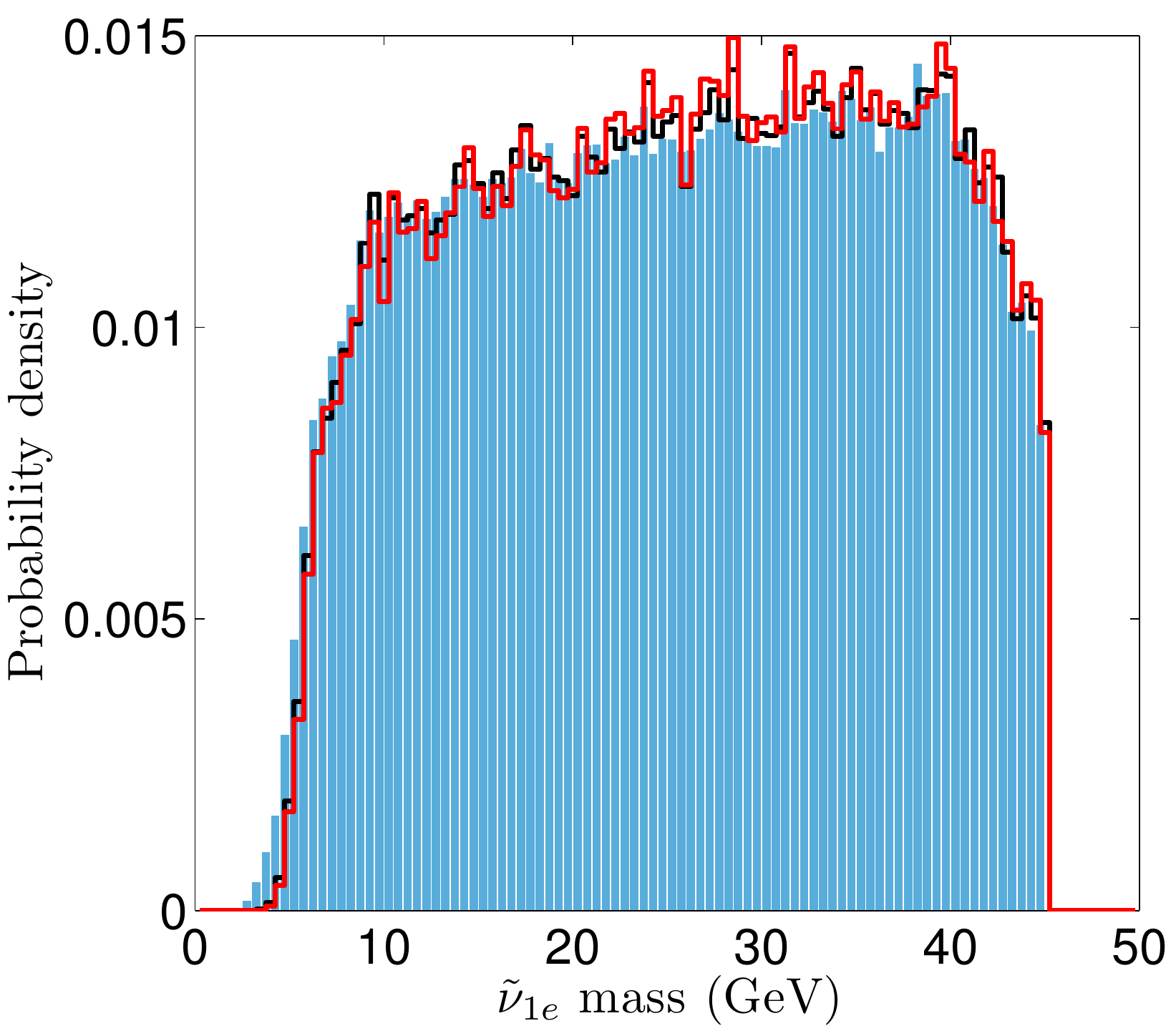}
   \includegraphics[width=4.96cm]{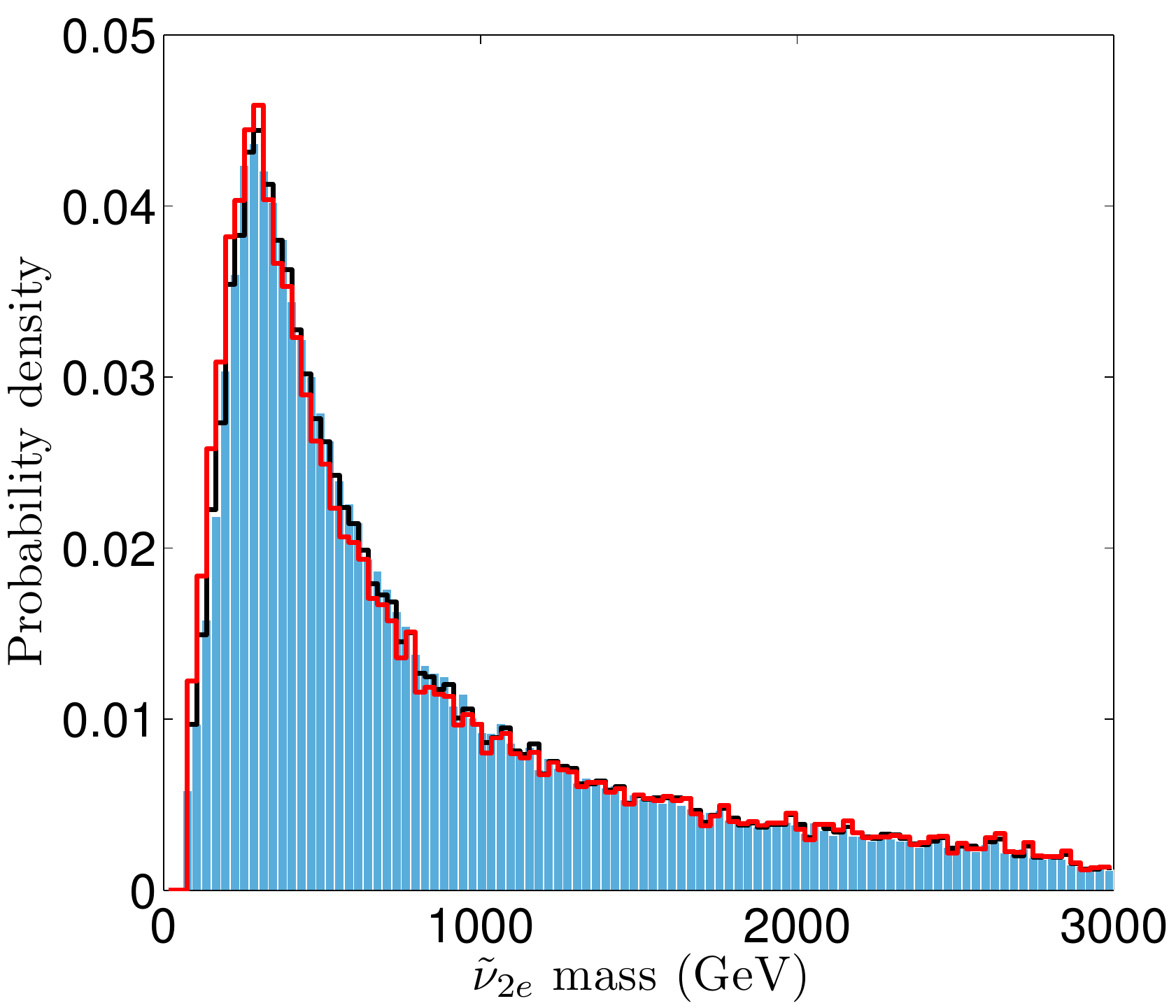}
   \includegraphics[width=4.96cm]{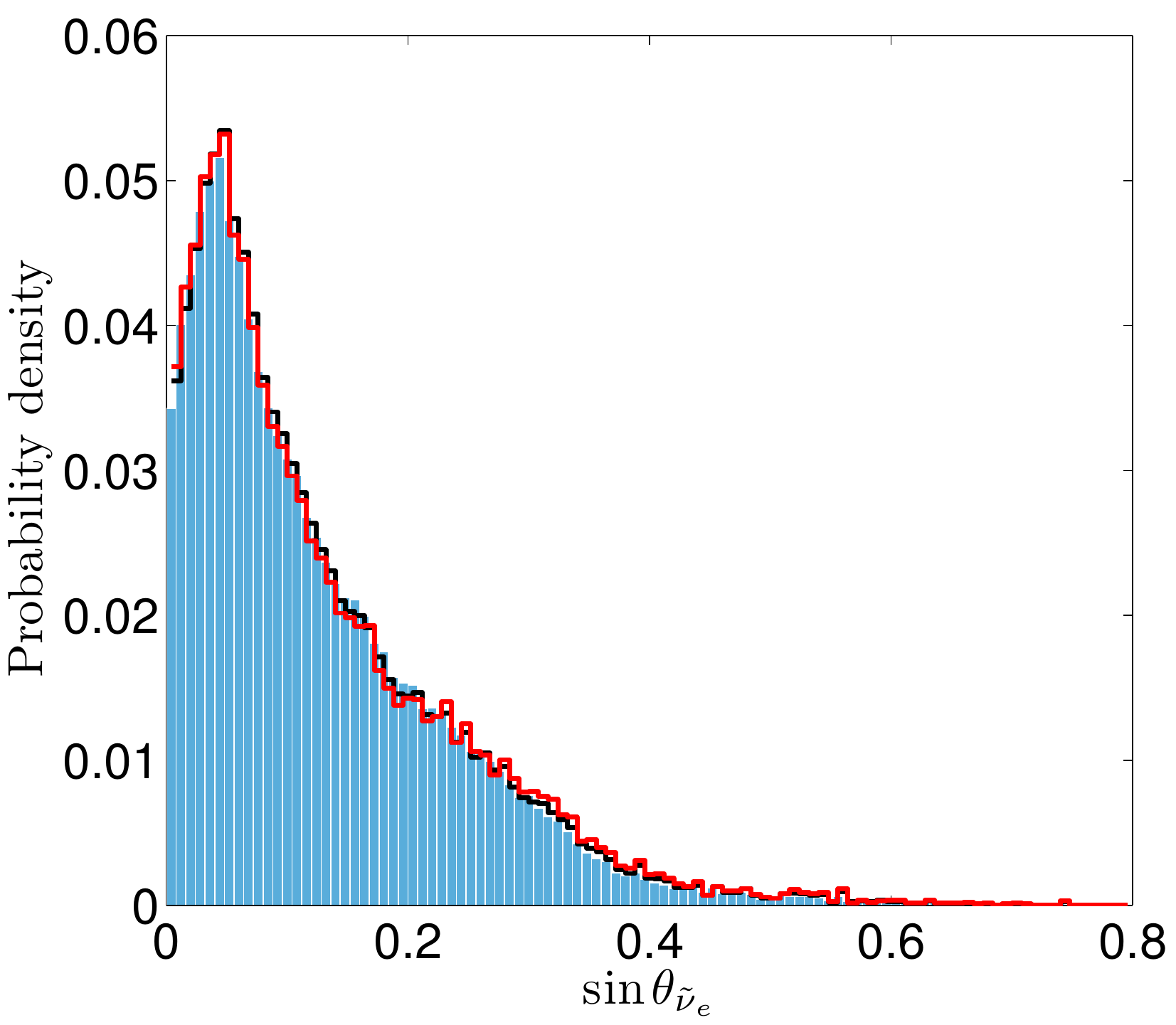}
   \includegraphics[width=4.96cm]{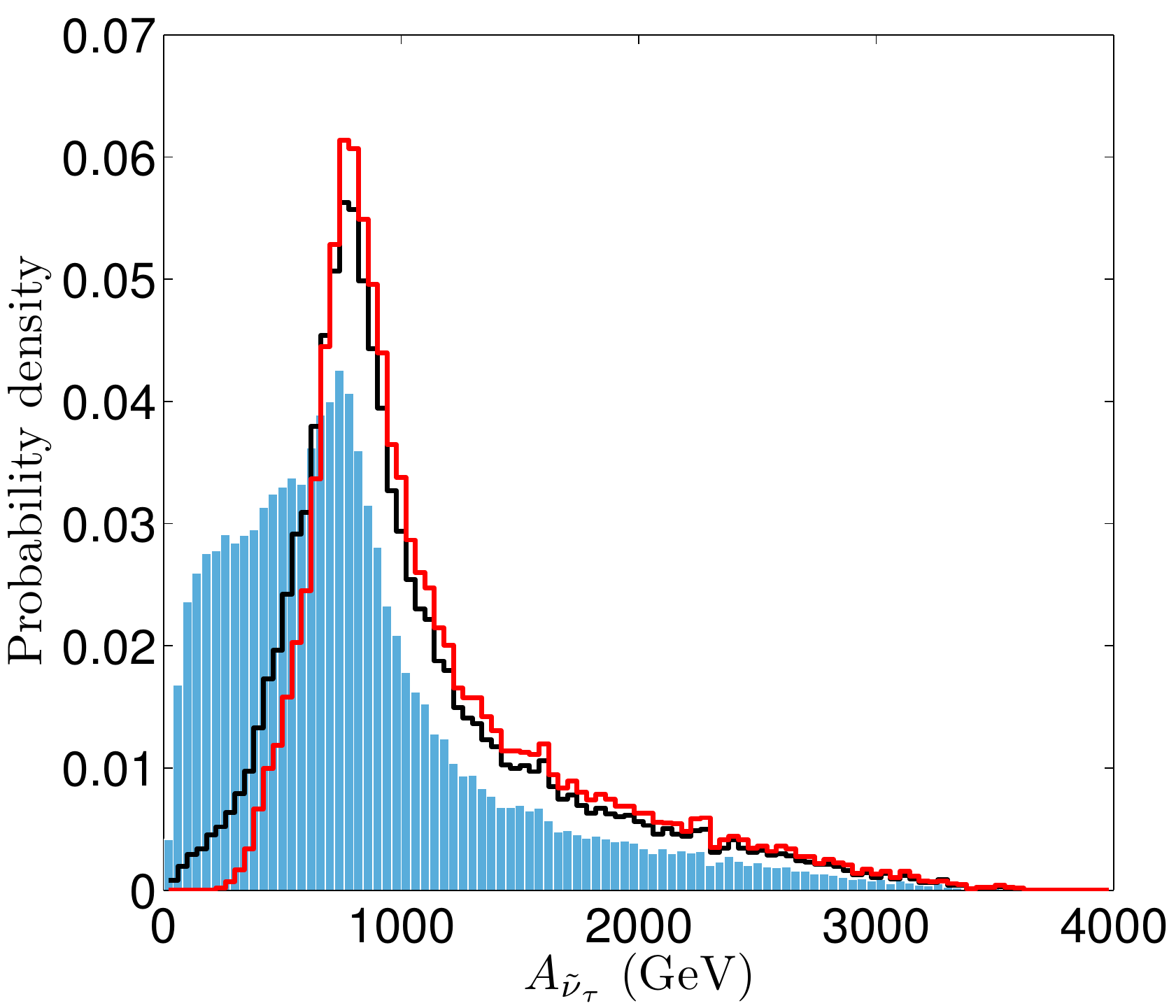}
   \includegraphics[width=4.96cm]{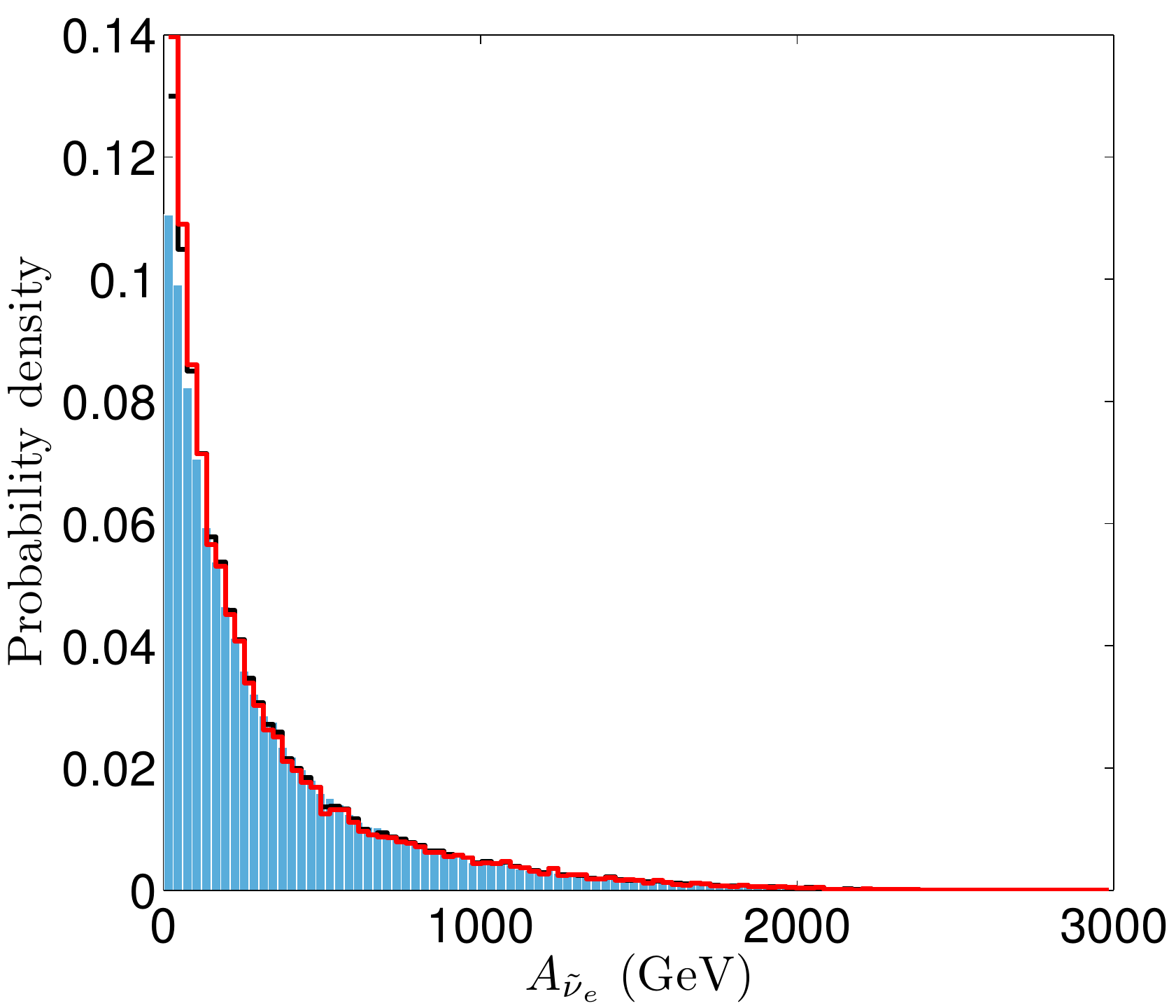}
   \includegraphics[width=4.96cm]{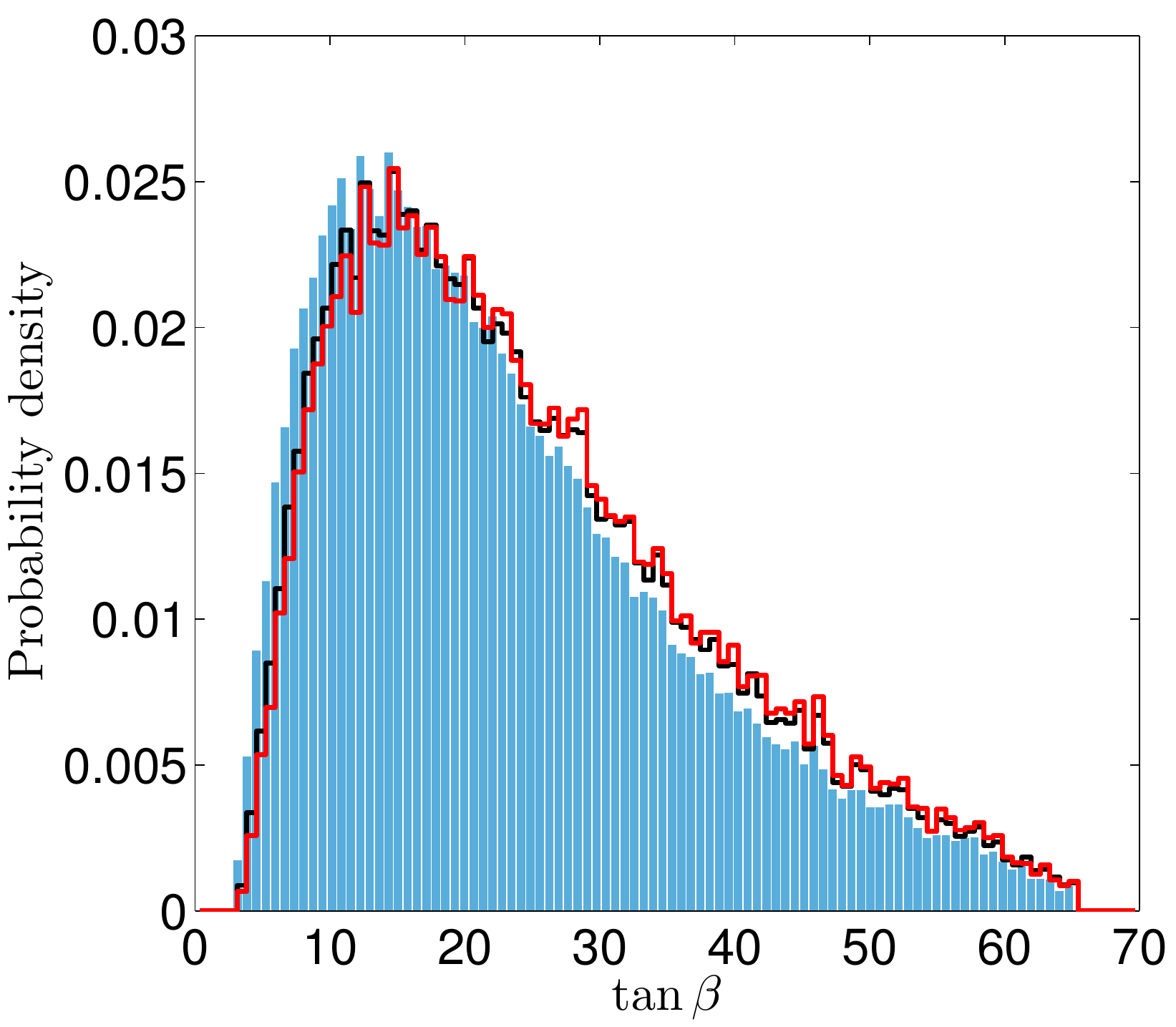}
   \includegraphics[width=4.96cm]{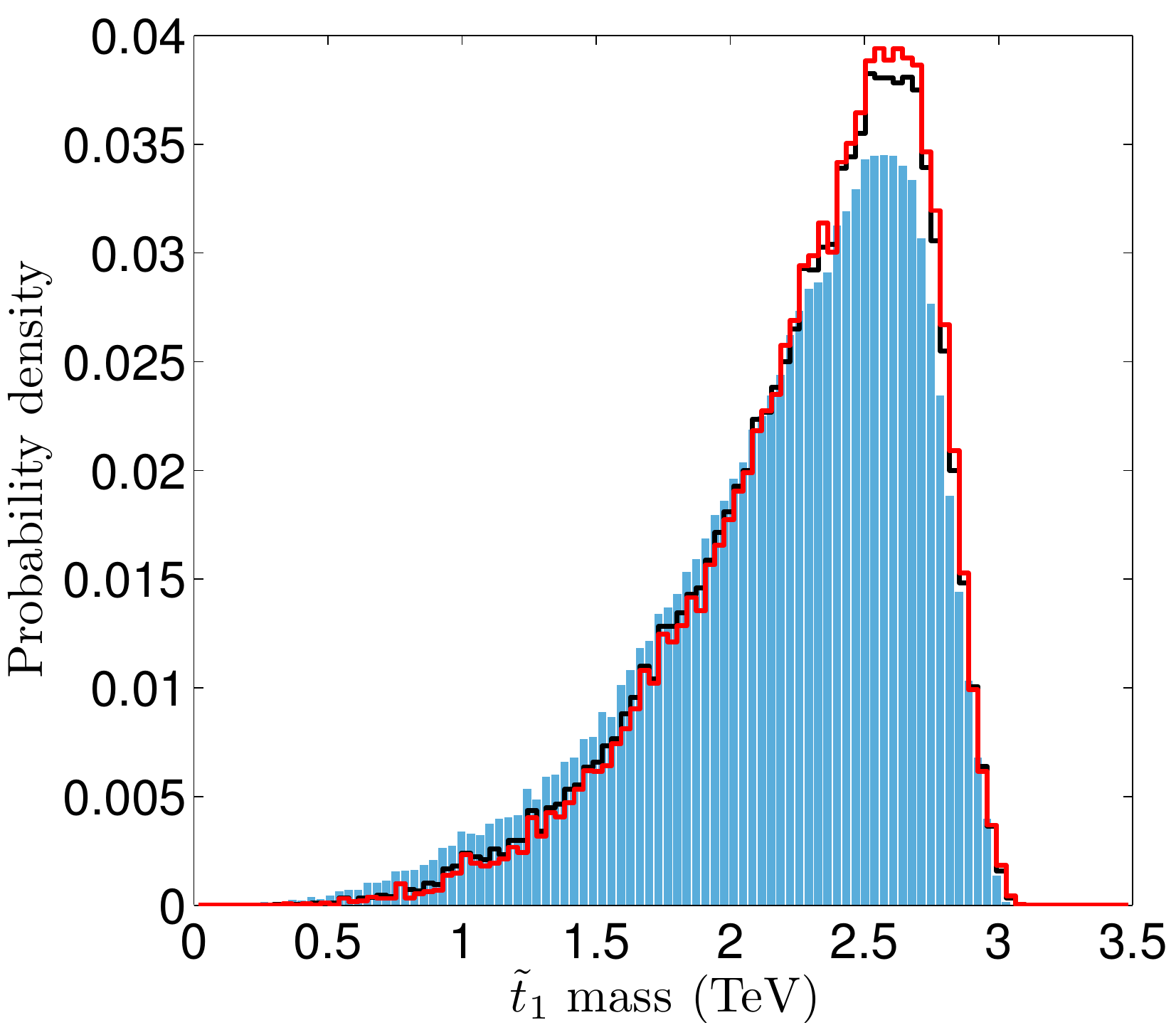}
   \includegraphics[width=4.96cm]{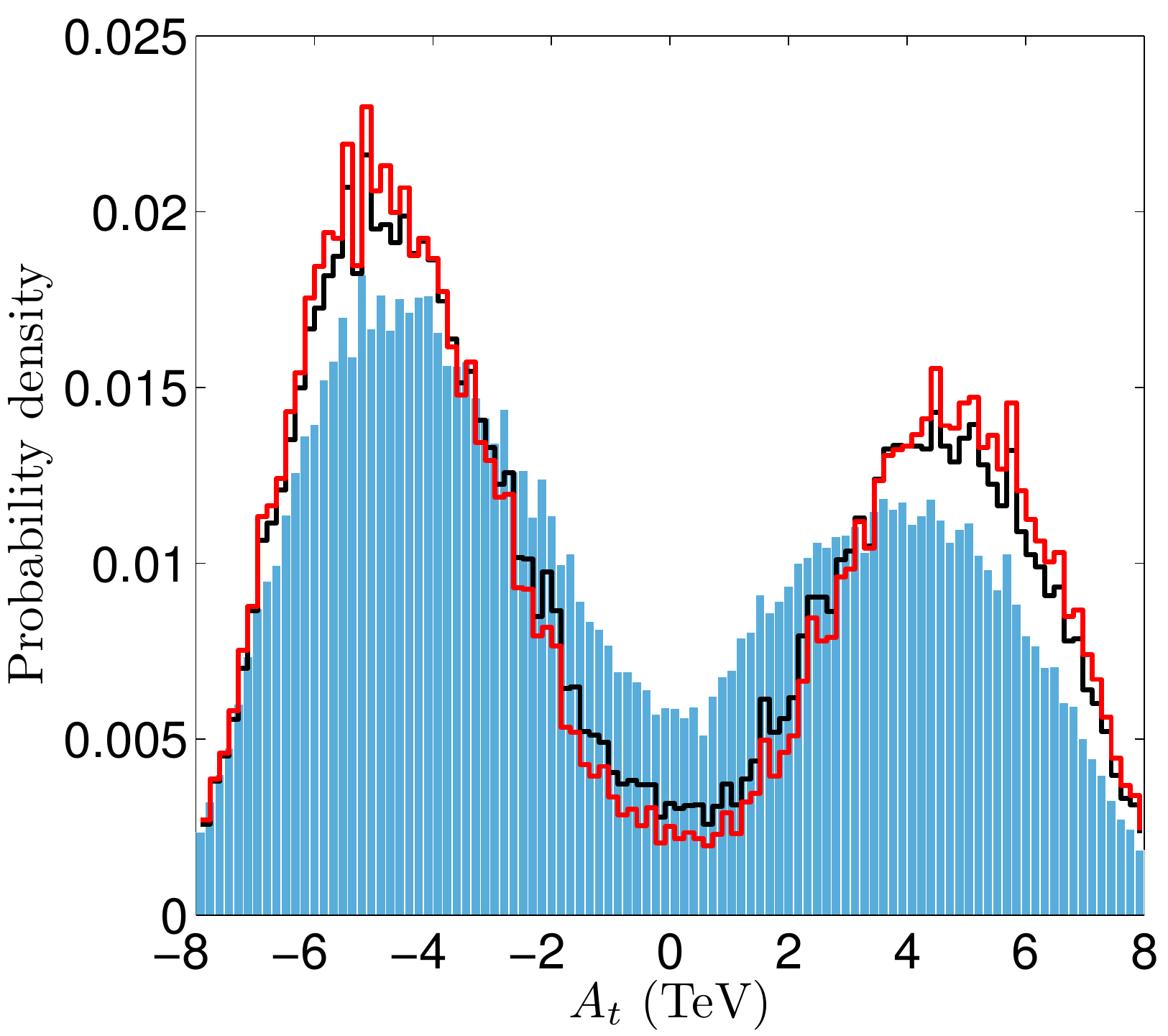}
   \includegraphics[width=4.96cm]{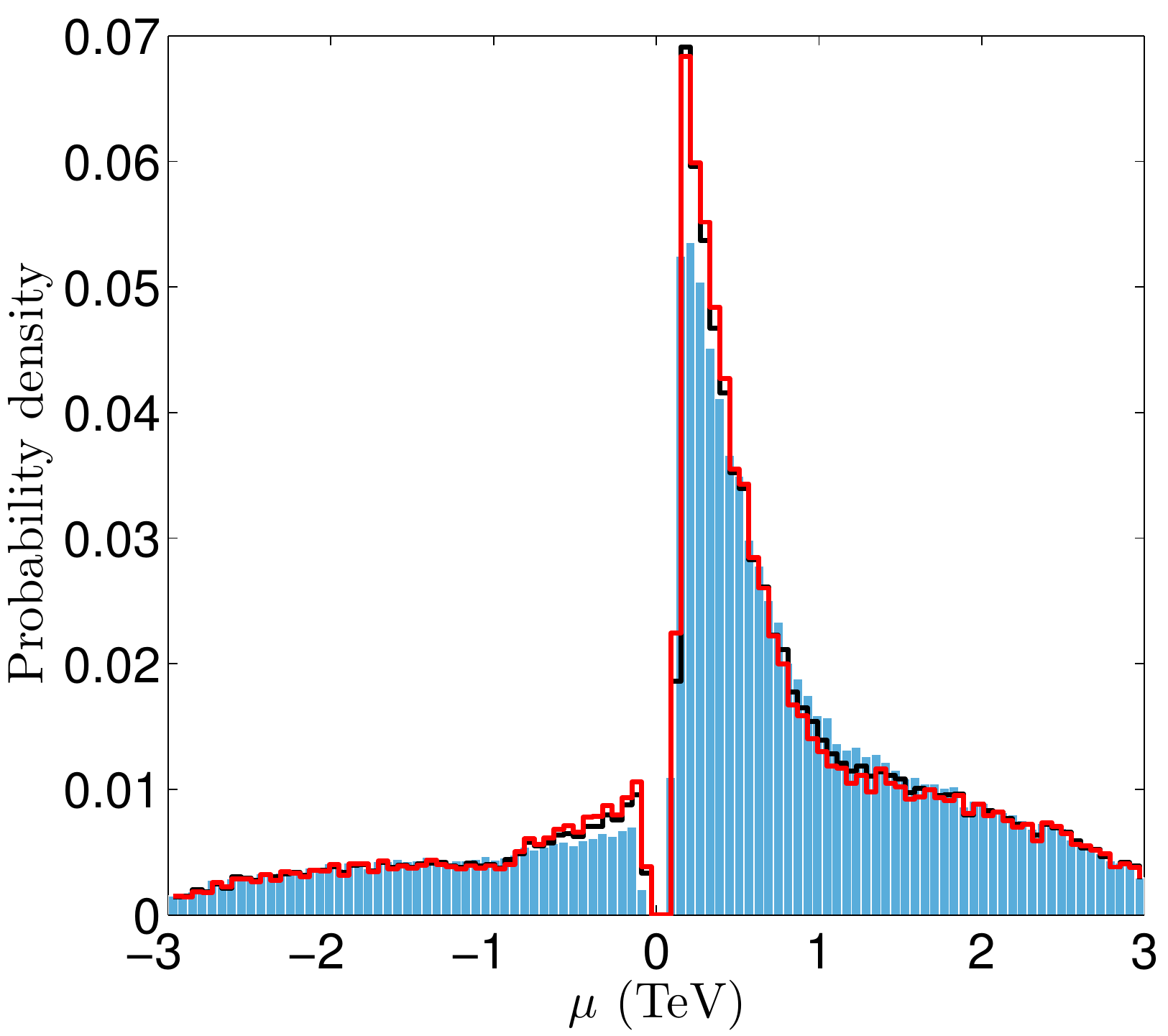}
   \includegraphics[width=4.96cm]{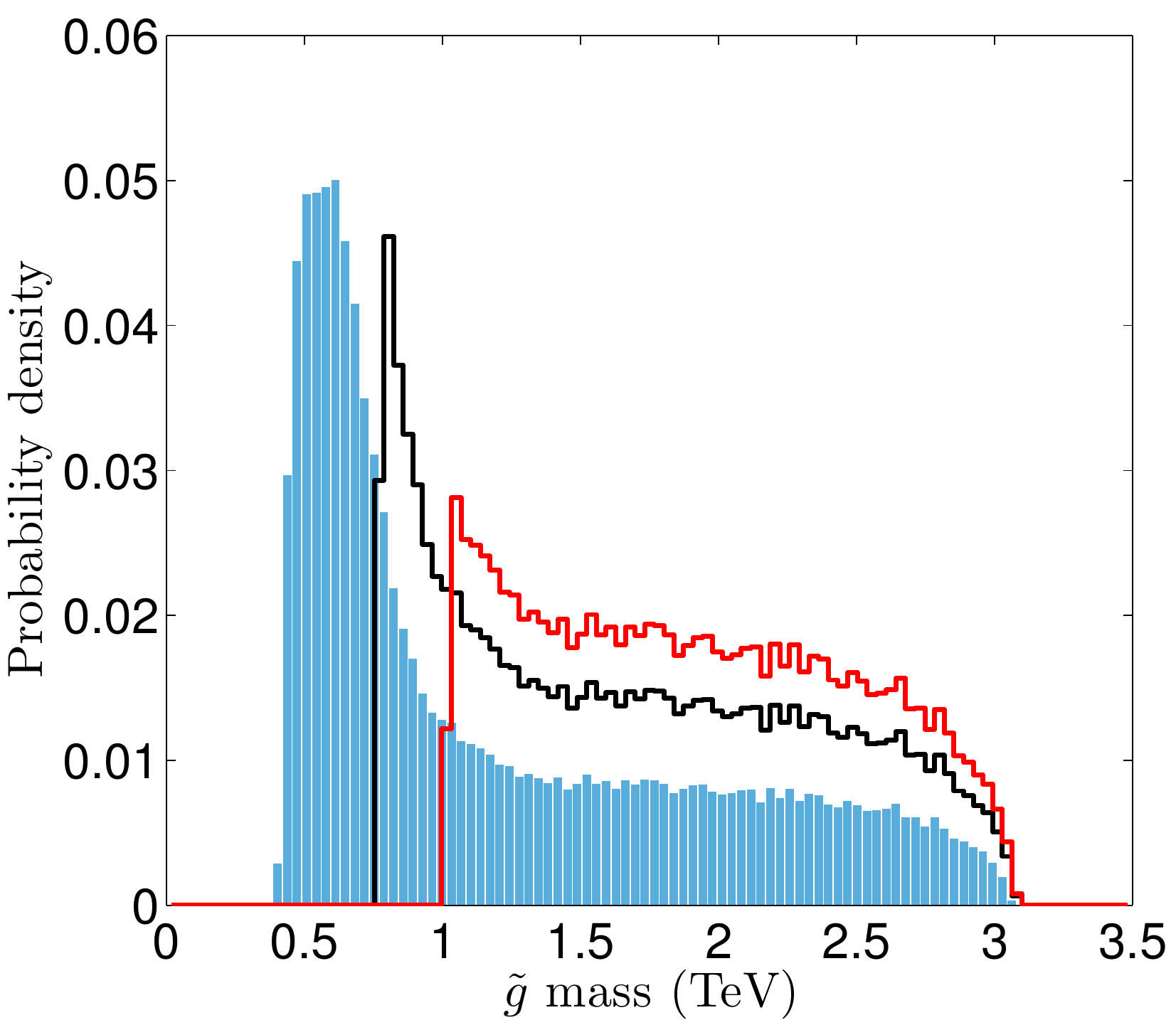}
   \includegraphics[width=4.96cm]{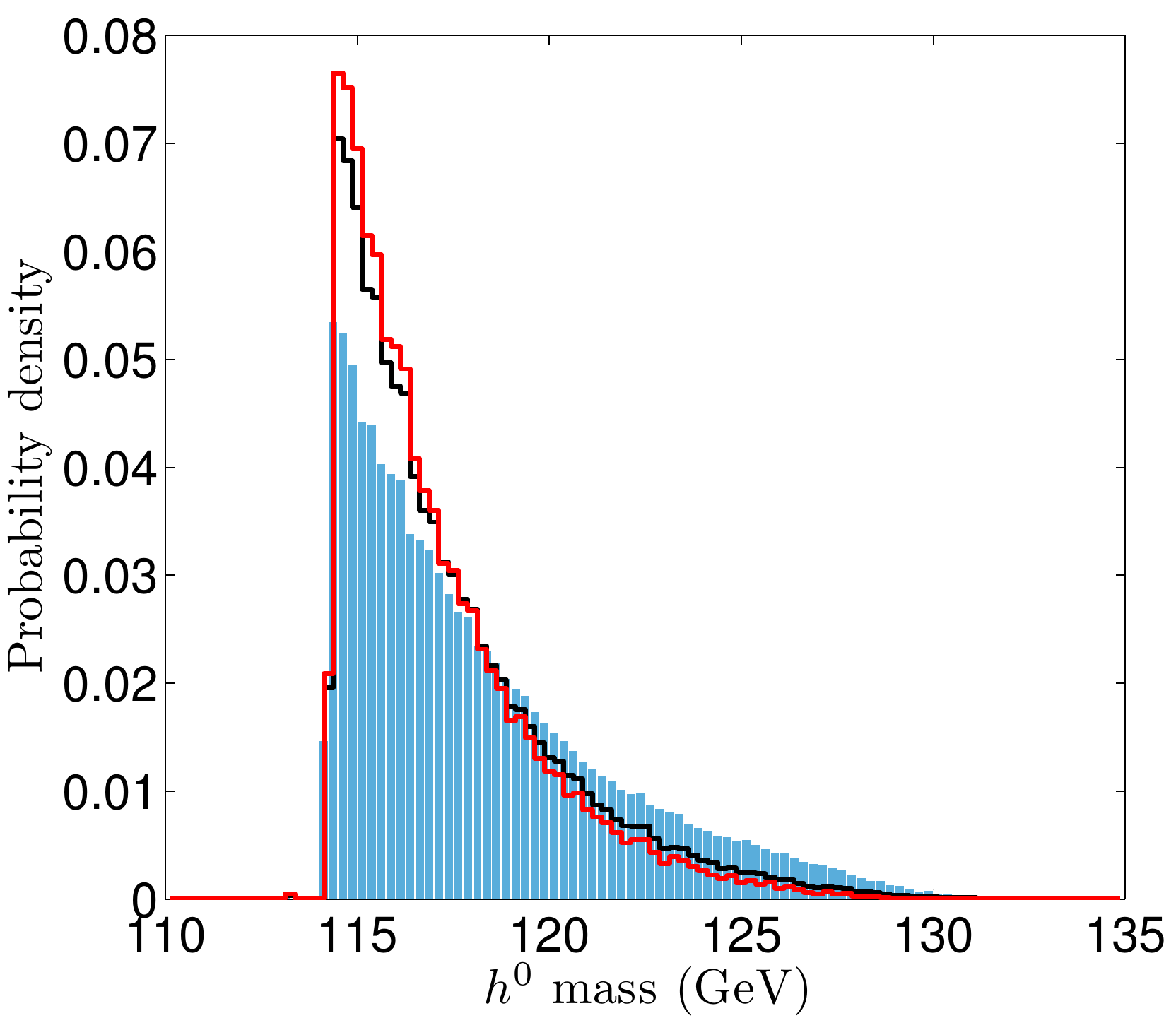}
   \includegraphics[width=4.96cm]{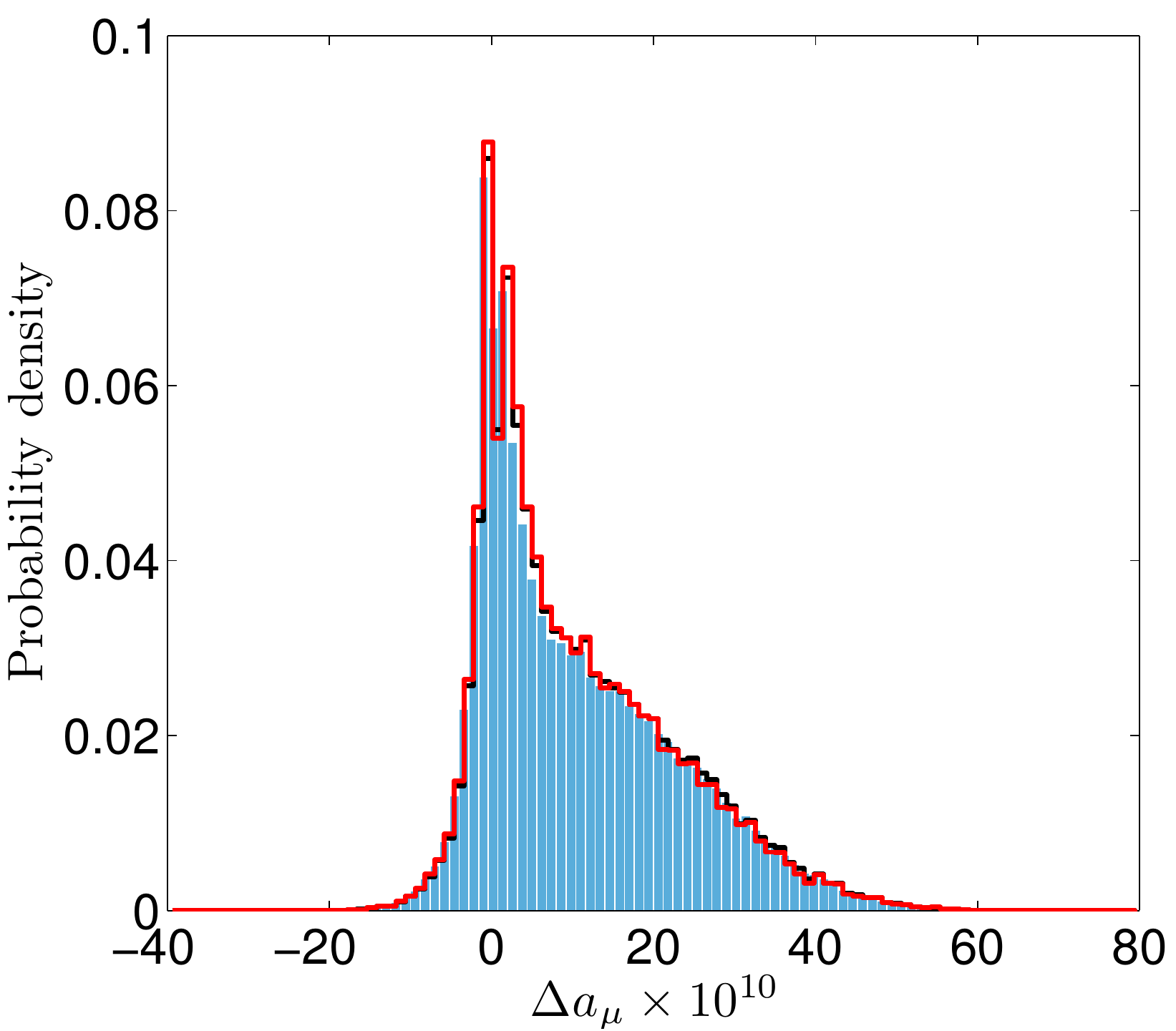}
   \caption{Posterior PDFs in 1D for the light sneutrino case. Specific values for best fit and quasi-mean points as well as the 68\% and 95\% BC intervals are given in Appendix~\ref{aTables}.}
   \label{fig:light-1d}
\end{figure}

We begin with the case of light sneutrinos that was previously studied by some of us in~\cite{Belanger:2010cd}.
Figure~\ref{fig:light-1d} shows the 1-dimensional (1D) marginalized posterior PDFs of various interesting quantities, in particular sneutrino masses and mixing angles, $A$ terms, squarks, gluino and Higgs masses, etc. 
The blue histograms are the posterior PDFs taking into account constraints 1--7 and 9--11 of Table~\ref{tab:const}, 
while the black (red) lines show the posterior distributions after requiring in addition that the gluino be heavier than 750 (1000)~GeV. 
Note that a lower bound on the gluino mass not only cuts the peak of the gluino distribution but  also leads to a lower bound  on the chargino and neutralino masses, since $ 6 m_{\tilde\chi_1^0}\approx 3 m_{\tilde\chi^+}\approx m_{\tilde g}$. 
(We do not show the $m_{\tilde\chi^0_1}$, $m_{\tilde\chi^0_2}$, $m_{\tilde\chi^\pm_1}$ posterior probabilities in Fig.~\ref{fig:light-1d}, because they follow completely the $m_{\tilde g}$ distribution.)

As can be seen, the DD limits, in particular from XENON10, require the sneutrino LSP to be lighter than about 7~GeV, with the distribution peaking around 4~GeV. (The shoulder at $4.5\mbox{--}5$~GeV is due to the onset of the $b\bar b$ annihilation channel.) 
For LSP masses below 4~GeV, the DD limits are not important.  Indeed the  largest cross section, obtained with  the
\clearpage
\noindent maximum value of $\sin\theta_{\tilde{\nu}_{\tau}}$ allowed by the $Z$ invisible width, is below the current experimental limits~\cite{Belanger:2010cd}.
The gluino mass bound from the LHC disfavors very light sneutrinos of about $1\mbox{--}3$~GeV, because the $\lsp\lsp\to \nu\nu$ and $\lsp^*\lsp^*\to \bar\nu\bar\nu$ annihilation channels get suppressed (recall that we assume GUT relations between gaugino masses).  This means one needs to rely on 
annihilation through $Z$ or Higgs exchange, as is reflected in the change of the $\sin\theta_{\tilde{\nu}_{\tau}}$
and $A_{\tilde\nu_\tau}$ probability densities in Fig.~\ref{fig:light-1d}.

The other distributions are basically unaffected by the gluino mass cut, the exceptions being $A_t$ and $m_{h^0}$.  Larger values 
of $A_t$ are  preferred for $m_{\tilde g}>1$~TeV, because it is needed to compensate the negative loop correction to $m_{h^0}$ from the larger $A_{\tilde\nu_\tau}$ in order to still have $m_{h^0}>114$~GeV. Regarding $m_{h^0}$, the distribution is shifted towards the lower limit of 114~GeV because of this negative loop correction. Finally, we note that the light Higgs decays practically 100\% invisibly into sneutrinos. 
Therefore, should the excess in events hinting at a Higgs near 125~GeV be confirmed, the light sneutrino DM scenario would be ruled out. 

Regarding the supersymmetric contribution to $\Delta a_\mu$, shown in the bottom right panel in Fig.~\ref{fig:light-1d}, 
this is peaked towards small values. Nevertheless, the probability of falling within the experimental $1\sigma$  band is sizable, 
$p(\Delta a_\mu=(26.1 \pm 12.8) \times 10^{-10})= 31\%$. The larger values  of  $\Delta a_\mu$ are obtained when there is a large  contribution from the sneutrino exchange diagram.

\begin{figure}[t] 
   \centering
   \includegraphics[width=7cm]{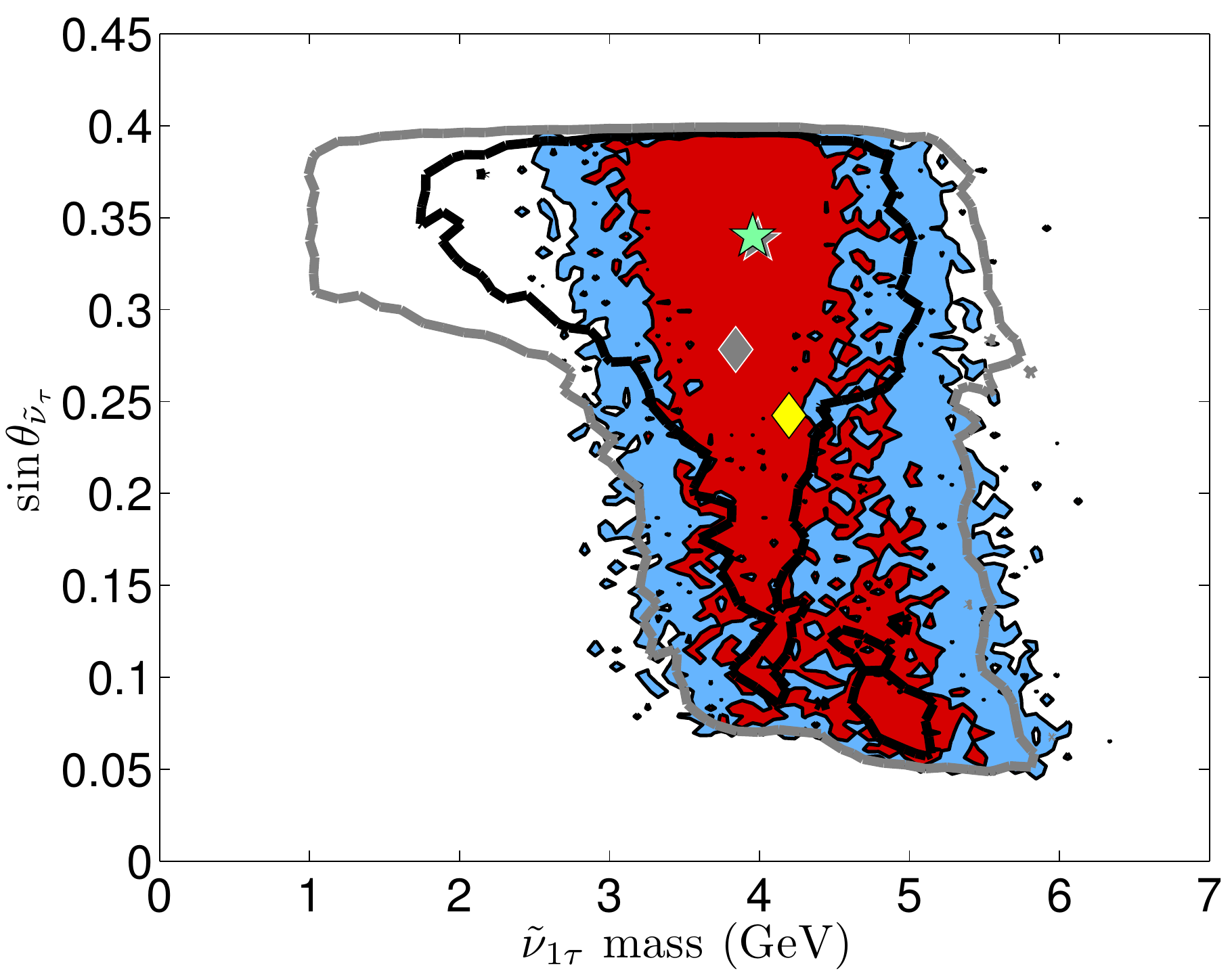}
   \caption{Posterior PDF of $\sin\theta_{\tilde{\nu}_{\tau}}$ versus $m_{\tilde\nu_{1\tau}}$ for the light sneutrino case. The black and grey lines show the 68\% and 95\% BCRs before gluino mass limits from the LHC. The red and blue regions are the 68\% and 95\% BCRs requiring $m_{\tilde g}>1$~TeV. The green star marks the bin with the highest posterior probability after the gluino mass limit, while the
 yellow diamond marks the  mean of the 2D PDF. The grey star/diamond are the highest posterior and  mean points before imposing the gluino mass limit.}
   \label{fig:light-2d-sinth}
\end{figure}

Our expectations regarding the relation between mass and mixing angle are confirmed in Fig.~\ref{fig:light-2d-sinth}, which shows the 2-dimensional (2D) posterior PDF of $\sin\theta_{\tilde{\nu}_{\tau}}$ versus $m_{\tilde\nu_{1\tau}}$. To be more precise, what is shown are the 68\% and 95\% Bayesian credible regions (BCRs) before and after a gluino mass cut of $m_{\tilde g}>1$~TeV. As can be seen, the region of $m_{\tilde\nu_{1\tau}}\approx 1\mbox{--}3$~GeV, which requires $\sin\theta_{\tilde{\nu}_{\tau}}\approx 0.3\mbox{--}0.4$ to be consistent with WMAP, gets completely disfavored by a heavy gluino.\footnote{To be more precise, it gets disfavored by a heavy wino, since $m_{\tilde g}>1$~TeV implies $m_{\tilde\chi^0_2}\gtrsim 300$~GeV in our model.} 

\begin{figure}[!t] 
   \centering
   \includegraphics[width=7cm]{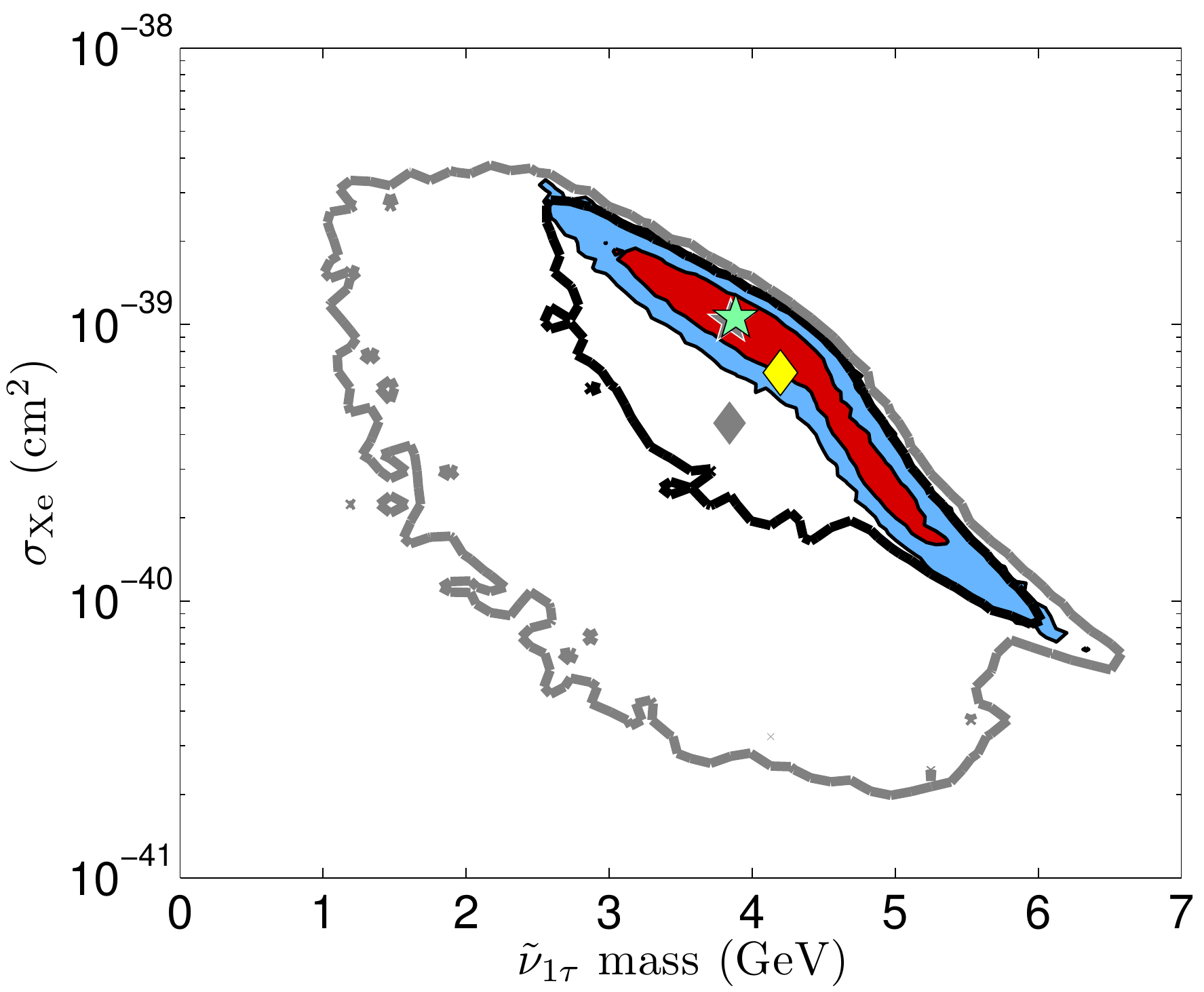}\quad
   \includegraphics[width=7cm]{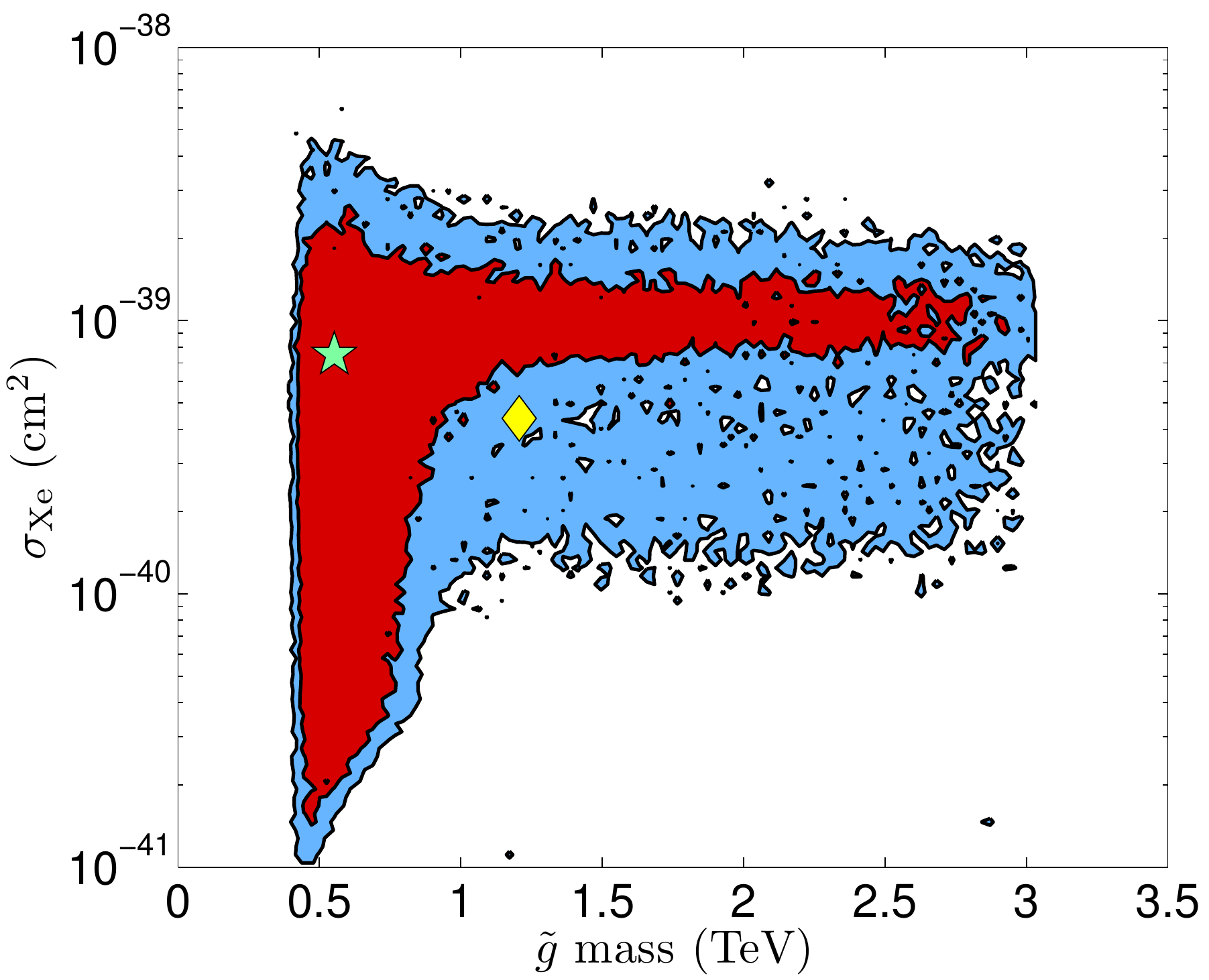}
   \caption{On the left, 2D posterior PDF of  $\sigma_{\rm Xe}$ versus $m_{\tilde\nu_{1\tau}}$ before and after imposing $m_{\tilde g}>1$~TeV; see the caption of Fig.~\ref{fig:light-2d-sinth} for the meaning of colors and symbols. On the right, correlation between $\sigma_{\rm Xe}$ and gluino mass; the red and blue areas are the  68\% and 95\% BCRs.}
   \label{fig:light-2d-sigXe}
\end{figure}

In Fig.~\ref{fig:light-2d-sigXe}, we show the influence of the gluino mass limit on the predicted DD cross section for Xenon (we display the Xenon cross section to directly compare with the best limit which comes from XENON10).
Imposing $m_{\tilde g}>1$~TeV has quite a striking effect, limiting $\sigma_{\rm Xe}$ to a small region just below the current limit. 
We recall that XENON10 only constrains the mass range above $\approx 4$~GeV; for lower $\lsp$ masses, the DD cross section is constrained from above by the $Z$ invisible width. We also note that there is a lower limit on the DD cross section \cite{Belanger:2010cd}, so that if a lower threshold can be achieved to probe masses below 4~GeV, in principle the light sneutrino DM case can be tested completely. (For $\mlsp\approx 4\mbox{--}6$~GeV, an improvement of the current sensitivity by about a factor 3 is sufficient to cover the 95\% region, while an improvement by an order of magnitude will completely cover this mass range.) 

\begin{figure}[!t] 
   \centering
   \includegraphics[width=7cm]{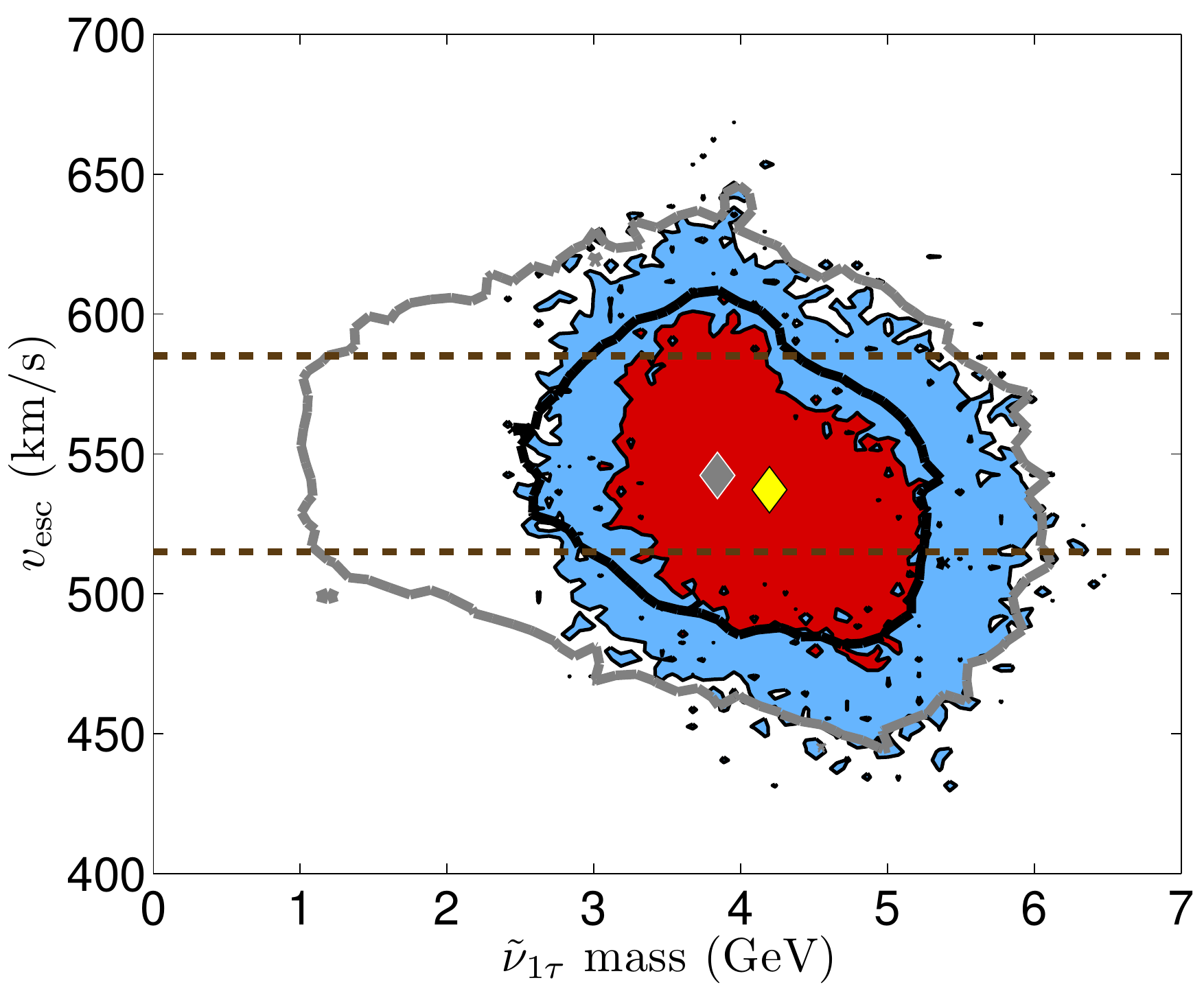}\quad
   \includegraphics[width=7cm]{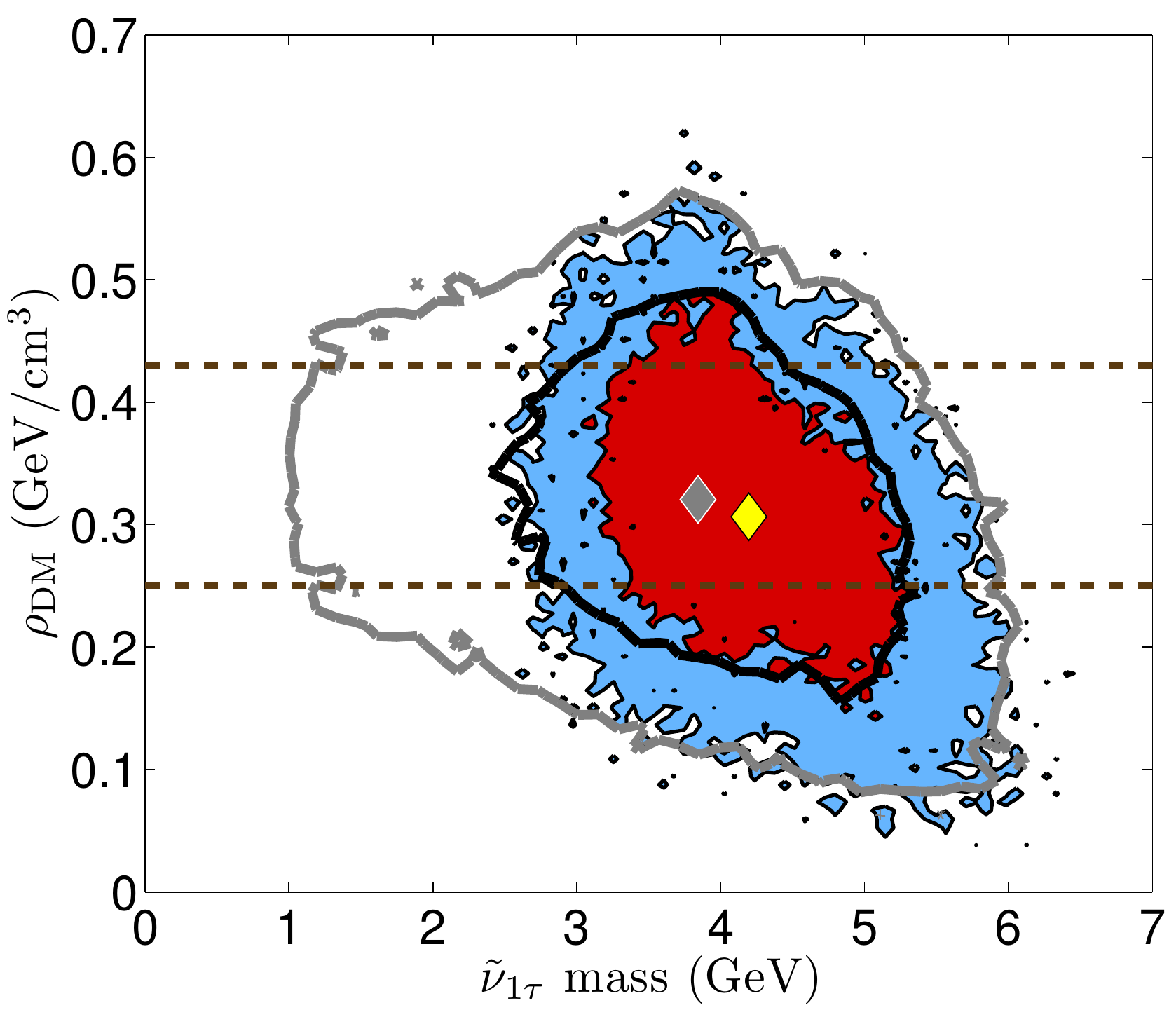}
   \caption{68\% and 95\% BCRs of $v_{\rm esc}$ versus $m_{\tilde\nu_{1\tau}}$ (left) and of $\rho_{\rm DM}$ versus $m_{\tilde\nu_{1\tau}}$ (right). The black (grey) contours are the 68\% (95\%) BCRs without gluino mass cut, while the red (blue) areas are the 68\% (95\%) BCRs for $m_{\tilde g}>1$~TeV. The dashed lines mark the $1\sigma$ experimental bounds for $v_{\rm esc}$ and $\rho_{\rm DM}$.}
   \label{fig:light-2d-nuisance}
\end{figure}

The influence of the nuisance parameters is also interesting. For example, a low local DM density can bring points with high DD cross section in agreement with the XENON10 limits. Likewise, a small mixing angle at sneutrino masses around 4~GeV allows for higher $\rho_{\rm DM}$, because the DD cross section is low. 
Analogous arguments hold for $v_0$ and $v_{\rm esc}$, since for light DM one is very sensitive to the tail of the velocity distribution. 
The effect is illustrated in Fig.~\ref{fig:light-2d-nuisance}.

\begin{figure}[!t] 
   \centering
   \includegraphics[width=7cm]{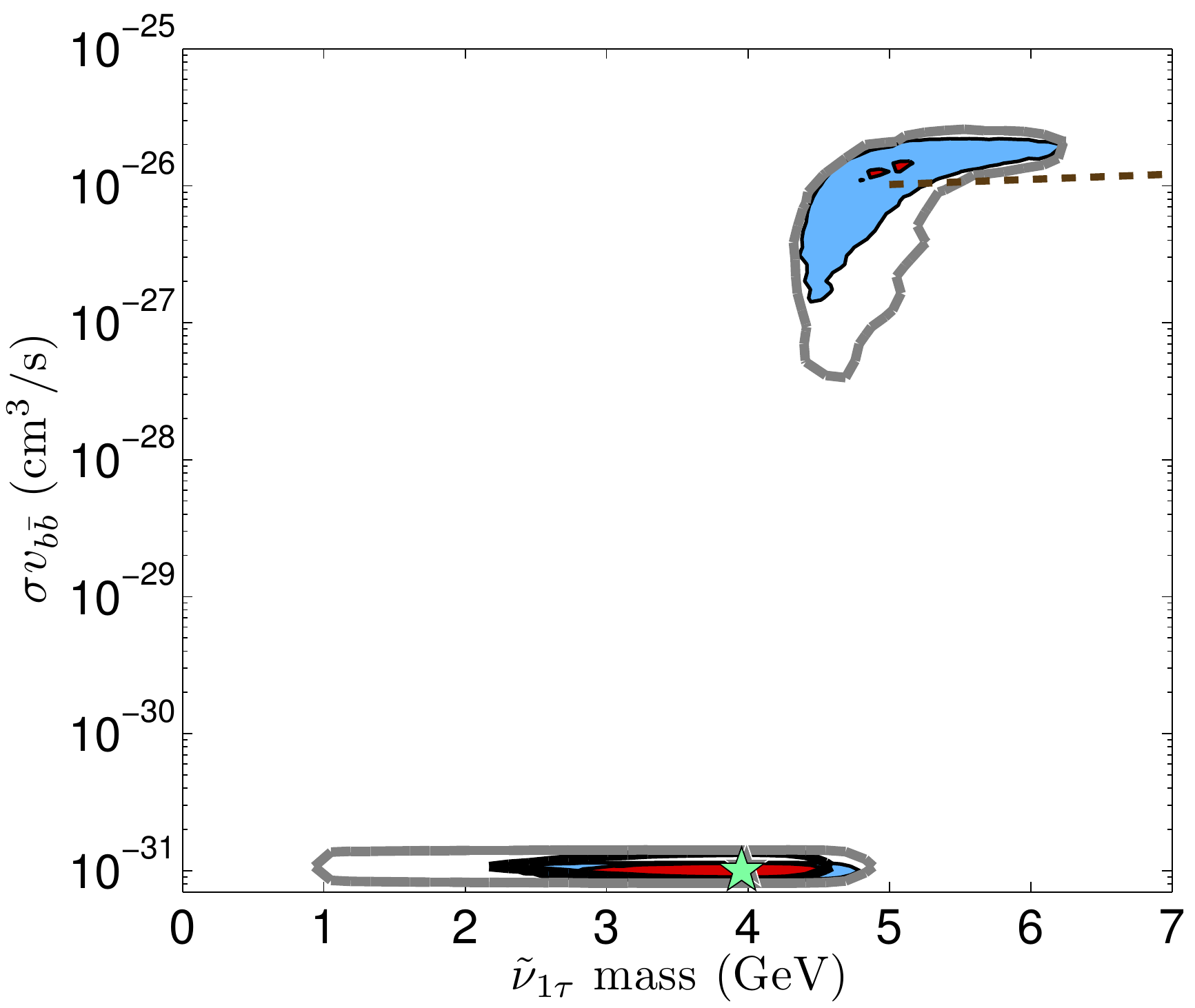}\quad 
   \includegraphics[width=7cm]{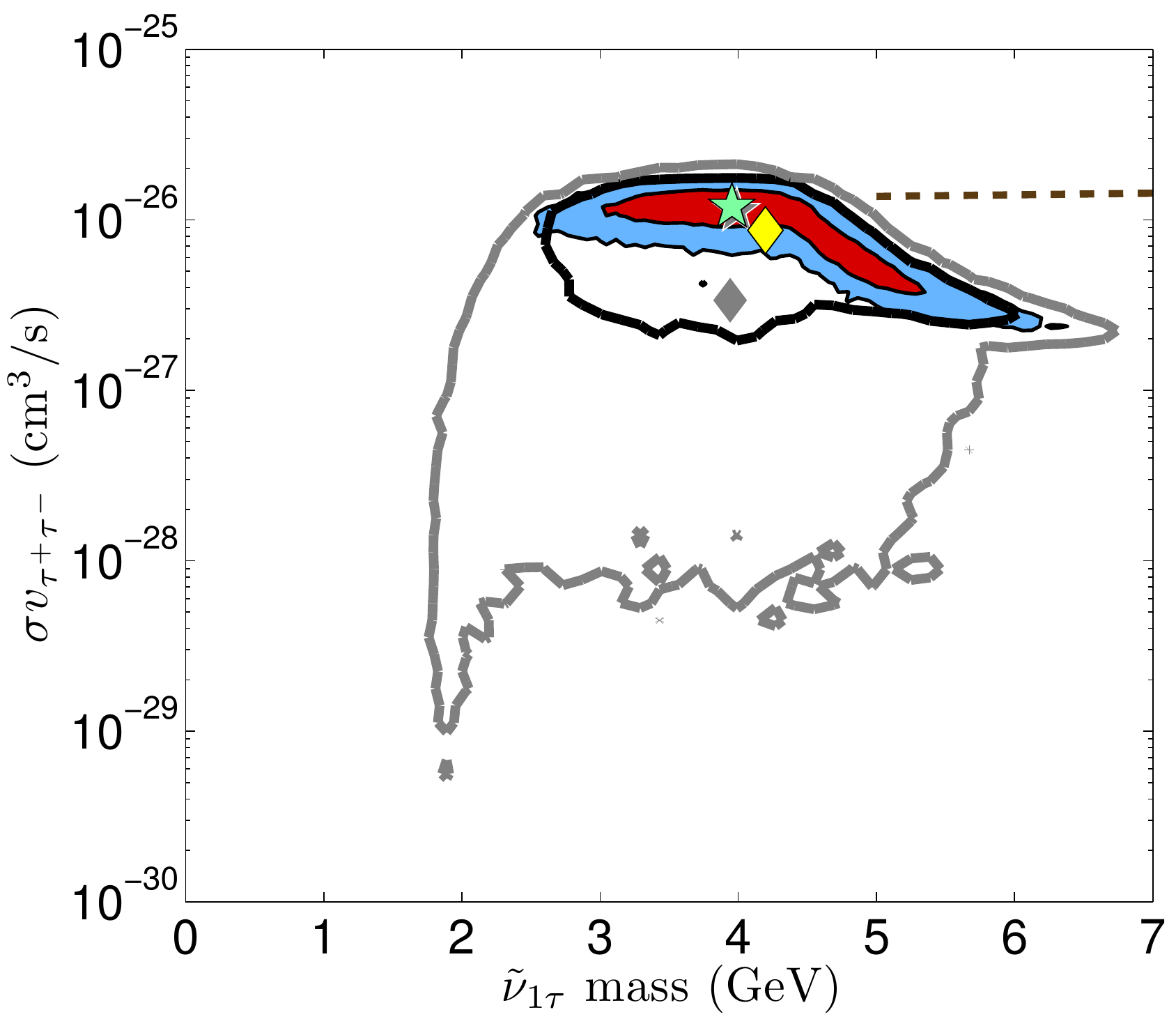}\\
   \includegraphics[width=7cm]{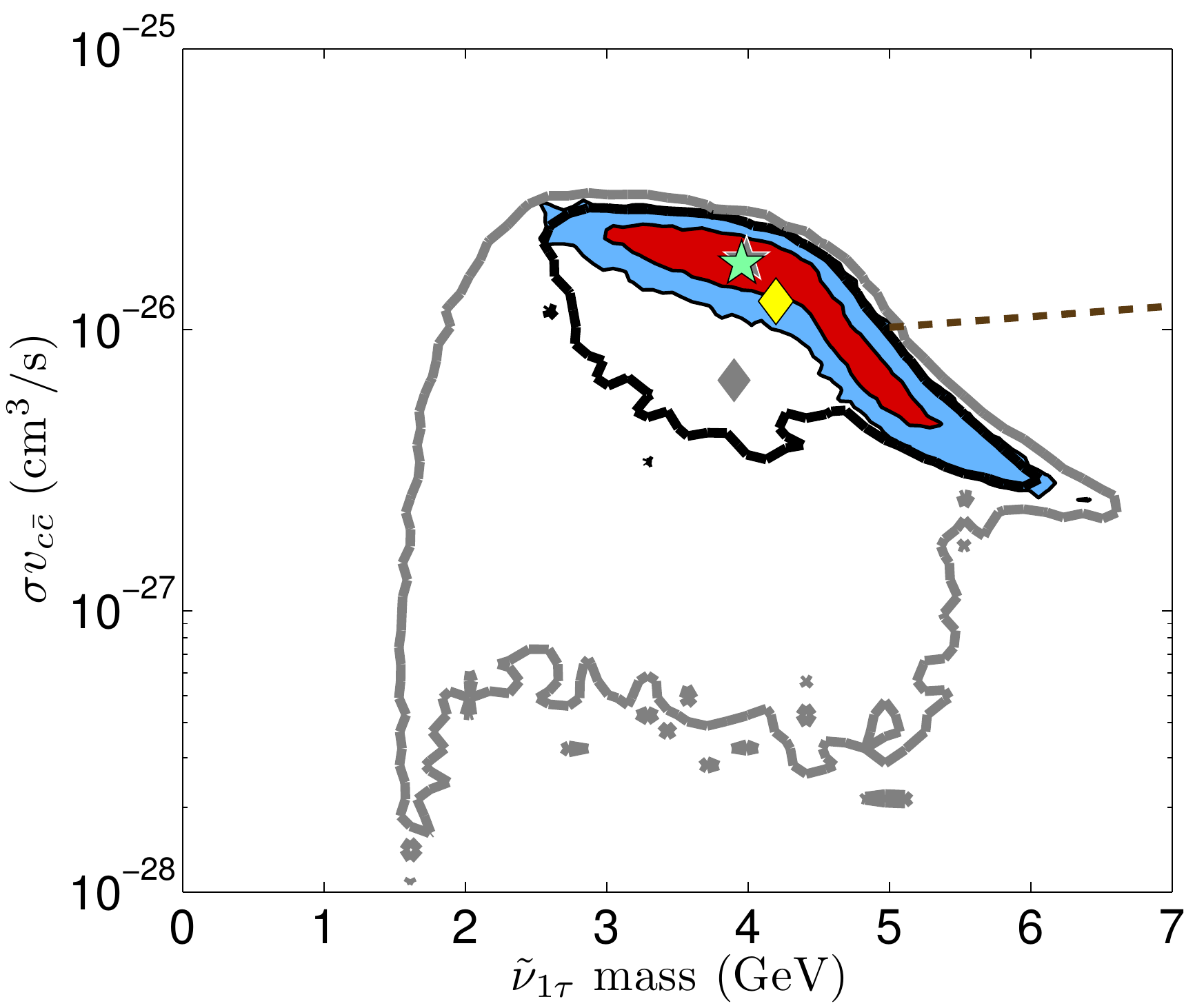}\quad
   \includegraphics[width=7cm]{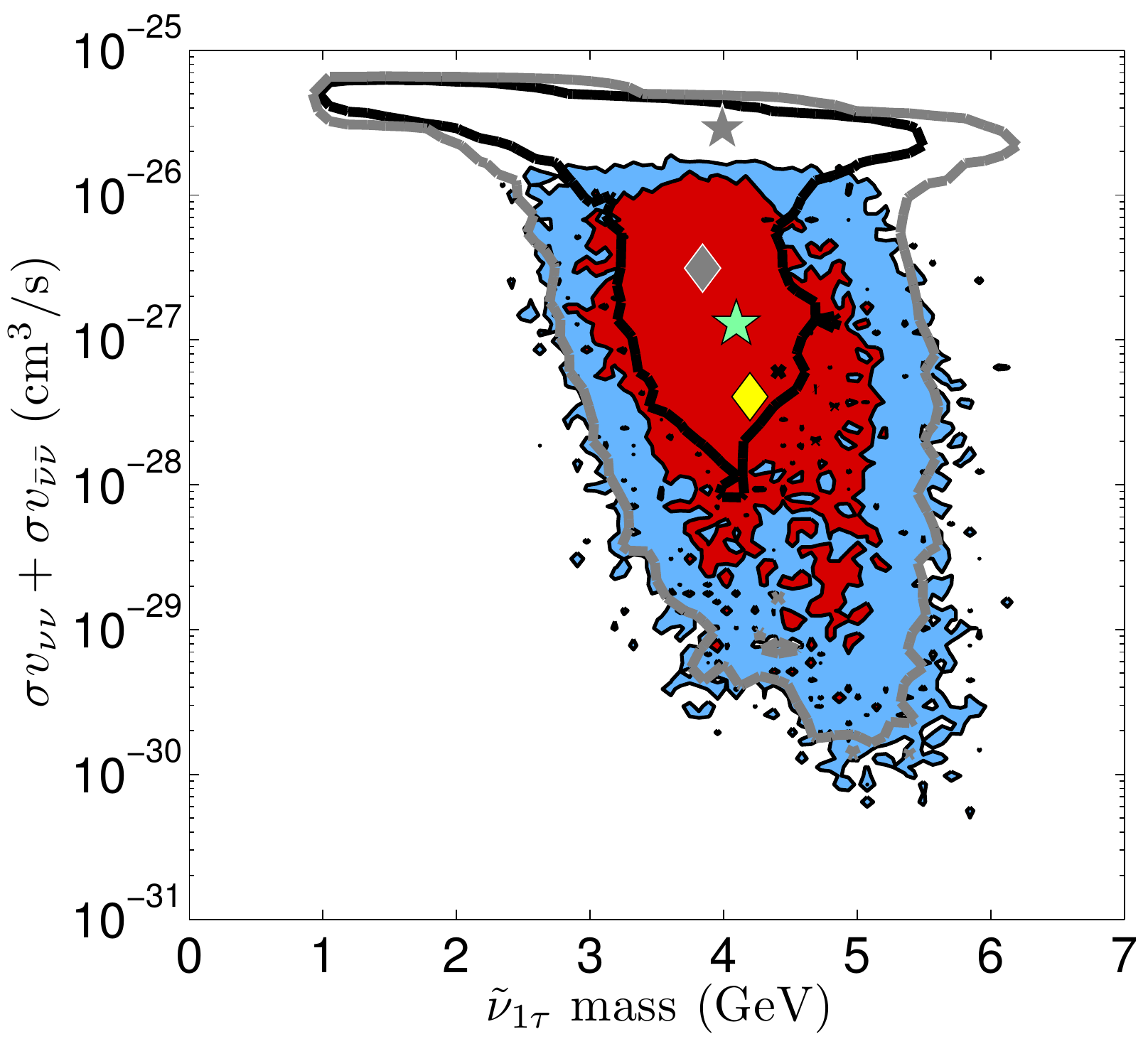}
   \caption{68\% and 95\% BCRs for $\sigma v$ versus sneutrino mass in various channels. Color code as in the previous figures. The dashed lines correspond to the {\it Fermi}-LAT limit~\cite{Ackermann:2011wa}, where  for $c\bar{c}$ we have used the same value as for $b\bar{b}$. Note that for $m_{\lsp}<m_b$,  the cross section is zero, however to display this region we have arbitrarily set it to $\sigma v_{b\bar b}=10^{-31}~\rm{cm}^3 /{\rm s}$.}
   \label{fig:light-2d-indirect}
\end{figure}

The MCMC approach also permits us to make predictions for the annihilation cross section  of light sneutrino dark matter into different final states, relevant for indirect DM searches, see Fig.~\ref{fig:light-2d-indirect}.
When $m_{\lsp}>m_b$,
the dominant DM annihilation channels are into $\nu\nu$ or   $b\bar{b}$ pairs. The latter will lead to a large photon flux---in fact the partial cross section into $b\bar{b}$  is always in the region constrained by {\it Fermi}-LAT when $m_{\lsp}>5.2$~GeV.

For lighter DM, the charged fermions final states  giving photons are  $c\bar{c}$ and  $\tau^+\tau^-$. Here note that for a given  LSP mass, imposing the lower limit on the gluino mass  selects the upper range for both $\sigma v_{c\bar{c}}$  and $\sigma v_{\tau^+\tau^-}$ while having only a mild effect on $\sigma v_{b\bar{b}}$. 
In particular the $c\bar{c}$ channel typically has a large cross section of $\sigma v_{c\bar c}\gtrsim 10^{-26}~\rm{cm}^3 /{\rm s}$ throughout the 95\% BCR when $m_{\tilde g}>1$~TeV. This could hence be probed if  the {\it Fermi}-LAT search was extended to a lower mass range. 

Regarding annihilation into neutrinos, as mentioned earlier,  the gluino mass limit  strongly constrains scenario where annihilation into neutrino pairs is dominant, leading to an upper limit of $\sigma v_{\nu\nu} + \sigma v_{\bar{\nu}\bar{\nu}} \lesssim 1\times 10^{-26}~\rm{cm}^3 /{\rm s}$, see the bottom-right panel in Fig.~\ref{fig:light-2d-indirect}. A discussion of the neutrino signal for light sneutrino DM can be found in~\cite{Belanger:2010cd}. 
As mentioned, we leave a more detailed analysis of neutrinos from the Sun for a future work.  

Dark matter annihilation in our galaxy can also lead to antiprotons. 
To illustrate the impact of the antiproton measurements on the parameter space of the model, we have computed the antiproton flux for some sample points and compared those to the flux measured by PAMELA~\cite{Adriani:2010rc}. To compute this flux we have used the  semi-analytical two-zone propagation model of~\cite{Maurin:2001sj,Donato:2001ms} with two sets of propagation parameters called MIN and MED, see~\cite{Belanger:2010gh}. For the background we have used the semi-analytical formulas of~\cite{Maurin:2006hy} with a solar modulation of $\phi=560~{\rm MeV}$,  which fit well the measured spectrum of PAMELA.

\begin{figure}[!b] 
   \centering
   \includegraphics[width=7cm]{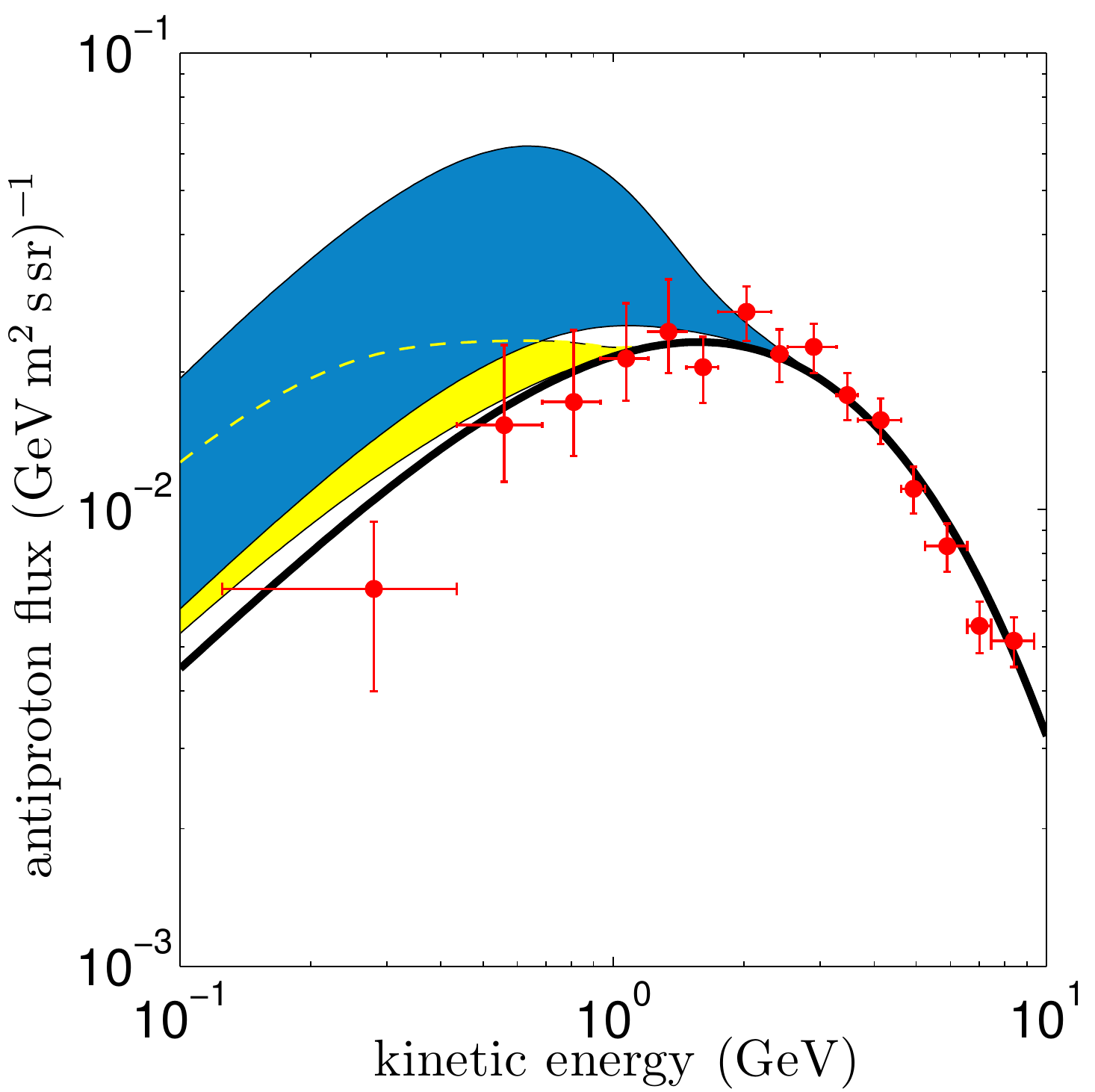}
   \caption{Antiproton flux as a function of the kinetic energy of the antiproton for two representative points as described in the text. The blue (yellow) band corresponds to $m_{\lsp}= 4.8$ (2.3)~GeV, with the upper curve corresponding to MED and the lower curve corresponding to  MIN propagation parameters. We also display the background only (black line) and the PAMELA data for energies below 10 GeV  (red crosses).}
   \label{fig:light-antiproton}
\end{figure}

The first sample point has  a DM  mass of 4.8~GeV and  is dominated by annihilation into $b\bar{b}$  with $\sigma v_{bb}=1.1 
\times 10^{-26}~{\rm cm}^3/{\rm s}$.
The resulting antiproton flux is displayed as the blue band in Fig.~\ref{fig:light-antiproton}.  A large excess is expected at 
energies below 1~GeV for MED propagation parameters, corresponding to the upper edge of the blue band. With MIN propagation parameters however,  the flux exceeds the $1\sigma$ range only in the lowest energy bin ($E_{\bar{p}}=0.28$~GeV). We therefore conclude that such sneutrino DM would  be compatible with the PAMELA measurements only for a restricted choice of propagation model parameters. Here note that the lowest energy bins are  the ones where the background is most affected by  solar modulation effects.

The second sample point has  lighter DM, $m_{\lsp}=2.3$~GeV,  and annihilation into c-quarks dominates the hadronic channels ($\sigma v_{c\bar{c}}=1.7 \times 10^{-26}~{\rm cm}^3/{\rm s}$) although the dominant annihilation channel is into neutrinos. The antiproton flux is therefore expected to be both lower and shifted towards lower energies as compared to the previous case. 
We find that the antiproton flux again exceeds  the measured spectrum by more than $1\sigma$ only in the first energy bin.  Such a sneutrino DM is therefore not constrained by the antiproton measurements unless one chooses propagation parameters that lead to large fluxes. In this respect note that we can of course get even larger fluxes than those displayed in Fig.~\ref{fig:light-antiproton} using the MAX set of propagation parameters.

\clearpage

%--------------------------------------------------------------------------------
\subsection{Heavy sneutrino DM}\label{sec:heavy}
%--------------------------------------------------------------------------------

Let us now turn to the case of heavy sneutrinos. We will first discuss the heavy non-democratic (HND) case, where the LSP is the $\tilde\nu_{1\tau}$, and then the heavy democratic (HD) case, where all three neutrinos are close in mass and any of them can be the LSP or co-LSP.

\begin{figure}[t] 
   \centering
   \includegraphics[width=4.96cm]{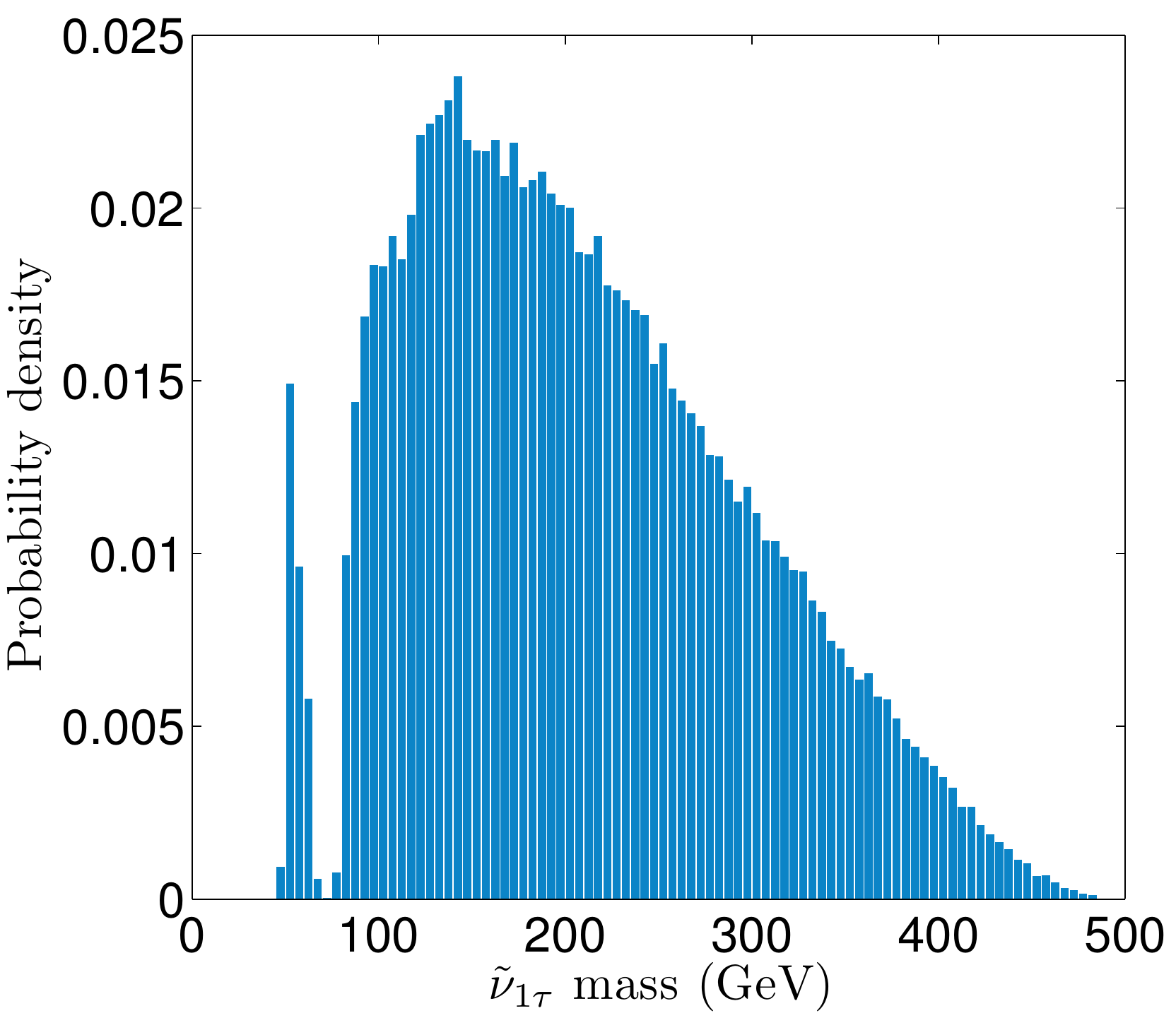}
   \includegraphics[width=4.96cm]{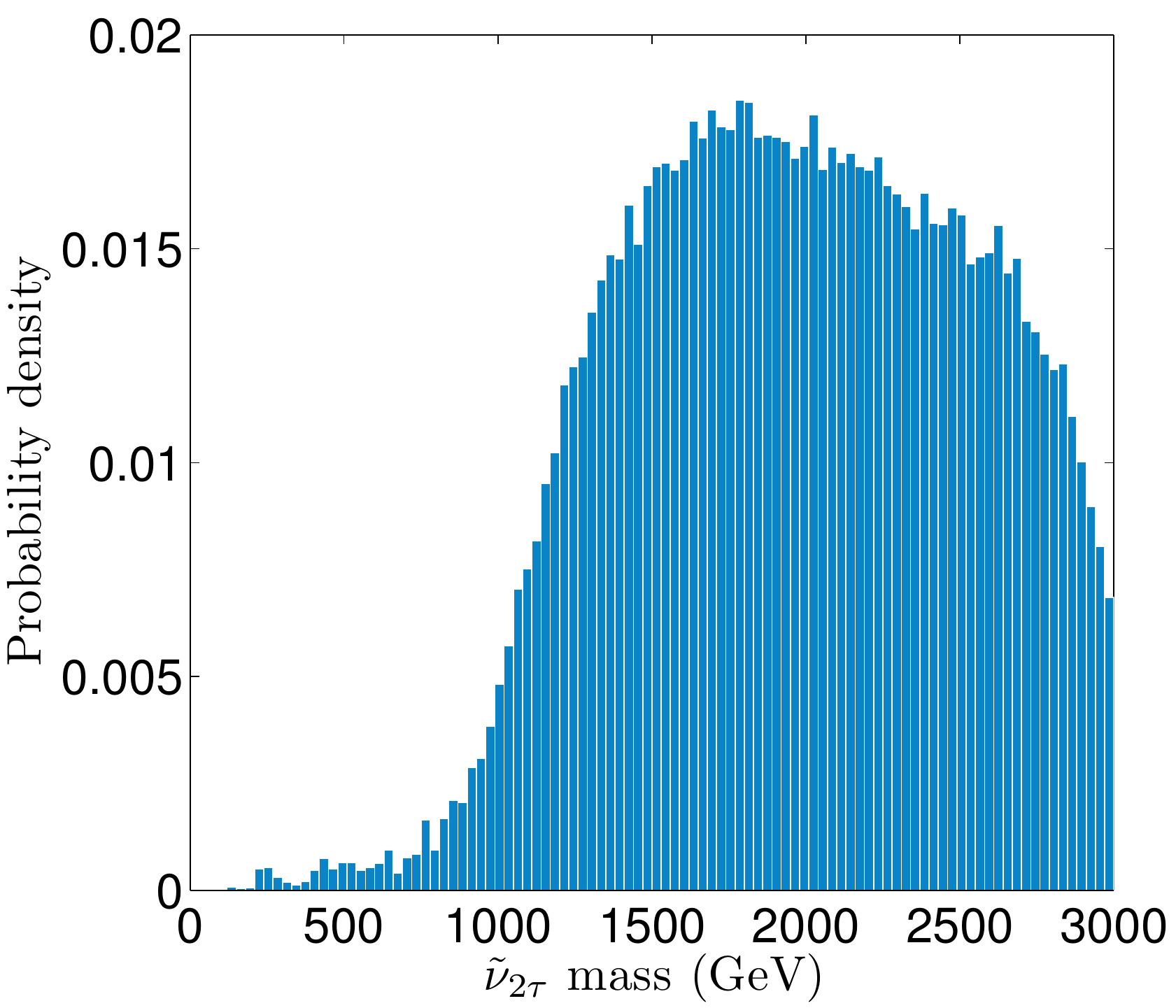}
   \includegraphics[width=4.96cm]{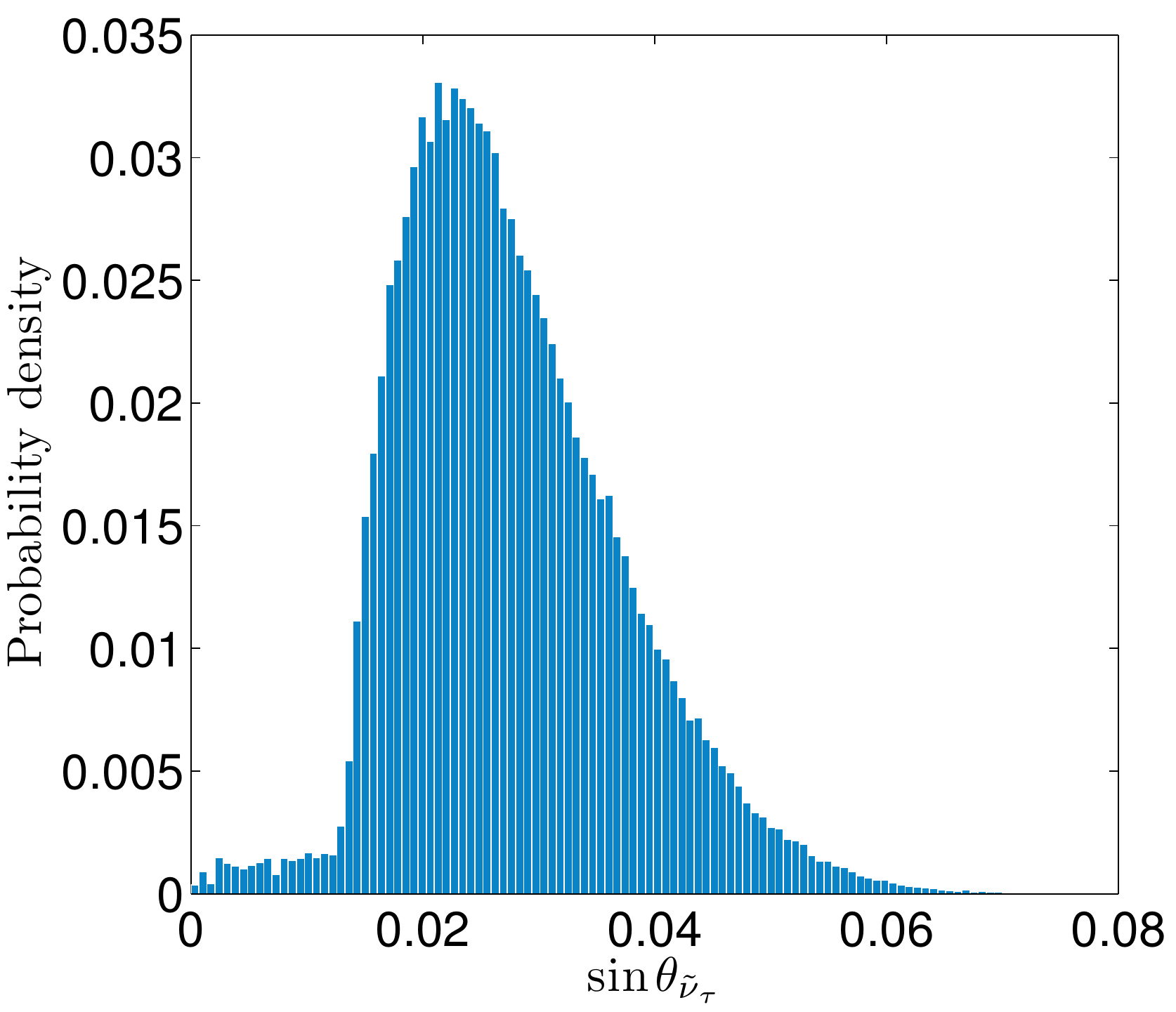}
   \includegraphics[width=4.96cm]{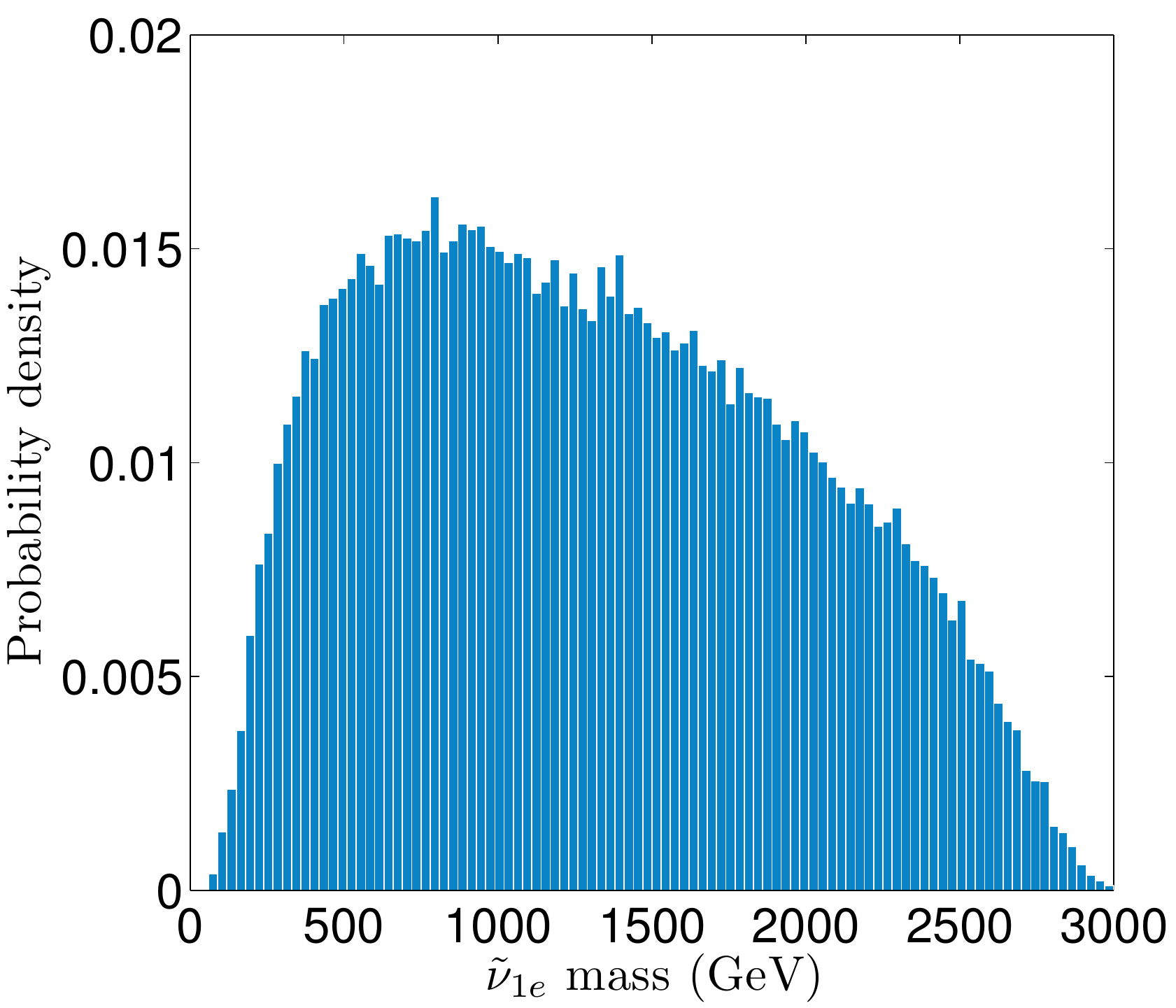}
   \includegraphics[width=4.96cm]{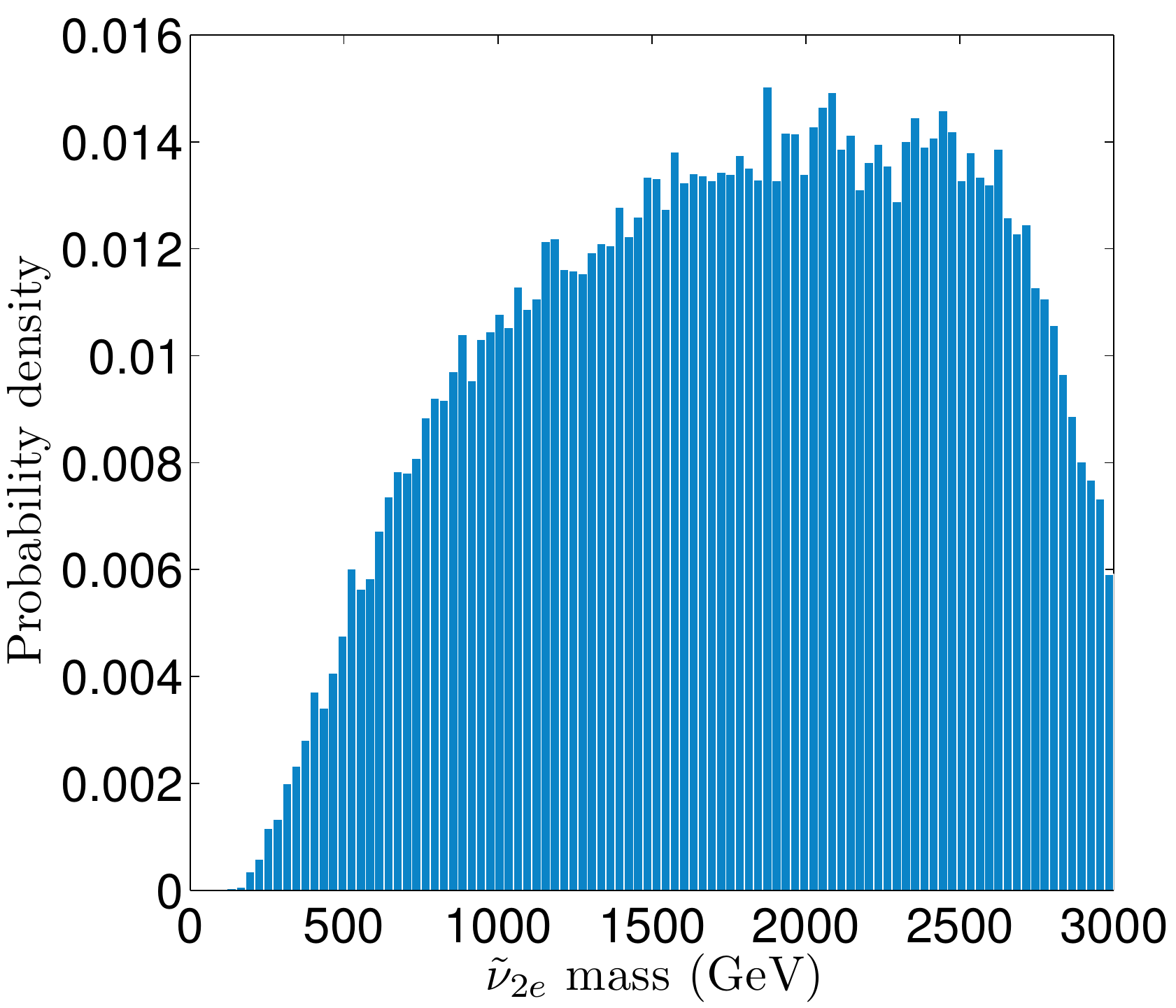}
   \includegraphics[width=4.96cm]{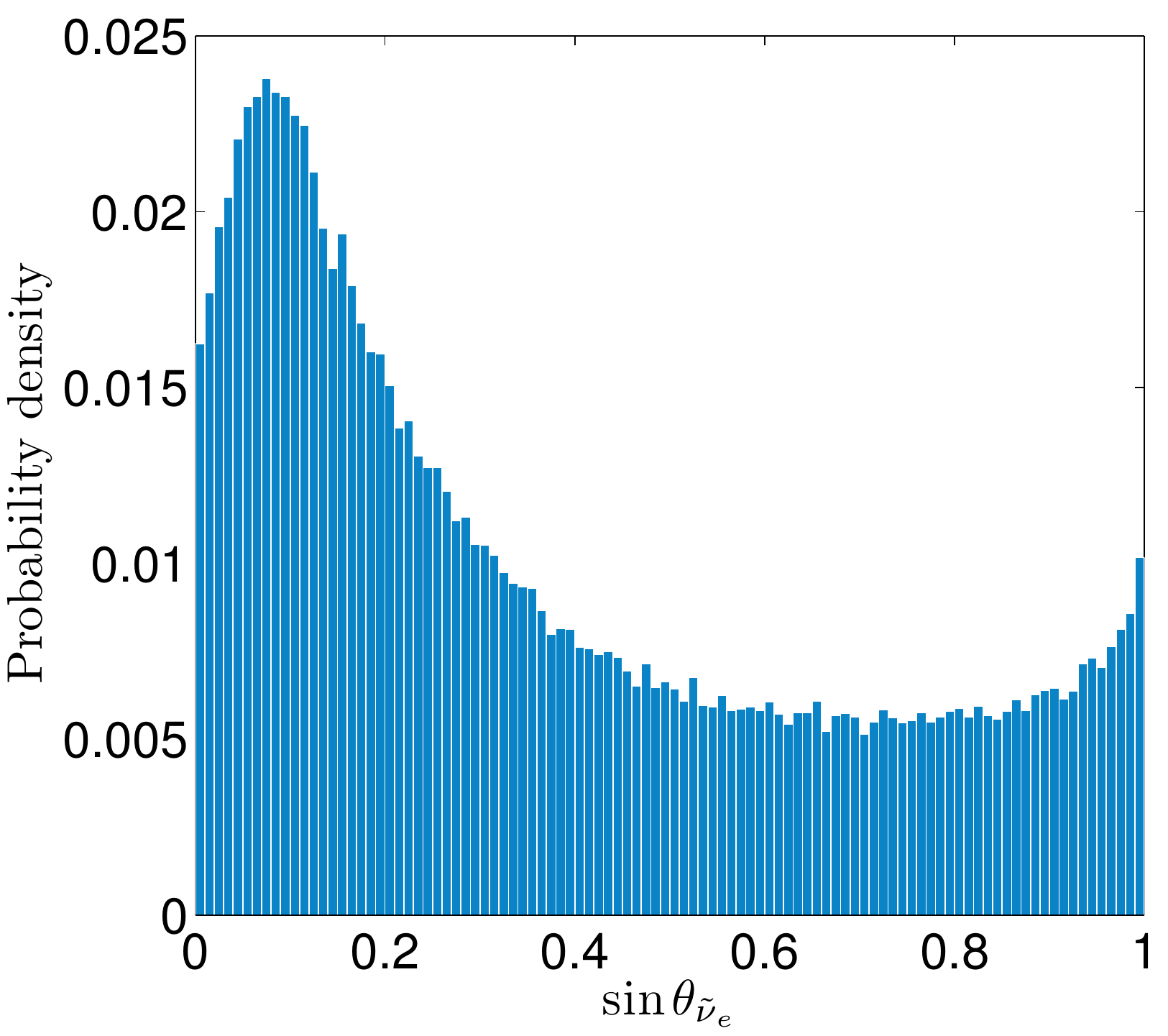}
   \includegraphics[width=4.96cm]{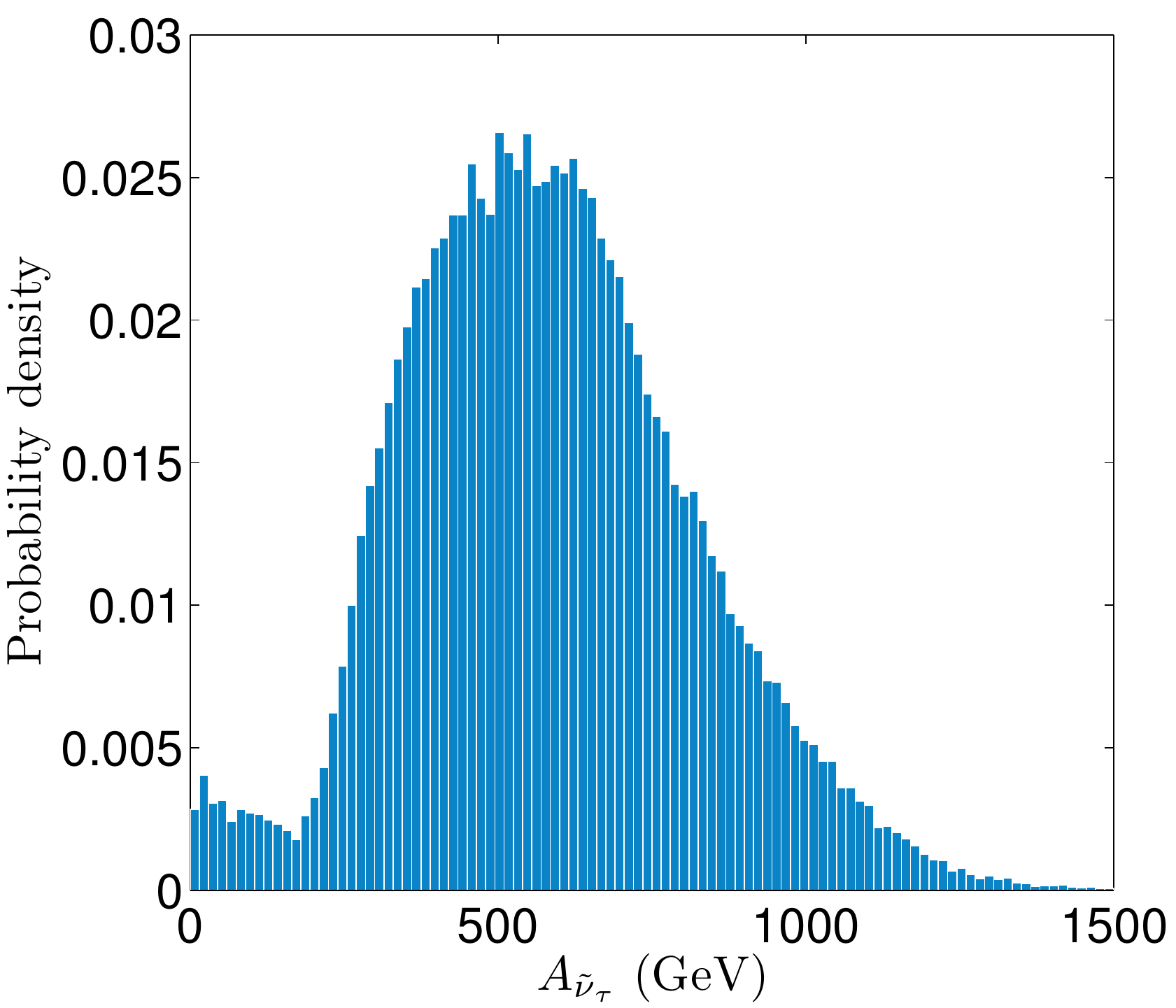}
   \includegraphics[width=4.96cm]{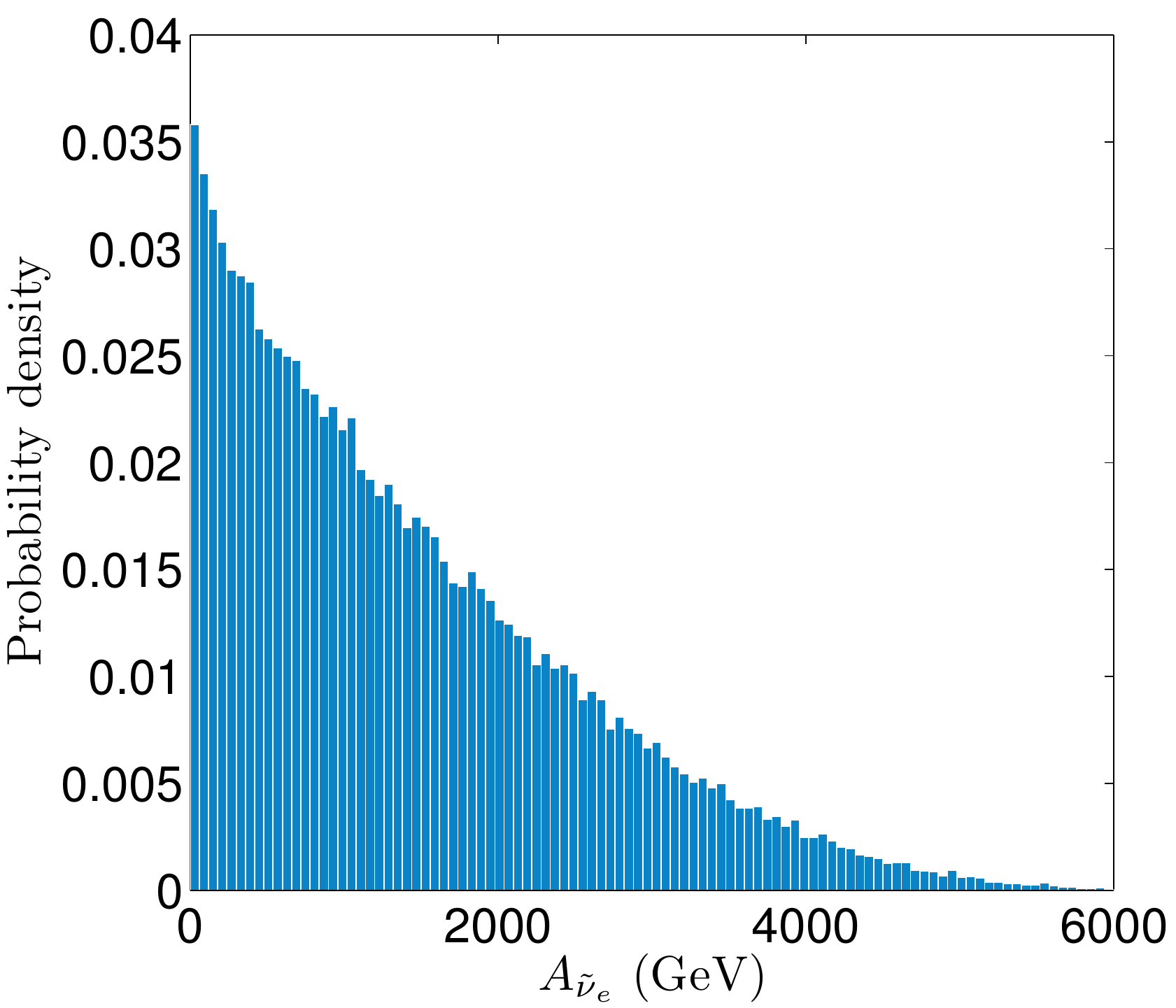}
   \includegraphics[width=4.96cm]{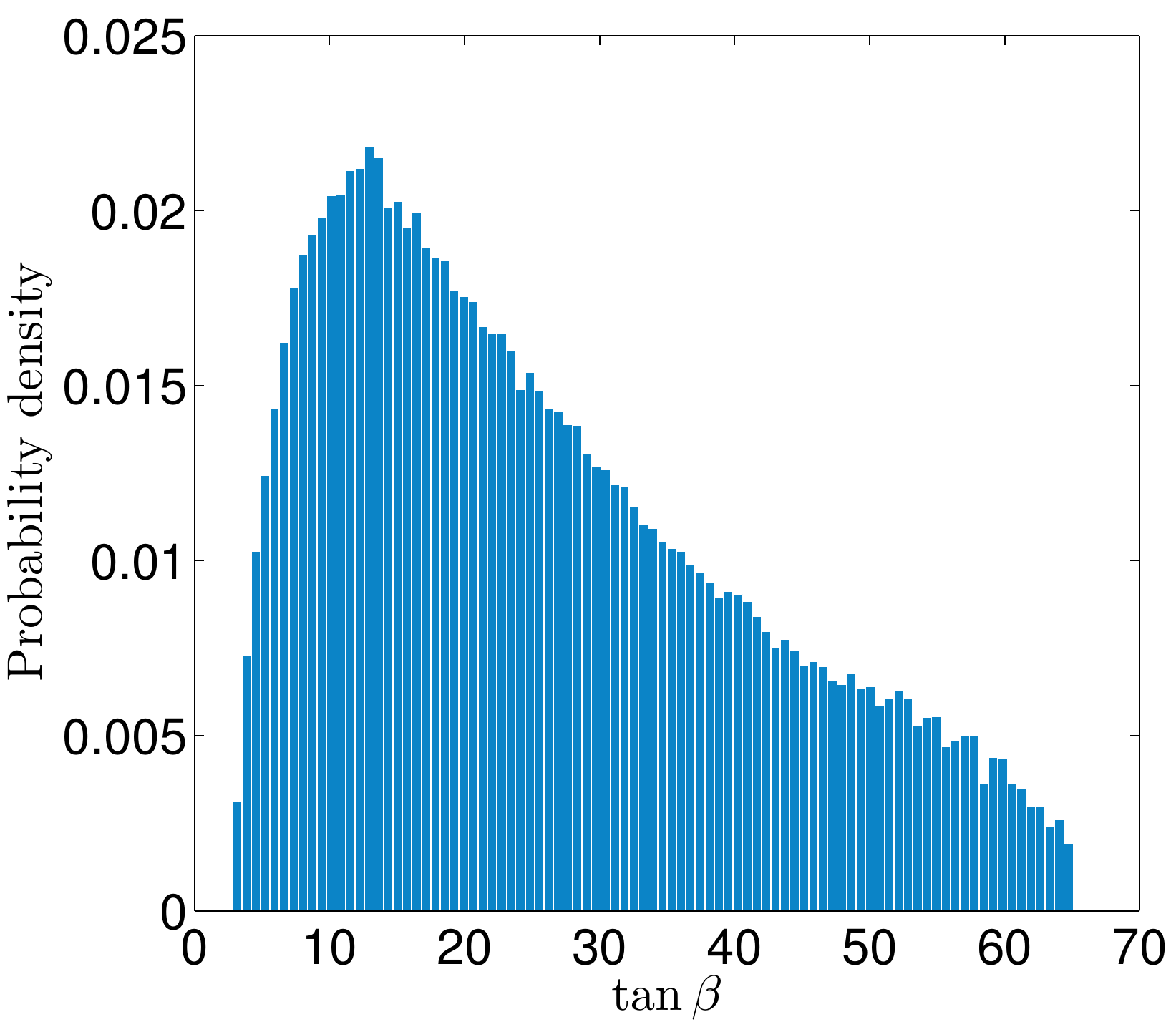}
   \includegraphics[width=4.96cm]{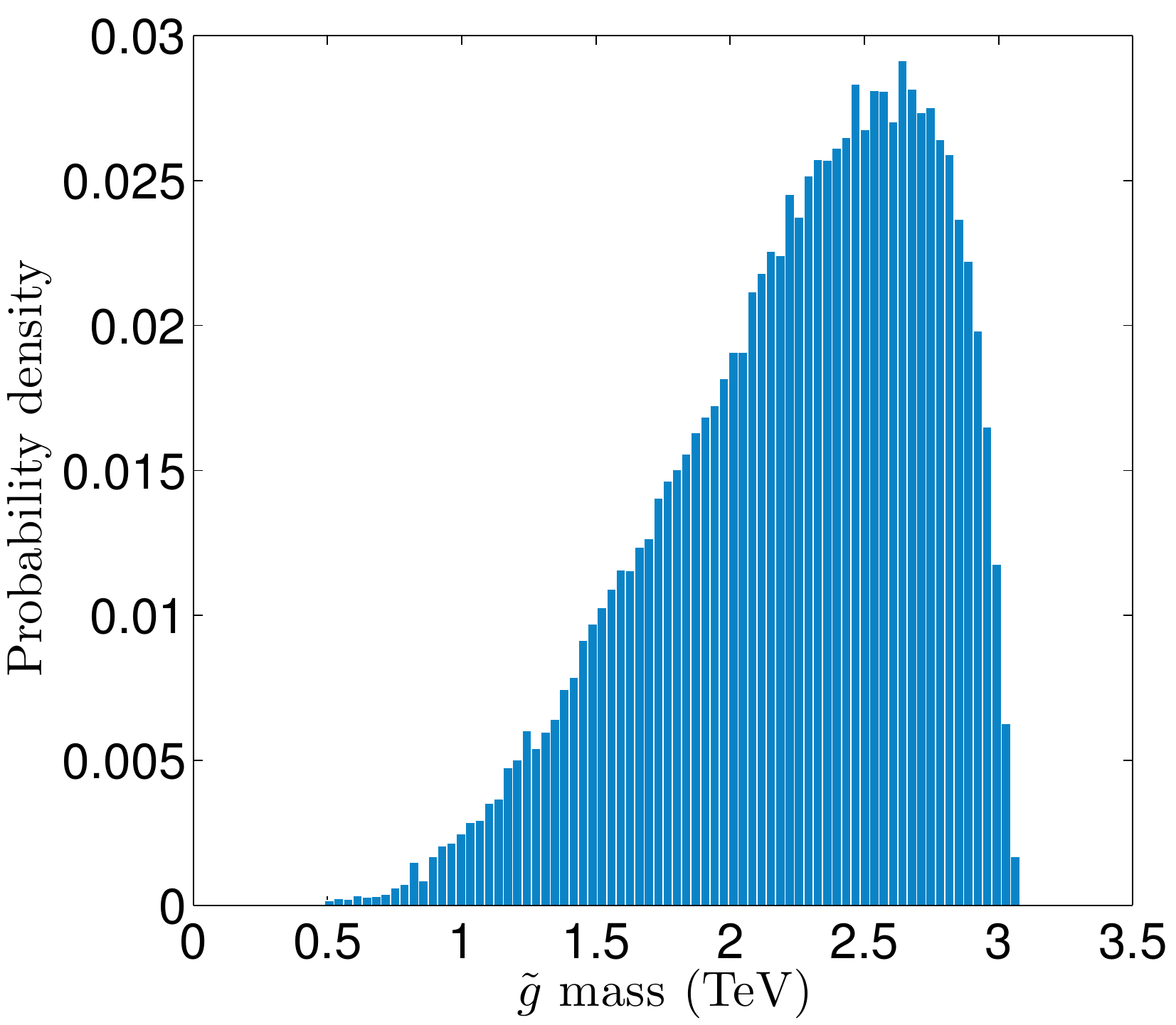}
   \includegraphics[width=4.96cm]{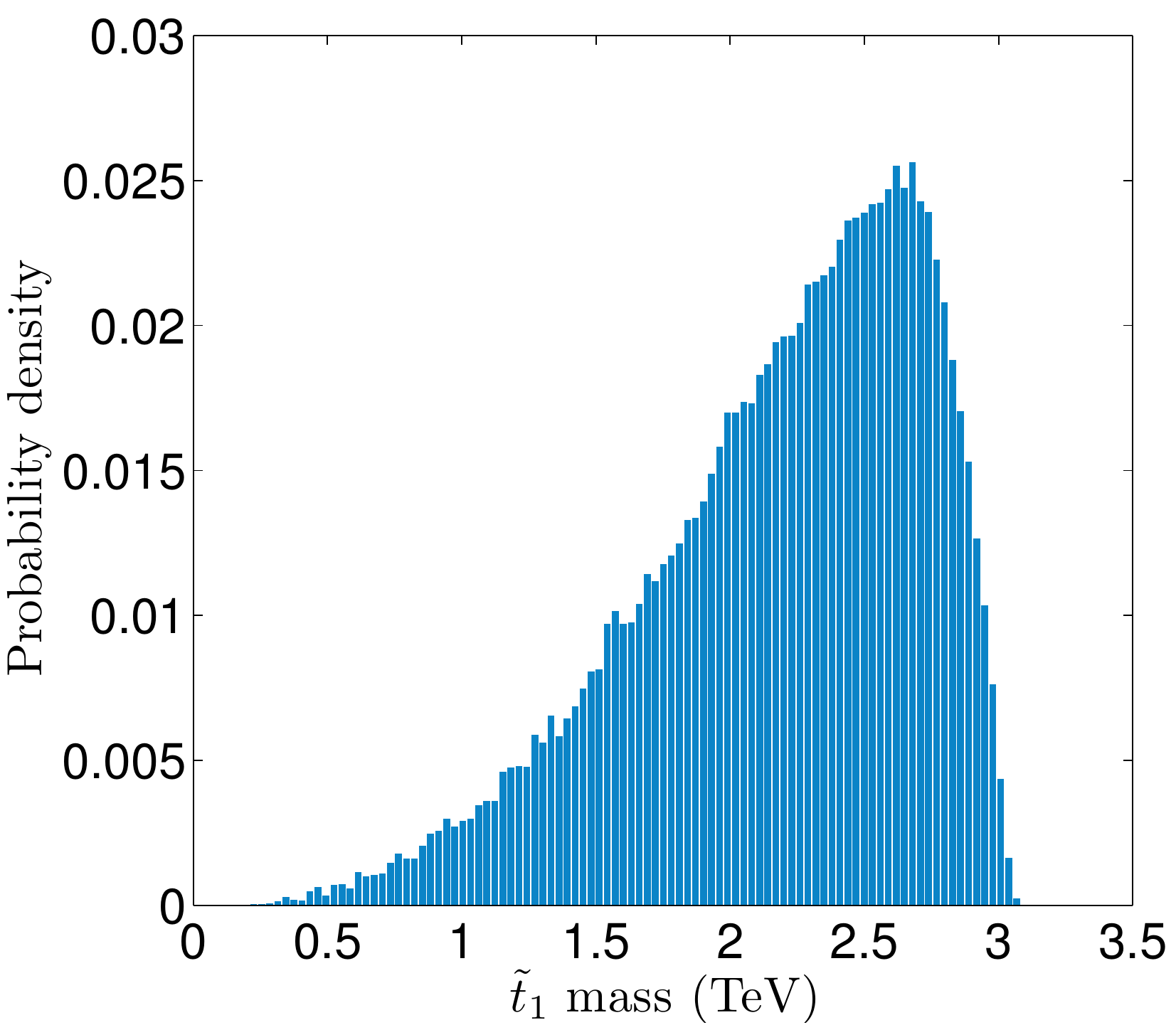}
   \includegraphics[width=4.96cm]{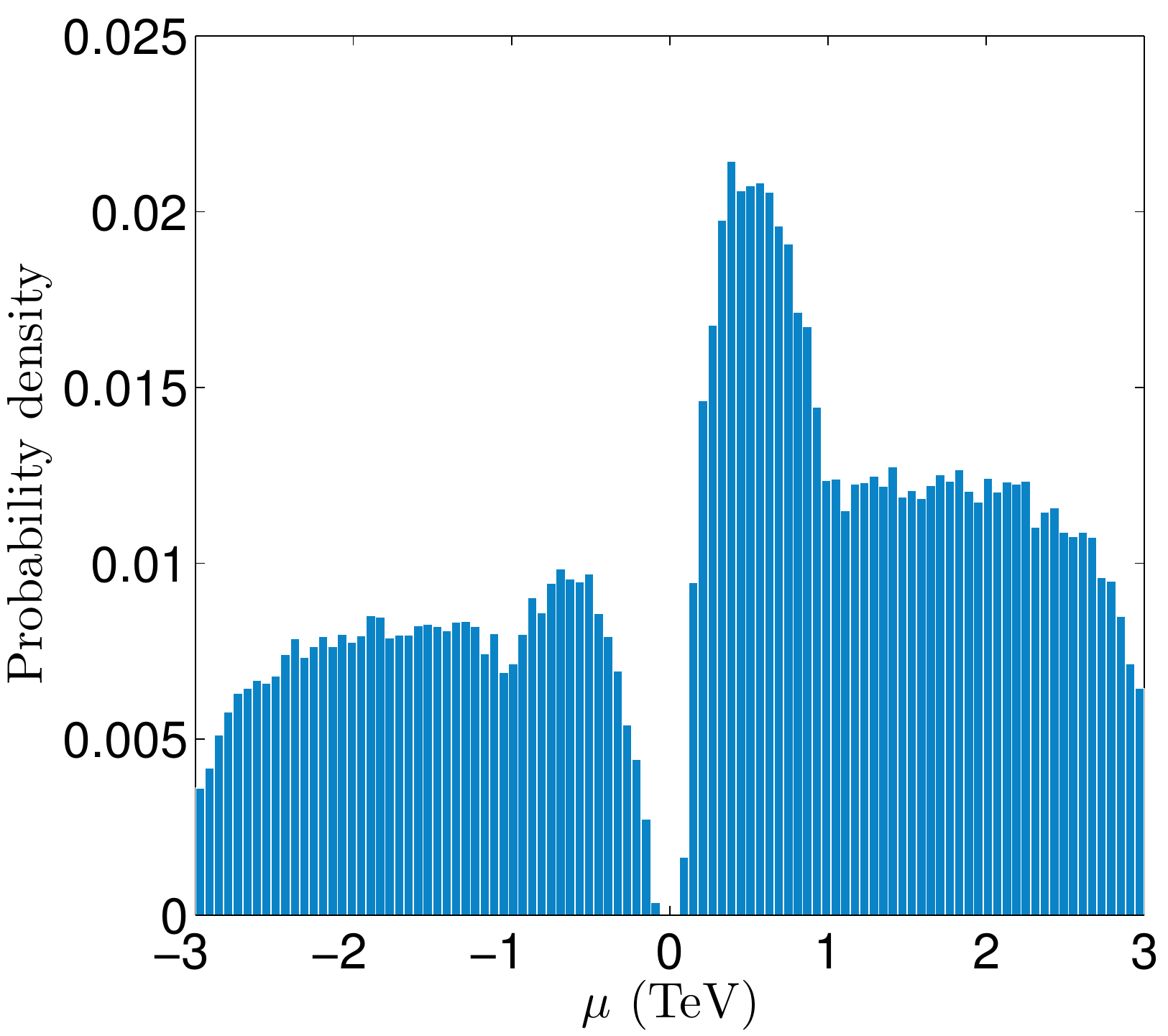}
   \includegraphics[width=4.96cm]{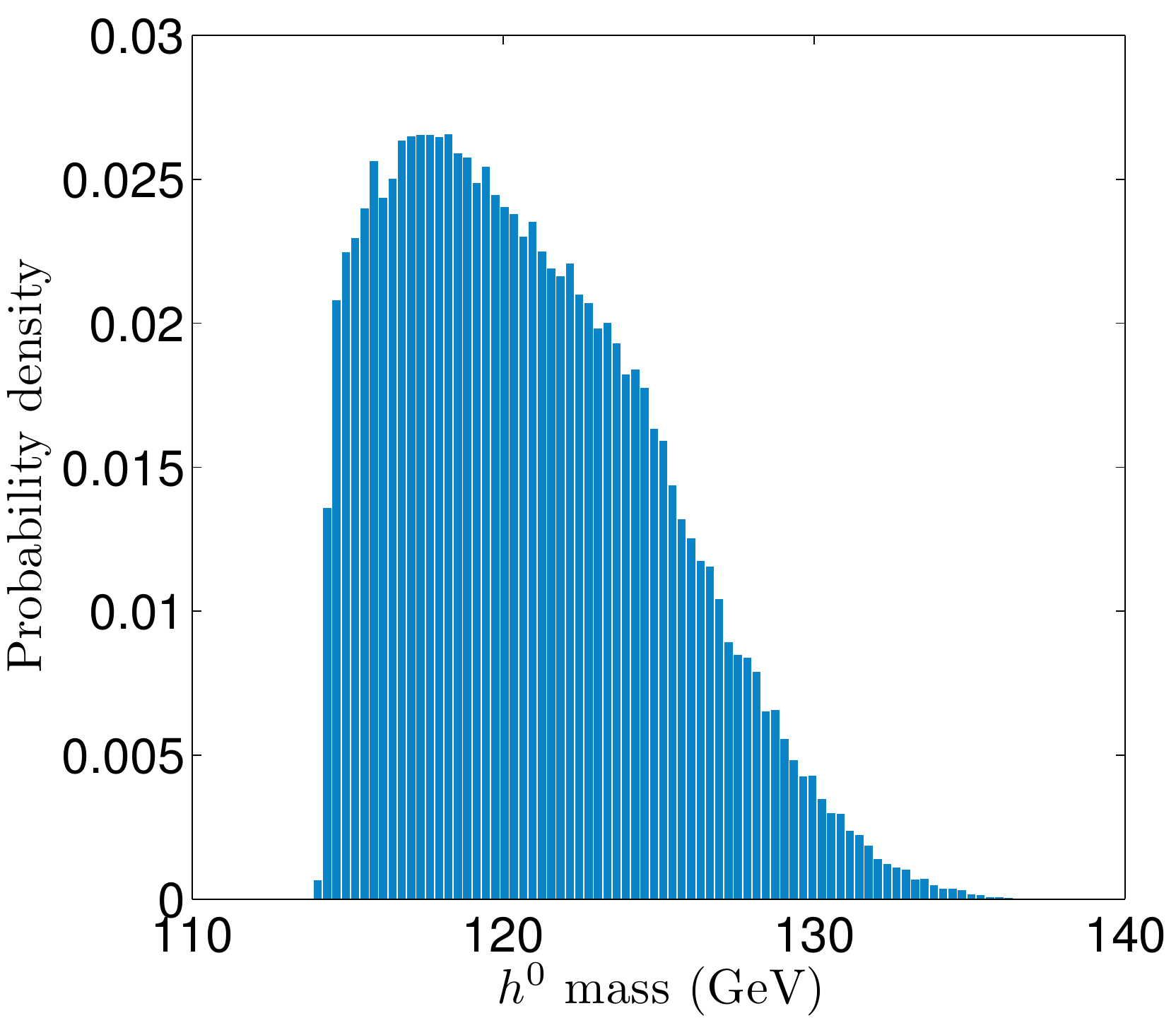}
   \includegraphics[width=4.96cm]{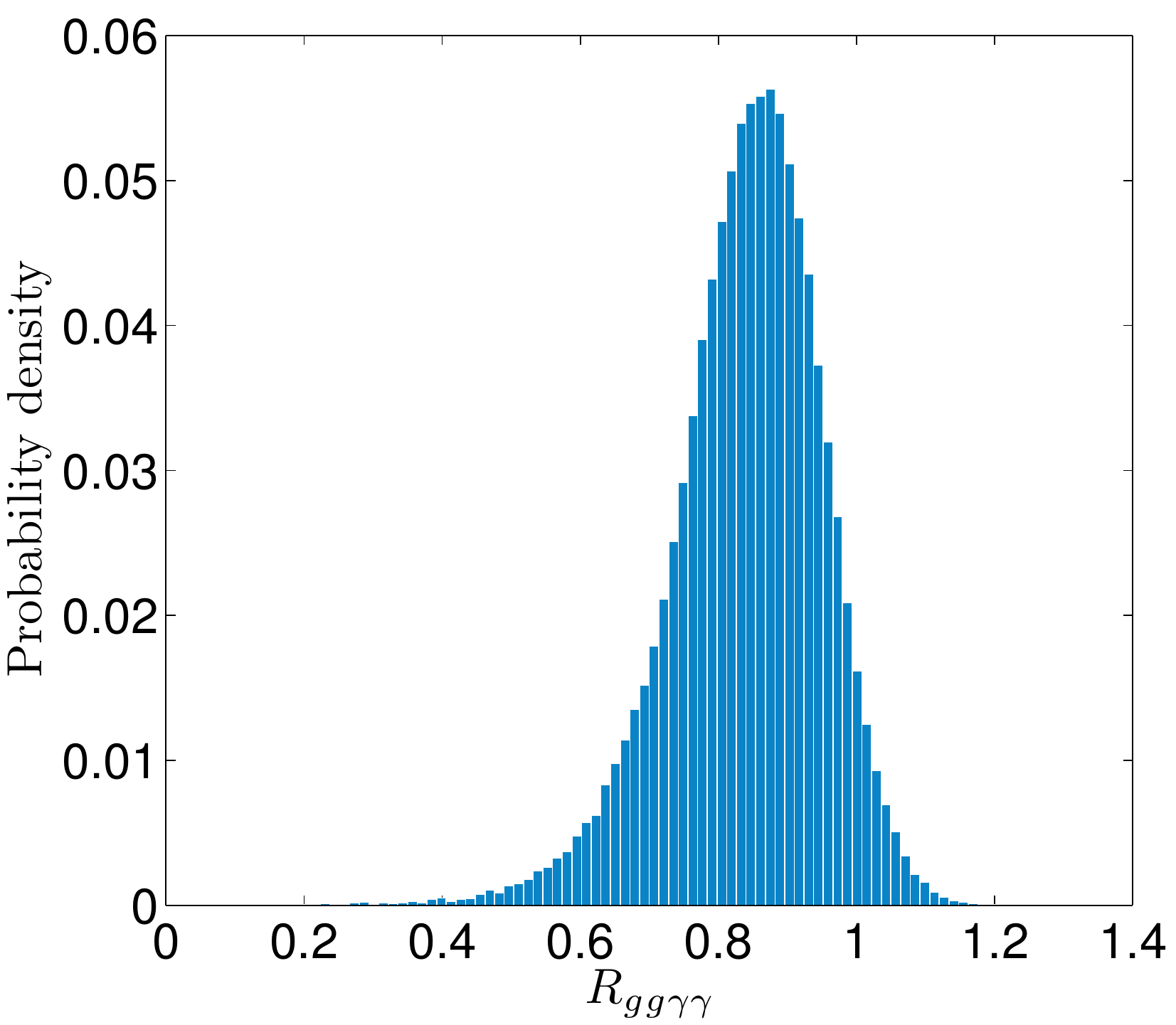}
   \includegraphics[width=4.96cm]{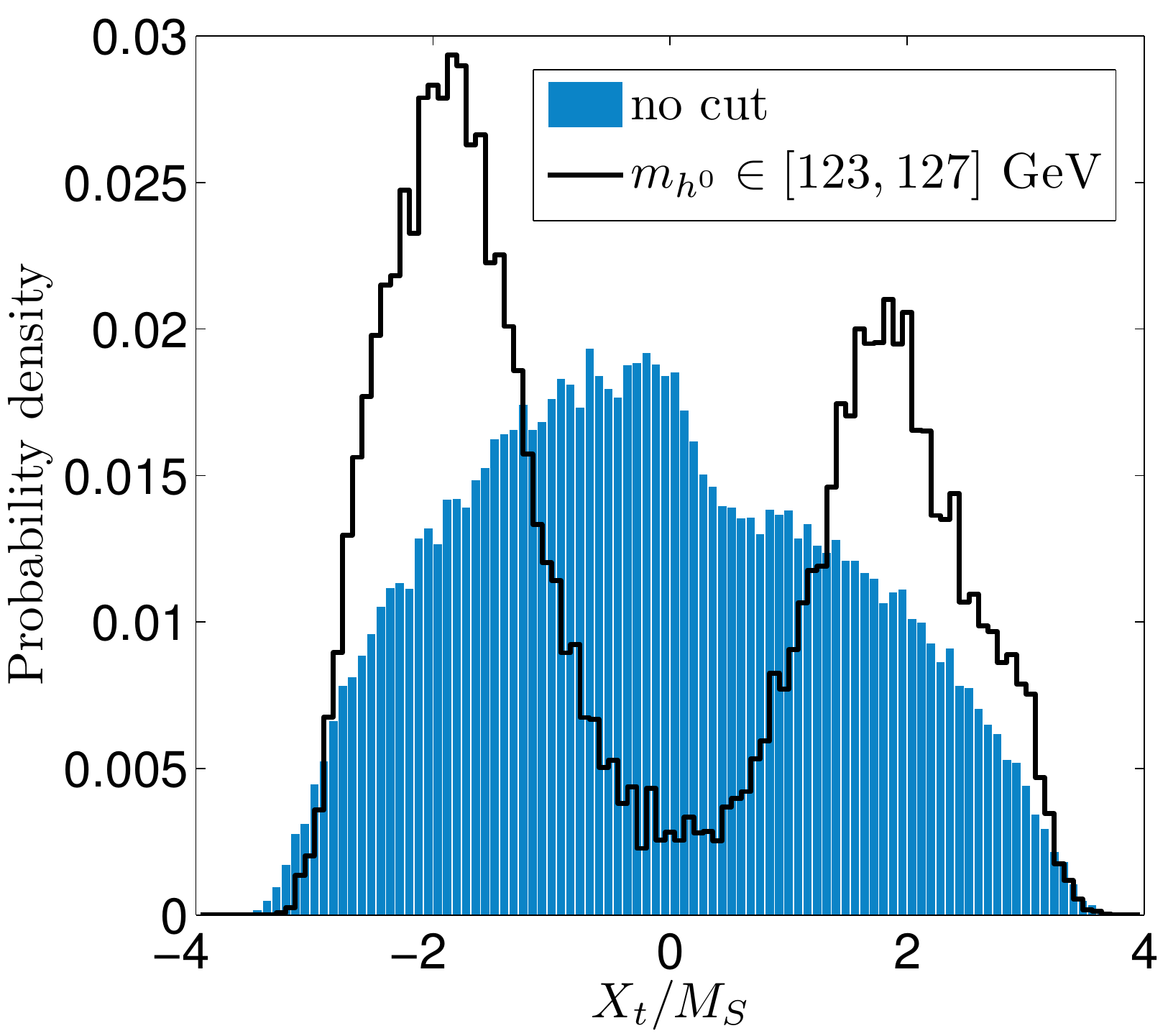}
   \caption{Posterior PDFs in 1D for the heavy non-democratic (HND) sneutrino case.}
   \label{fig:heavy-1d}
\end{figure}

The posterior PDFs in 1D for the HND case are shown in Fig.~\ref{fig:heavy-1d}. Here, we do not superimpose the distributions with $m_{\tilde g}>750$ or $1000$~GeV, because the gluino automatically turns out heavy, with 99\% probability above 1~TeV.
The $\tilde\nu_{1\tau}$ masses now range from 90 to 255 (80 to 375)~GeV at 68\% (95\%) Bayesian credibility.
There is also a small region near $m_{\tilde\nu_{1\tau}}\approx 60$~GeV, where the sneutrino annihilates through the light Higgs resonance; this region has 3\% probability.\footnote{As mentioned in Section~\ref{sec:relic}, the sneutrino can also annihilate through the heavy scalar (not the pseudoscalar!) Higgs resonance. We have checked that this process does occur in our chains. However, it  turns out that it is statistically insignificant and does not single out any special region of parameter space.} 
See Table~\ref{table:bci-hnd} in Appendix~\ref{aTables} for more details.  
The $\tilde\nu_{2\tau}$ mass is typically very heavy, above 1~TeV, and the mixing angle is required to be very small to evade the DD limits, cf.\ the discussion in Section~\ref{sec:directdetection}. Interestingly, the mixing can be almost vanishing; 
this happens either when $m_{\tilde\nu_{1\tau}}\simeq m_{h^0}/2$ so that the annihilation is on resonance, or when co-annihilation channels are important. In the first case, the $\anu$ term must be very small, otherwise the annihilation cross section would be too large and $\Omega h^2$ too small.  
Note that the upper limit on the sneutrino LSP mass is determined by the range for the gluino mass used in the scan which in turn sets an upper bound of 500~GeV on the lightest neutralino and hence on the sneutrino LSP.

The light Higgs mass is not much affected by radiative corrections from a heavy sneutrino, so the posterior PDF of $m_{h^0}$ is like in the conventional MSSM. (See the bottom row of Fig.~\ref{fig:heavy-1d} for Higgs-related quantities.) A light Higgs in the $123\mbox{--}127$~GeV mass range has 21\% probability in this case. As in the MSSM, this mass range requires large mixing, see the distribution for $X_t/M_S$.\footnote{$X_t=A_t-\mu/\tan\beta$ and $M_S^2=m_{\tilde t_1}m_{\tilde t_2}$. In fact the distribution of $A_t$ is the only one that is significantly changed by requiring $m_{h^0} \in [123,127]$~GeV, see also \cite{Brooijmans:2012yi}.}  
The signal strength in the $gg\to h\to\gamma\gamma$ channel relative to SM expectations ($R_{gg\gamma\gamma}$) is also just like in the MSSM~\cite{Brooijmans:2012yi}, with the highest probability being around $R_{gg\gamma\gamma}\approx 0.9$. 
In this scenario, it is much more difficult to reach larger values of $\Delta a_\mu$ as the sneutrino contribution is never large.  
We find $\Delta a_\mu \le 8.6 \times 10^{-10}$ at 95\% BC.

\begin{figure}[t] 
   \centering
   \includegraphics[width=7cm]{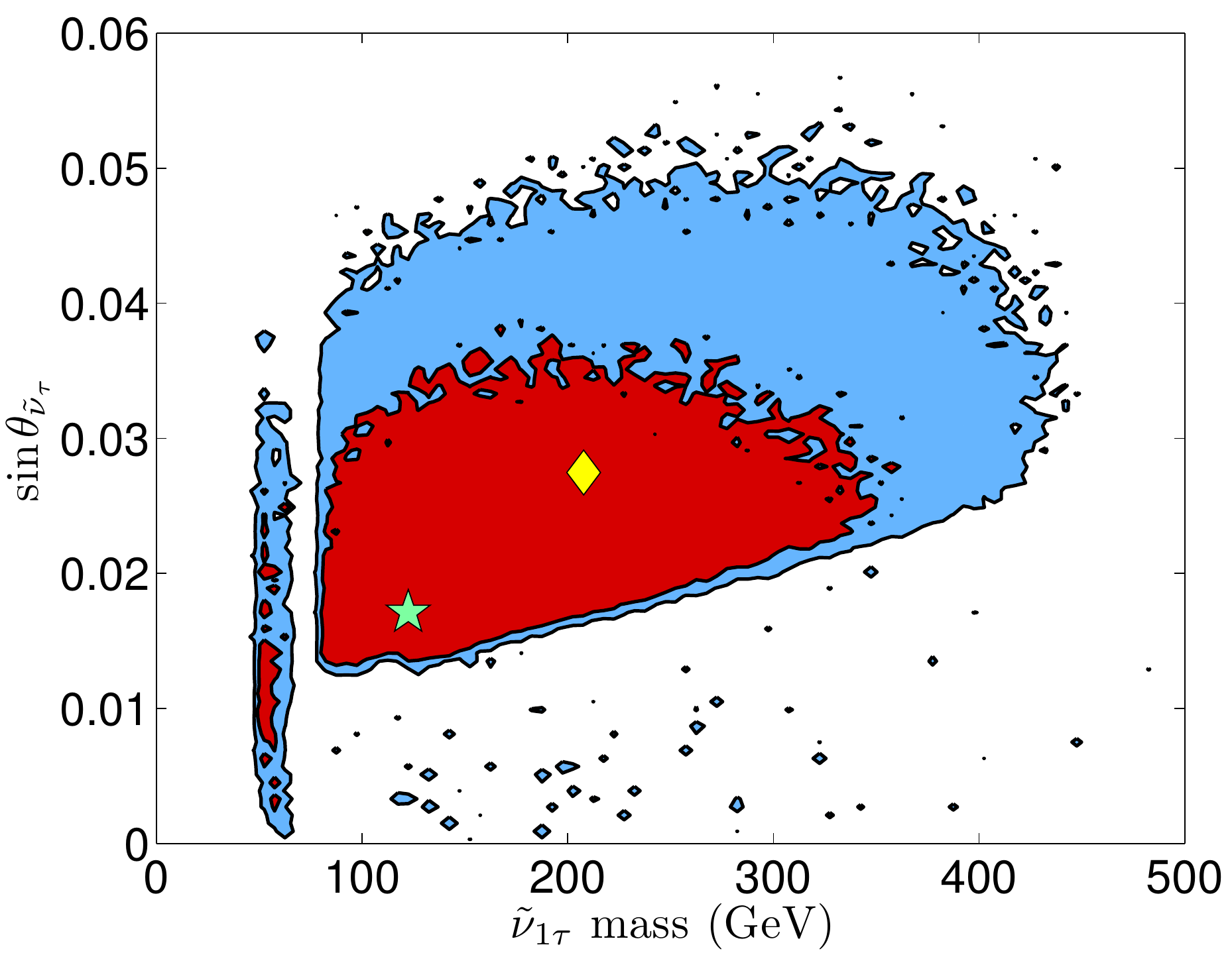} \quad
   \includegraphics[width=7cm]{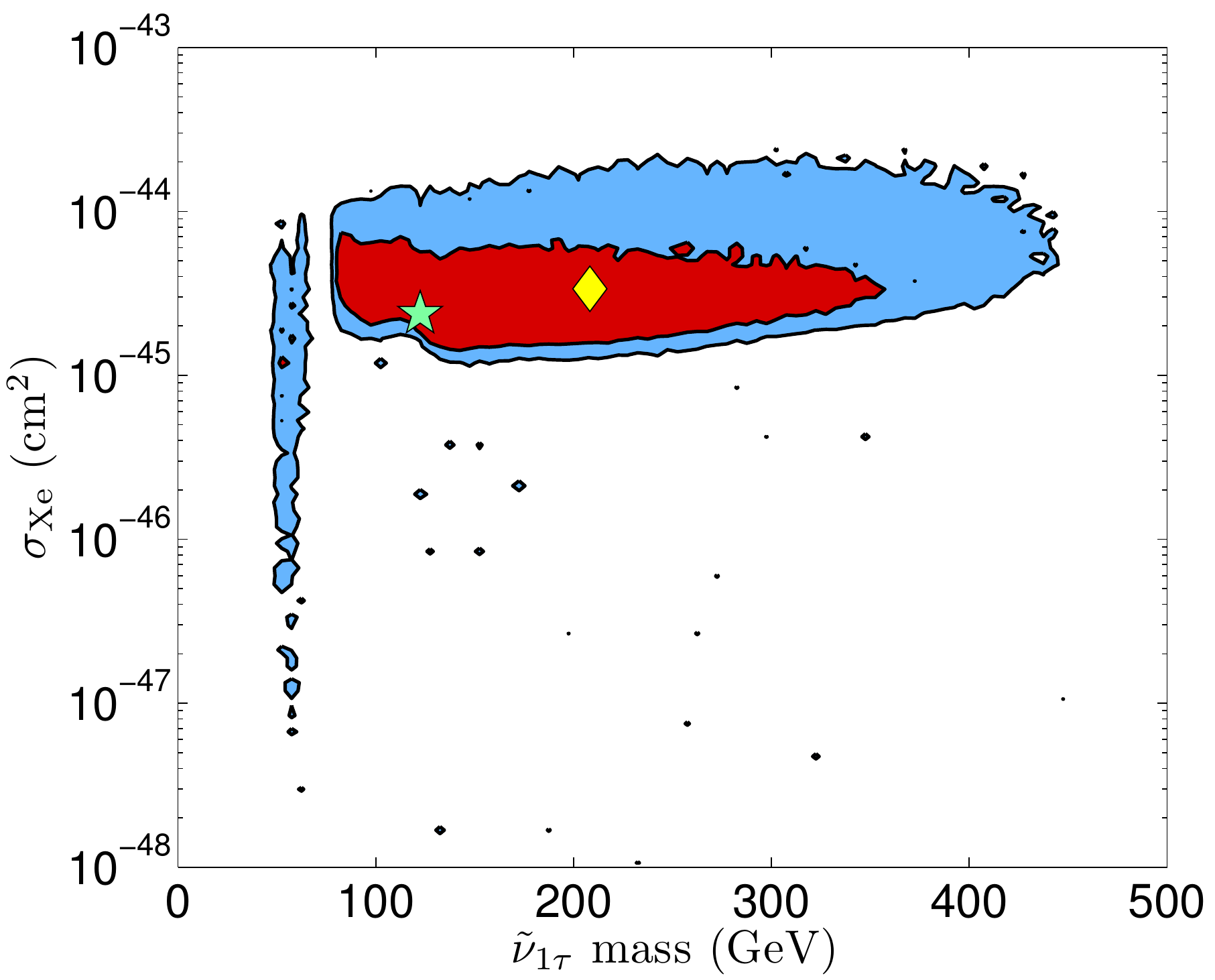}
   \caption{Posterior PDFs in 2D of $\sin\theta_{\tilde{\nu}_{\tau}}$ (left) and $\sigma_{\rm Xe}$ (right) versus $m_{\tilde\nu_{1\tau}}$ for the HND case. The red and blue areas are the  68\% and 95\% BCRs, respectively. The green stars mark the highest posterior, while the yellow diamonds mark the  mean of the  PDF.}
   \label{fig:hnd-2d}
\end{figure}

In Fig.~\ref{fig:hnd-2d}, we show the 2-dimensional posterior PDF of $\sin\theta_{\tilde{\nu}_{\tau}}$ versus $m_{\tilde\nu_{1\tau}}$. 
As can be seen, the mixing angle is always in the $\sin\theta_{\tilde{\nu}_{\tau}}\approx 0.01 -0.05$ region  except when $m_{\lsp}\approx m_{h^0}/2$ or for a few scattered points with heavier LSP masses. The latter correspond to cases where the co-annihilation of pairs of  NLSPs nearly degenerate with the sneutrino LSP helps to increase the effective annihilation cross section, so that the  relic density of the sneutrino is in agreement with WMAP. 
The NLSP can be either a neutralino or a slepton. For the bulk of the points, however, the minimal value of the mixing increases with the sneutrino mass. 
 
The predictions for the SI cross section  are within one order of magnitude of the XENON and CDMS bounds except when $m_{\tilde\nu_{1\tau}}\simeq m_{h^0}/2$ and for the scattered point where coannihilation dominates,  see the right panel in Fig.~\ref{fig:hnd-2d}. Indeed,  when the annihilation in the early Universe is enhanced by a resonance effect, the coupling of the LSP to the Higgs has to be small, hence  one needs a small mixing angle. This also means that the sneutrino coupling to the $Z$ is small, leading to  a small SI cross section. 

\clearpage 

\begin{figure}[t] 
   \centering
   \includegraphics[width=7cm,height=6cm]{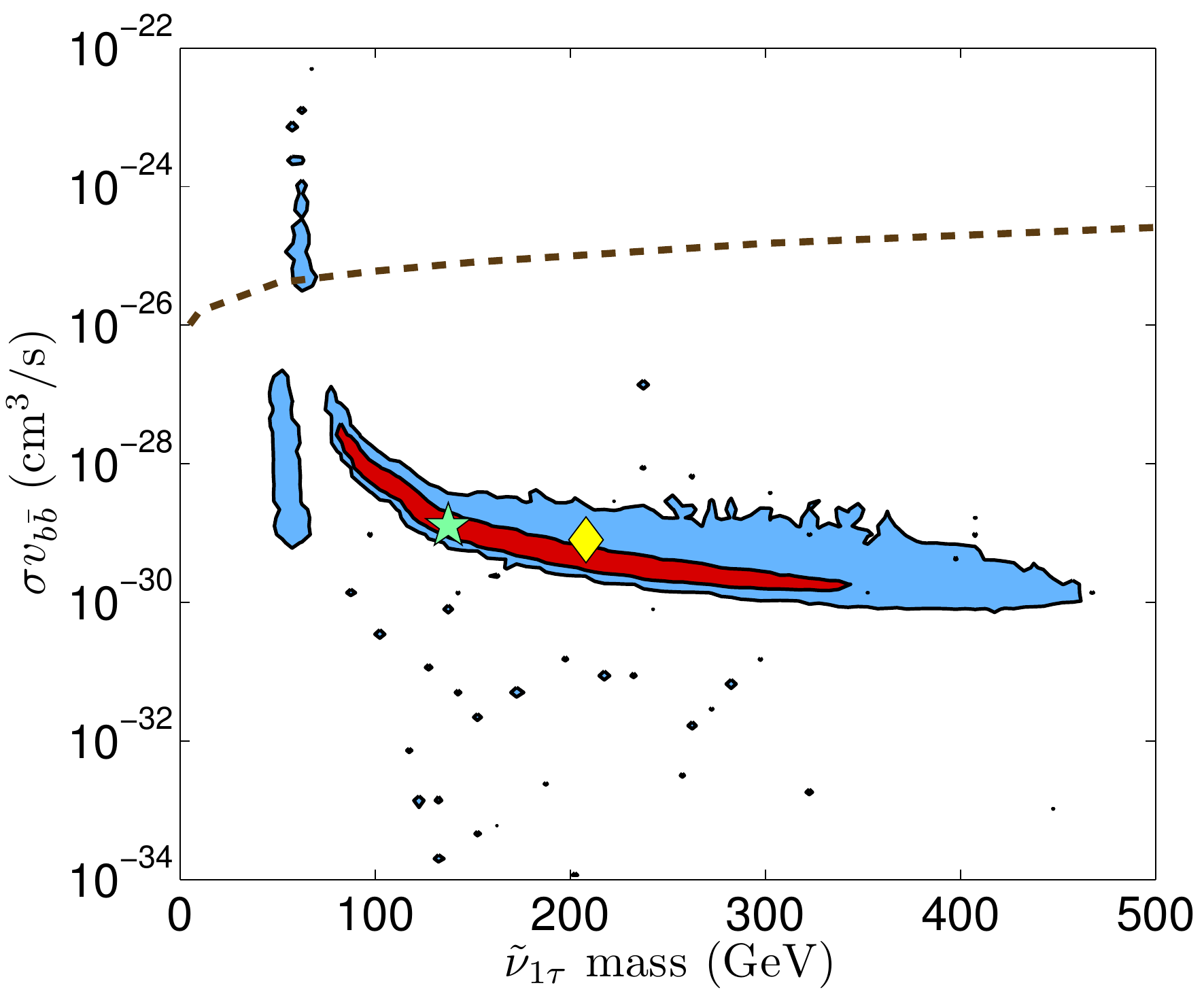} \quad
   \includegraphics[width=7cm,height=5.7cm]{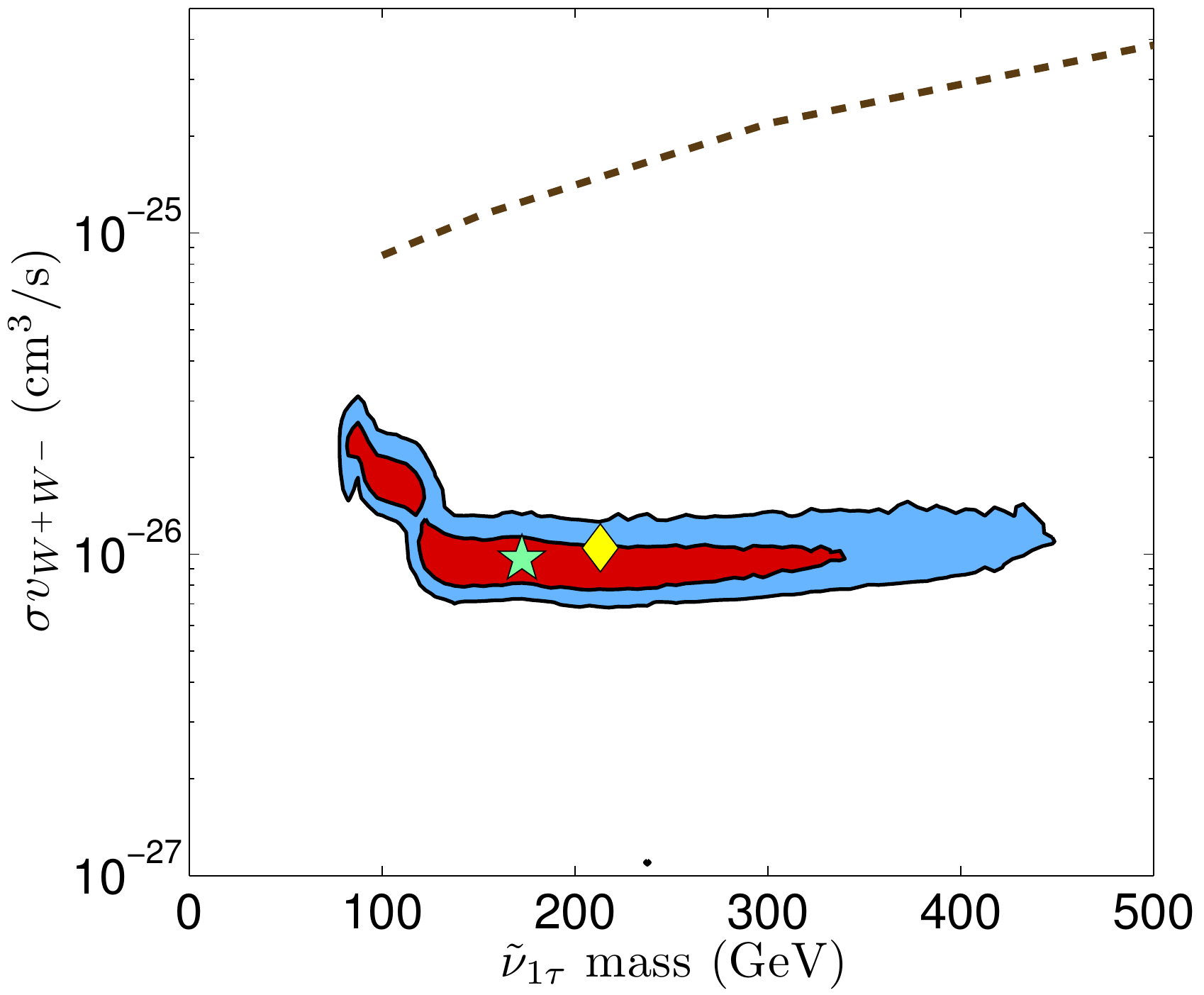}\\
   \includegraphics[width=7cm,height=6cm]{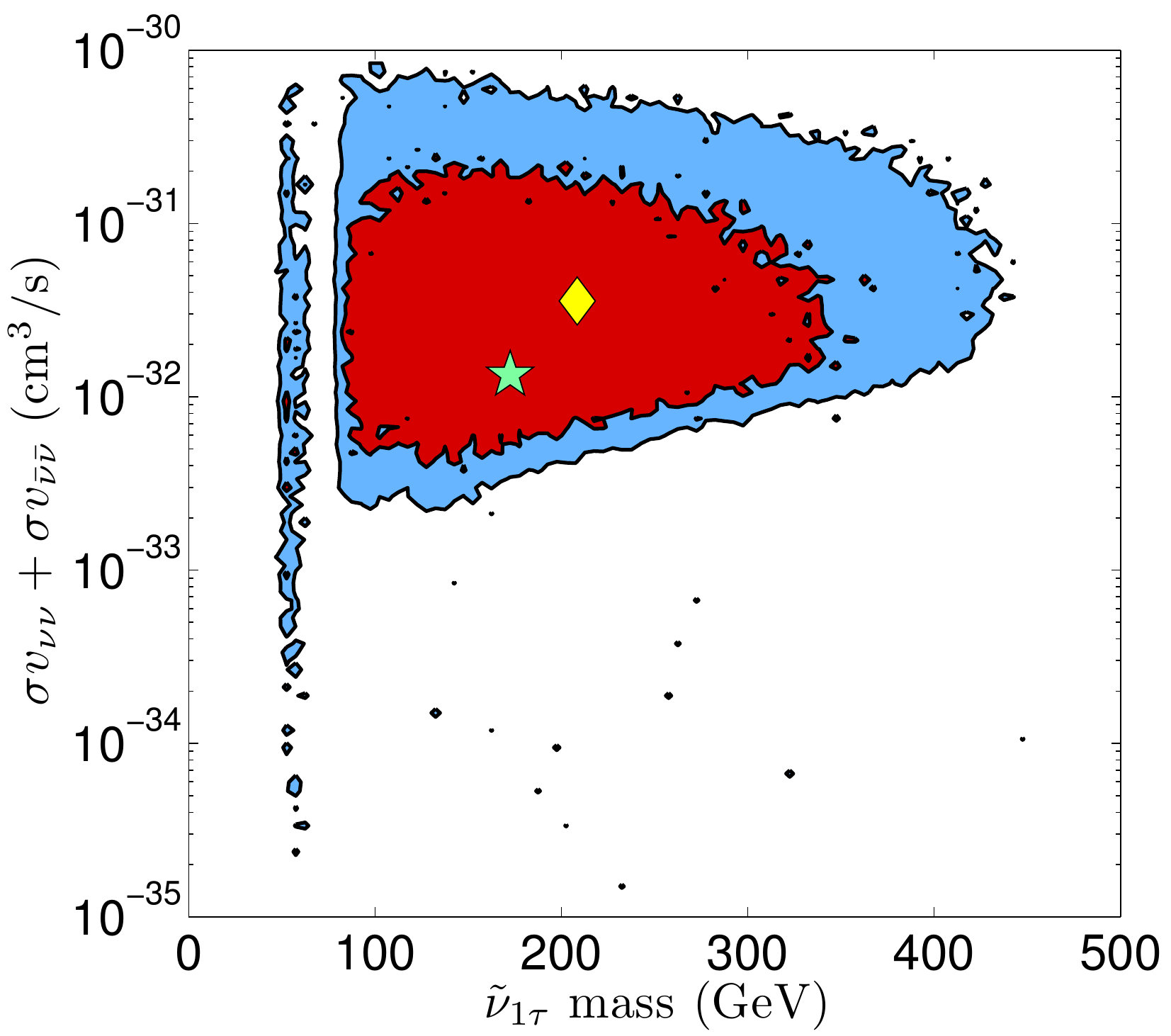} \quad 
   \includegraphics[width=7cm,height=6cm]{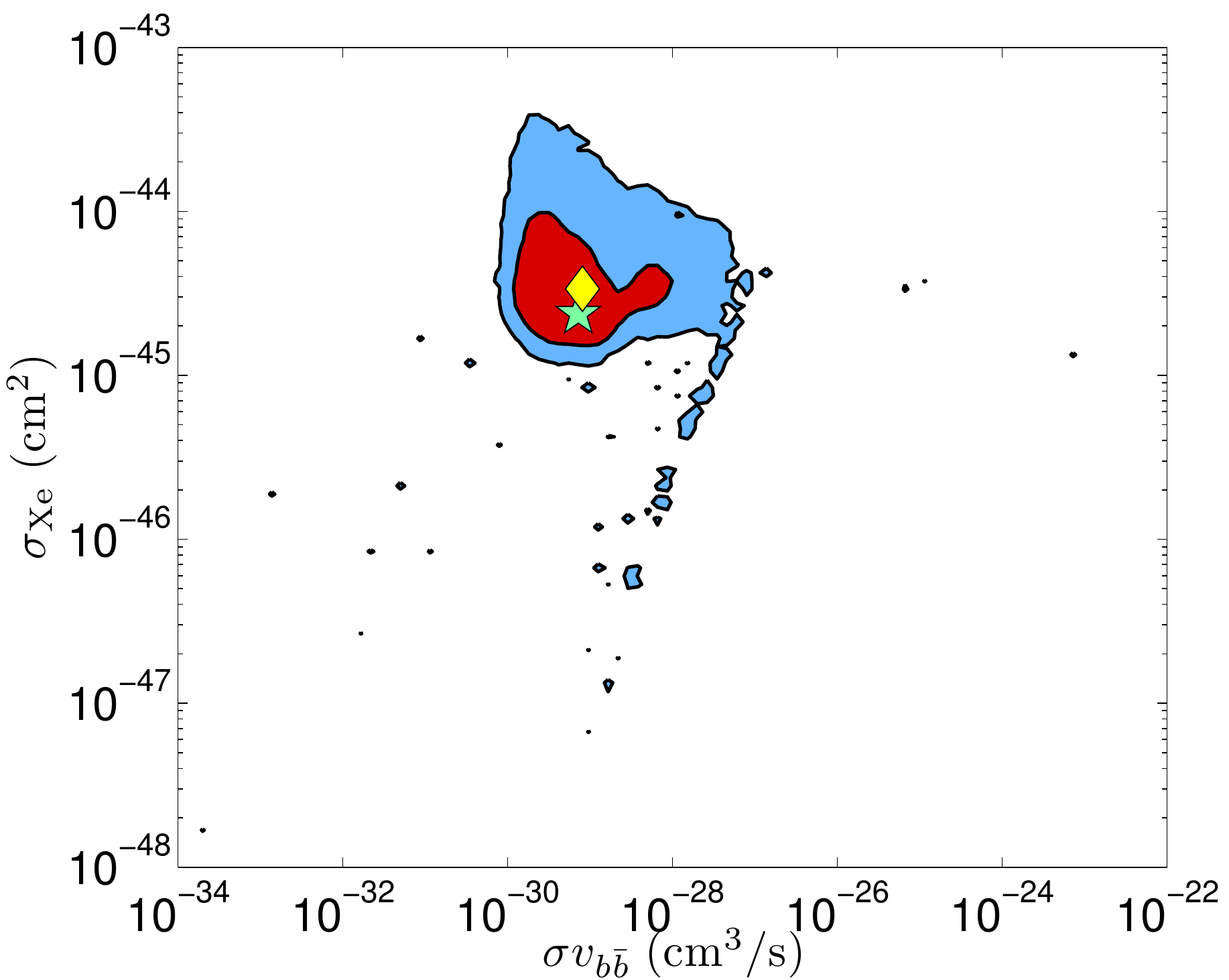}
   \caption{2D posterior PDFs for the HND case relevant for indirect DM detection; color codes etc.~as in Fig~\ref{fig:hnd-2d}.}
   \label{fig:hnd-2d:ann}
\end{figure}

The precise relation between the LSP mass and the Higgs mass has important consequences when we consider annihilation channels in the galaxy.
In some cases, such annihilations can be strongly enhanced with respect to their values in the early Universe. This Breit-Wigner enhancement can occur when the annihilation proceeds through a s-channel exchange of a  Higgs particle near resonance, the cross section is then sensitive to the thermal kinetic energy: at small velocities, one gets the full resonance enhancement while at $v\approx c$, one only catches the tail of the resonance~\cite{Feldman:2008xs,Ibe:2008ye,Bi:2009uj,AlbornozVasquez:2011js}. 
This occurs when $1-m_{h^0}^2/4m^2_{\tilde\nu_{1\tau}}\ll 1$, thus when the annihilation is primarily into $b\bar{b}$. In the left panel in Fig.~\ref{fig:hnd-2d:ann}, a small region at 95\% BC has a photon flux above the limit imposed by {\it Fermi}-LAT. Away from this special kinematical configuration, the annihilation cross section into $b\bar{b}$ is usually two orders of magnitude below the present limit. The dominant annihilation channel is rather into $W$-boson pairs.  Even for this channel, the predictions are  at least one order of magnitude below the {\it Fermi}-LAT limit except when $m_{\tilde\nu_{1\tau}}\approx 100$~GeV, where the predictions are only a factor 2--3 below the limit. The annihilation into neutrino pairs is always subdominant for heavy sneutrinos, with $\sigma v_{\nu\nu} + \sigma v_{\bar{\nu}\bar{\nu}} < 10^{-30}~\rm{cm}^3 /{\rm s}$.

Note that even after removing the points that are excluded by {\it Fermi}-LAT in the $b\bar{b}$ channel, the predictions for $\sigma_{\rm Xe}$ extend to small values. Indeed for these points  there is no large enhancement of the annihilation rate in the early Universe, hence no need to have small  couplings of the LSP to the Higgs. Therefore the predictions for the SI cross section covers a wide range and is not correlated with $\sigma v_{b\bar{b}}$, see the bottom right plot in Fig.~\ref{fig:hnd-2d:ann}.

We have also computed the predictions for the antiproton flux for the heavy sneutrino case. The largest fluxes are expected for DM masses around 100 GeV where the annihilation cross section can reach $3\times 10^{-26}{\rm cm^3}/{\rm s}$.  We found that with the MED propagation parameters the flux is barely above the background and always within the 1$\sigma$ experimental error bars. Note that a large flux is also expected  for the few points that have a large annihilation into $b\bar{b}$, these points are however already excluded by {\it Fermi}-LAT as discussed above.

\begin{figure}[ht] 
   \centering
   \includegraphics[width=4.96cm]{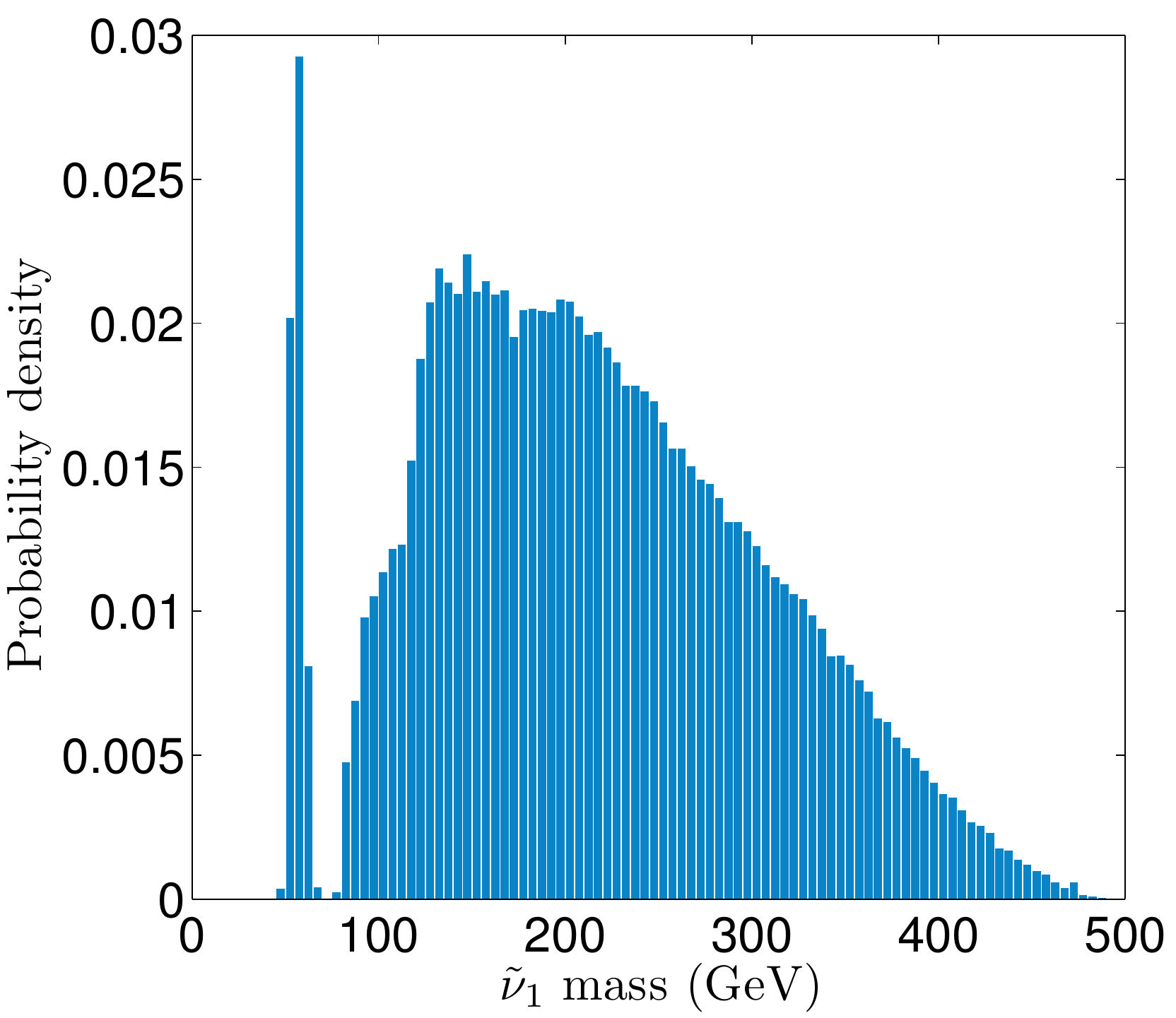}
   \includegraphics[width=4.96cm]{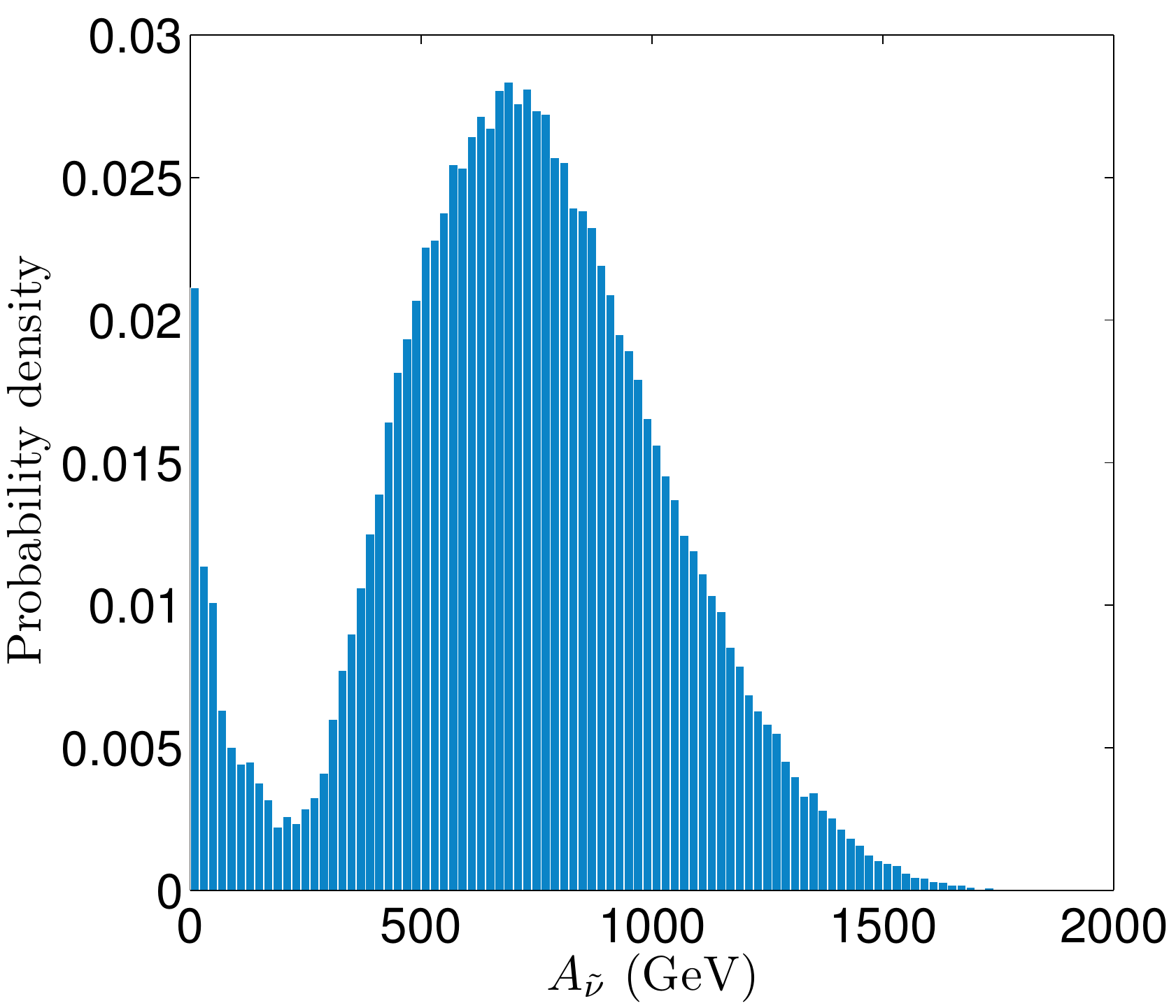}
   \includegraphics[width=4.96cm]{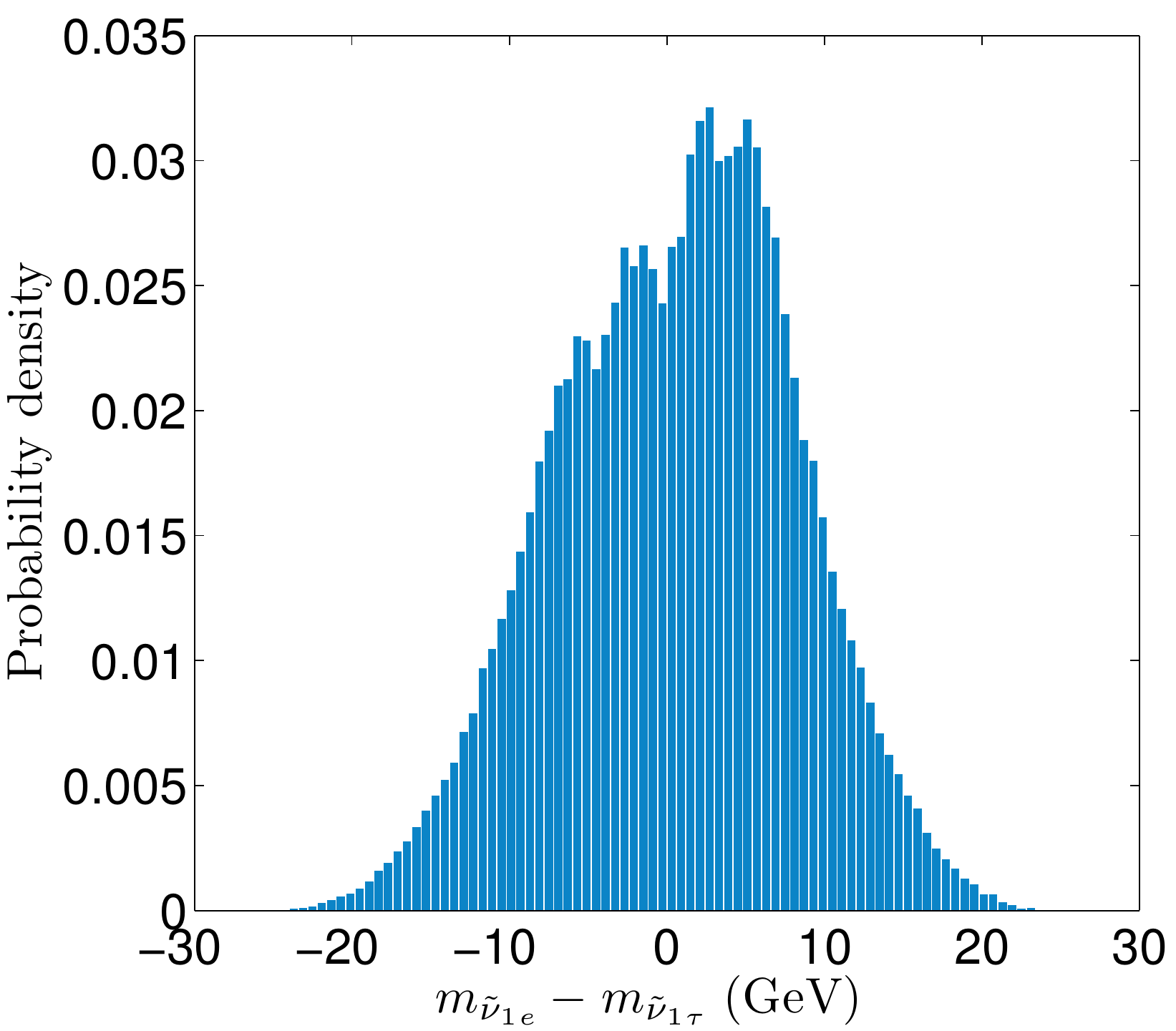}
   \caption{Posterior PDFs in 1D for the heavy democratic (HD) sneutrino case. All other distributions are practically the same as in the HND case.}
   \label{fig:heavy-1d-demo}
\end{figure}

The results discussed above for the HND case also hold for the HD sneutrino case. 
In fact, most of the distributions in the HD case are practically the same as in the HND case. 
The only differences are observed for the LSP mass, and for the associated $A_{\tilde\nu}$, see Fig.~\ref{fig:heavy-1d-demo}. 
We note a slightly higher probability of 6\% to be on the $h^0$ pole. Correspondingly, also small $\anu$ and small mixing angles have somewhat higher probability than in the HND case. Regarding the flavor of the LSP, we find that a $\tau$-sneutrino LSP has 55\% probability and is thus, as expected, somewhat preferred over $e/\mu$ sneutrino co-LSPs (45\% probability), see the right-most panel in Fig.~\ref{fig:heavy-1d-demo}. 
The fact that the $\tilde\nu_{1e}$--$\tilde\nu_{1\tau}$ mass difference peaks within $\pm 10$~GeV is however just a consequence of our prior assumption for the HD case. 

%================================================================================
\section{Conclusion}\label{sec:conclusions}
%================================================================================

We performed a global MCMC analysis of a sneutrino DM model with Dirac neutrino masses originating from supersymmetry breaking. The main feature of this model is a mainly RH mixed sneutrino as the LSP, which has a large coupling to the Higgs fields through a weak-scale trilinear $A$ term which is not suppressed by small Dirac-neutrino Yukawa couplings.  
We demonstrated that such a RH sneutrino can be an excellent cold dark matter candidate over a wide range of masses. In particular, it can be consistent with all existing constraints for masses around 3--6~GeV, as well as for masses of about 50--500~GeV (the upper limit coming from the fact that we consider gluino masses only up to 3 TeV). 

Direct detection limits in particular from XENON10 heavily constrain the low mass range. The DD cross section however sensitively depends on, e.g., the escape velocity in the light DM case. We therefore took special care to account for uncertainties arising from astrophysical parameters, like $v_0$,  $v_{\rm esc}$ and the local DM density $\rho_{\rm DM}$. Moreover, we accounted for  uncertainties from the quark contents of the nucleon, relevant for the Higgs exchange contribution to the DD cross section. 

Our main results are posterior probability distributions of parameters, masses, and derived observables---in particular the LSP mass and the direct and indirect detection cross sections. Assuming gaugino-mass unification, the recent LHC limits on the gluino mass exclude the very light sneutrino DM region below about 3~GeV, where the DD limits are not efficient. To be precise, requiring $m_{\tilde g}>1$~TeV  leads to $2.9<\mlsp< 5.6$~GeV at 95\% BC.  For heavy sneutrinos of the order of 100~GeV, the gluino is always heavy so that the LHC limits have no effect on sneutrino DM. 

Regarding the prospects for probing light sneutrino DM, we found that covering the 95\% BC region requires about a factor three  increase in sensitivity in direct detection for DM masses around 5 GeV, as well as  a lower threshold to be able to probe masses below 4~GeV. Similarly, the prospects for indirect detection through photons and antiprotons are promising if the sensitivity of experiments can be extended to lower masses. The crucial test however comes form the LHC: in the light sneutrino DM scenario the Higgs decays dominantly invisibly into sneutrinos. Therefore if the Higgs-like excess around 125 GeV is confirmed,  the light sneutrino model is ruled out. 

The heavy sneutrino scenario can also be probed by DD experiments, this requires an increase in sensitivity of roughly one order of magnitude over the current limits. Only  a small region where the sneutrino has about half the mass of the light Higgs would remain out of reach in this case. Such a scenario should lead to a Higgs signal that is compatible with the SM. Prospects of indirect detection are more challenging for the heavy sneutrino. 

Both, light and heavy, sneutrino scenarios offer distinctive LHC SUSY phenomenology. In particular neutralinos (typically $\tilde{\chi}^0_1$ and $\tilde{\chi}^0_2$) appearing in squark and gluino cascades can decay invisibly into the LSP. Indeed the probability for a 90\% invisible decay of the lightest (next-to-lightest) neutralino is about 80\% (50\%) for the light sneutrino scenario, and 
close to 100\% (30--40\%) in the heavy sneutrino scenarios, see Table~\ref{tab:branching}. This implies that there can be up to three different invisible 
sparticles in an event. The dominant decay of charginos (with a branching fraction larger than 0.5) is  into a charged lepton and the LSP 
with roughly 50\% probability.  The charged lepton is typically a $\tau$  for the light and heavy non-democratic scenarios or a $e$/$\mu$ for the heavy democratic scenario. 
The cascade decays of squarks, $\tilde q_R\rightarrow q \tilde{\chi}^0_1\rightarrow q \tilde\nu_1 \nu$, 
$\tilde q_L\rightarrow q \tilde{\chi}^0_2\rightarrow q \tilde\nu_1 \nu$, 
$\tilde q_L\rightarrow q' \tilde{\chi}^+_1\rightarrow q' l \tilde\nu_1$ therefore give different amount of missing energy as compared to the MSSM.
Furthermore the cascade decays of gluinos, $\tilde g\rightarrow \tilde{\chi}^0_i jj$ will also give a large contribution to the jets plus missing $E_T$ channel while the decay of gluino pairs via a chargino will give about the same amount of same-sign and opposite-sign lepton pairs.
Note that the alternative  dominant decay mode of the chargino  is $\tilde{\chi}^\pm_1\rightarrow W^\pm \tilde{\chi}^0_1$; in this case the mass of the invisible particle could be much larger than the DM mass.  
Probabilities for $\tilde{\chi}^0_1$, $\tilde{\chi}^0_2$ and $\tilde{\chi}^\pm_1$ decays in the light, HND and HD scenarios are summarized in Table~\ref{tab:branching}.

\begin{table}[t]
\begin{center}
\begin{tabular}{|l|c|c|c|}
\hline
       & light & HND & HD \\
\hline
${\cal B}(\tilde{\chi}^0_1\to {\rm inv})>0.9$ & 79\% & 96\% & 98\% \\
\hline
${\cal B}(\tilde{\chi}^0_2\to {\rm inv})>0.9$ & 53\% & 29\% & 42\% \\
\hline
${\cal B}(\tilde{\chi}^\pm_1\to \ell^\pm\,\tilde\nu_{1\ell})>0.5$ &
7\% & 9\% & 48\% \\
\hline
${\cal B}(\tilde{\chi}^\pm_1\to \tau^\pm\,\tilde\nu_{1\tau})>0.5$ &
46\% & 46\% & 10\% \\
\hline
\end{tabular}
\caption{\label{tab:branching}
Probabilities for neutralino and chargino decays in the
light, HND and HD sneutrino DM cases, requiring $m_{\tilde g}>1$~TeV.} 
\end{center}
\end{table}

Distinctive features of the light sneutrino scenario at the LHC were investigated in ~\cite{Belanger:2011ny} and  it was shown 
that distributions such as lepton and jet number as well as same-sign/opposite-sign dilepton rates could give a distinctive signature of a light sneutrino at the LHC.
A detailed analysis of the LHC sensitivity in the heavy sneutrino DM model, based on publicly available ATLAS and/or CMS results for different signal topologies, is underway. 

\bigskip 
{\bf Note added:} After the completion of this work, ATLAS and CMS announced a $\sim 5 \sigma$
discovery consistent with a SM-like Higgs boson around 125--126~GeV \cite{HiggsATLAS, HiggsCMS}. This
excludes the light sneutrino scenario discussed in our work, while the heavy
sneutrino case remains viable.
Moreover, new XENON100 results~\cite{XENON2012} became available, constraining
the heavy sneutrino case for $\sin \theta_{\tilde{\nu}}$ above 0.03.
For flat priors, the probability to obey the new 90\% CL exclusion
limit in the HND case is 62\% (54\% with logarithmic priors). Our
conclusions remain unchanged.

%================================================================================
\acknowledgments
%================================================================================

We gratefully acknowledge helpful discussions with C.~Arina (on MCMCs),  
J.~Cohen-Tanugi (on the $Fermi$-LAT limits), 
M. Kakizaki (on RGEs) and A.~Pukhov (on questions regarding \texttt{micrOMEGAs}). 
The work of B.D., S.K., and T.S. is partially supported from the  European Union FP7  ITN INVISIBLES (Marie Curie Actions, PITN-GA-2011-289442).

\clearpage

%================================================================================
\appendix
\section{Discussion of $T_{\rm QCD}/g_{\rm eff}/h_{\rm eff}$}
%================================================================================

\label{aTQCD}

In the standard freeze-out picture, $\Omega h^2$ is inversely proportional to the number of effective degrees of freedom, $g_{\rm eff}$. At the temperature where the QCD confinement occurs, around $T_{\rm QCD}\approx 300{\rm \ MeV}$, $g_{\rm eff}$ starts to drop and $\Omega h^2$ increases. This is relevant for DM masses below ca.\ 7~GeV, where the freeze-out temperature $T_f \approx m_{\rm DM}/20$ is of the order of $T_{\rm QCD}$.  

The uncertainty in the equation of state (actually in the effective degrees of freedom $g_{\rm eff}$ and $h_{\rm eff}$ contributing to the energy and entropy densities of the SM) at temperatures around $T_{\rm QCD}$ induces a non-negligible uncertainty in the calculated 
$\Omega h^2$ for light DM. In  \cite{Hindmarsh:2005ix}, Hindmarsh and Philipsen estimated this uncertainty  to be at around the 15\% level. They also provided five tables (A, B, B2, B3 and C) describing the evolution of the effective degrees of freedom in the early Universe using different parameters for the equation of state and different values for the temperature at which there is a sharp switch between quarks and gluons, and hadrons and their resonances. 

\begin{figure}[htb] 
   \centering
   \includegraphics[width=7cm]{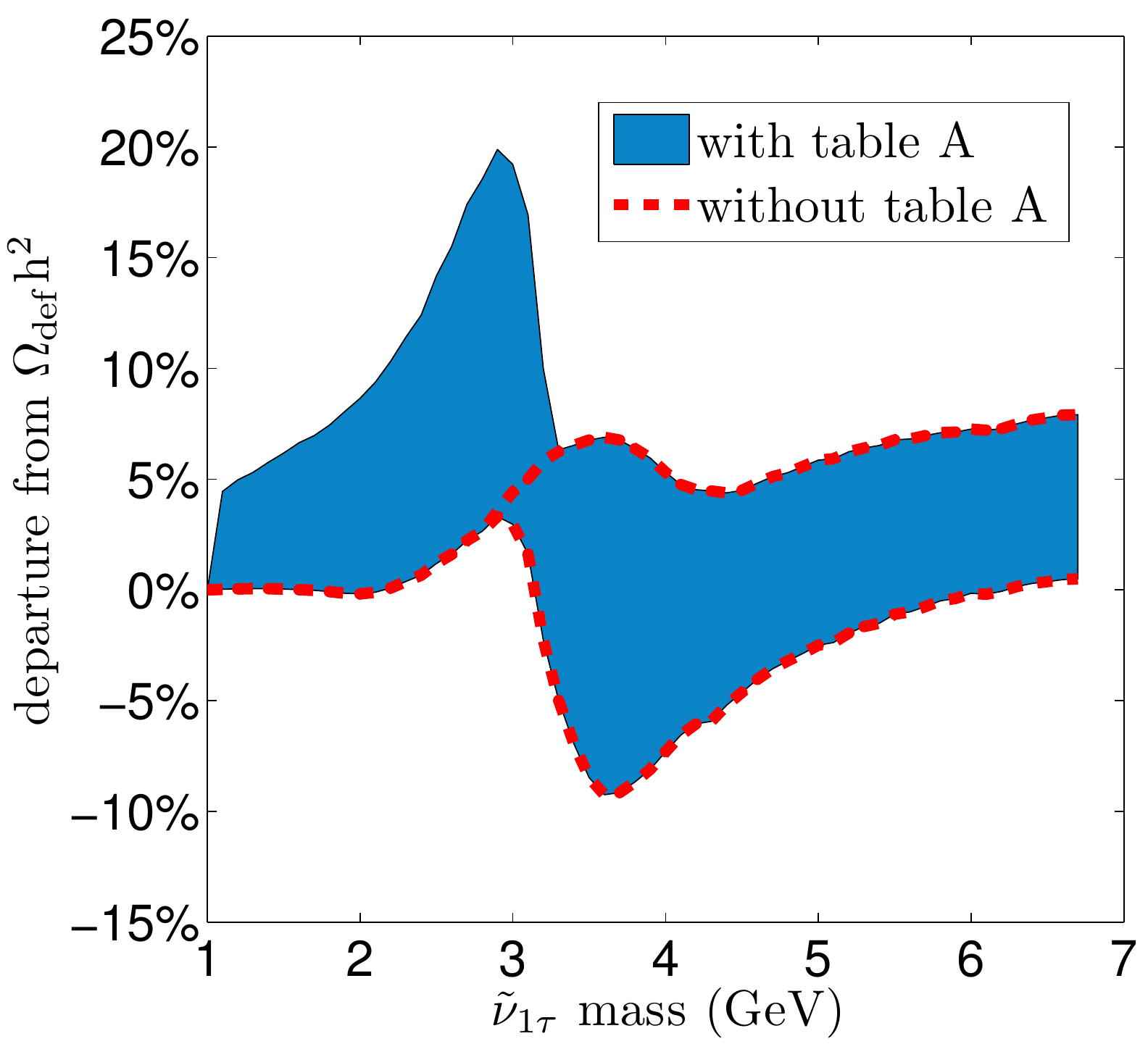}
   \caption{Maximum variations of $\Omega h^2$ for light sneutrinos, due to different evolution of the effective degrees of freedom from the tables of~\cite{Hindmarsh:2005ix} as compared to the default table in \texttt{micrOMEGAs}. Unlike the other tables, Table A ignores hadrons in the confined phase.}
   \label{fig:omega-TQCD}
\end{figure}

Taking 5000 sample points from our light sneutrino sample 
we compute $\Omega h^2$ using the five tables of \cite{Hindmarsh:2005ix} and compare it to the ``default'' value $\Omega_{\rm def} h^2$ 
obtained with the default \texttt{micrOMEGAs} table. The result is shown in Fig.~\ref{fig:omega-TQCD}.
We note that  the variations in the computed $\Omega h^2$ (relative to the default value of \texttt{micrOMEGAs}) can be as large as 20\%. 
The main upward fluctuation is due to table~A, which corresponds to a case where hadrons are ignored in the confined phase. 

We do not take this into account as additional uncertainty in our analysis, as a somewhat larger uncertainty in $\Omega h^2$ for DM masses below about 5~GeV would not sensitively impact our results. However, we note that the situation is unsatisfactory and would merit further study.

\clearpage

%================================================================================
\section{Tables: Bayesian credible intervals for parameters and observables}
%================================================================================

\label{aTables}

We provide in Tables~\ref{tab:bci-light}, \ref{table:bci-hnd} and \ref{table:bci-hd} the 68\% and 95\% BCIs (Bayesian credible intervals), which we define as highest posterior density intervals, for several parameters and observables, in the case of light, HND (heavy non-democratic) and HD (heavy democratic) sneutrinos. For each case, we also provide information on the  ``best fit point'' (the point with highest likelihood in the MCMC sampling), and a so-called ``quasi-mean point'' which is close to the mean of our parameters. Due to correlations between parameters in order to respect the constraints, as well as asymmetric and multimodal distributions, the mean point itself is very unlikely. 
Thus we pick in our samples the closest point to the mean with a good likelihood as an example of a typical point.
The SLHA files for these points are available as supplementary material on the {\tt arXiv}.

\begin{table}[h]
\begin{center}\vspace*{4mm}
\begin{tabular}{|c|c|c|c|c|}
\hline
 & 68\% BCI & 95\% BCI & best fit & quasi-mean \\
 &          &          & point    & point \\
\hline
$m_{\tilde{\nu}_{1\tau}}$~(GeV) & $[3.4,\,4.9]$ & $[2.9,\,5.6]$ & 3.0 & 4.4 \\
$m_{\tilde{\nu}_{2\tau}}$~(GeV) & $[480,\,1080]$ & $[420,\,2250]$ & 660 & 480 \\
$\sin\theta_{\tilde{\nu}_{\tau}}$ & $[0.20,\,0.39]$ & $[0.07,\,0.40]$ & 0.35 & 0.38 \\
$A_{\tilde\nu_\tau}$~(GeV) & $[480,\,1240]$ & $[360,\,2450]$ & 818 & 463 \\
$m_{\tilde{\nu}_{1e}}$~(GeV) & $[16.7,\,41.0]$ & $[8.0,\,44.5]$ & 30.3 & 31.2 \\
$m_{\tilde{\nu}_{2e}}$~(GeV) & $[60,\,870]$ & $[60,\,2400]$ & 514 & 313 \\
$\sin\theta_{\tilde{\nu}_{e}}$ & $[0,\,0.15]$ & $[0,\,0.34]$ & 0.01 & 0.10 \\
$A_{\tilde\nu_e}$~(GeV) & $[0,\,300]$ & $[0,\,1050]$ & 18 & 55 \\
$\tan \beta$ & $[7.0,\,30.8]$ & $[4.2,\,52.0]$ & 46.8 & 25.7 \\
$\mu$~(GeV) & $[-240,\,-120]$ & $[-2000,\,-60]$ & 334 & 485 \\
            & $\cup\ [60,\,1800]$ & $\cup\ [60,\,2950]$ & & \\
$A_t$~(GeV) & $[-7100,\,-2700]$ & $[-7900,\,-1100]$ & $-4404$ & $-319$ \\
            & $\cup\ [3400,\,5900]$ & $\cup\ [1200,\,7700]$ & & \\
$M_A$~(GeV) & $[1400,\,2900]$ & $[670,\,3000]$ & 892 & 1877 \\

$m_{h^0}$~(GeV) & $[114,\,119]$ & $[114,\,126]$ & 114.4 & 115.4 \\

$m_{\tilde{g}}$~(GeV) & $[1000,\,2200]$ & $[1000,\,2800]$ & 1117 & 1021 \\
$m_{\tilde{t}_1}$~(GeV) & $[2050,\,2800]$ & $[1450,\,2950]$ & 2677 & 2298 \\

$m_{\tilde{e}_R}$~(GeV) & $[90,\,870]$ & $[90,\,2300]$ & 520 & 321 \\
$m_{\tilde{\tau}_1}$~(GeV) & $[400,\,1050]$ & $[300,\,2300]$ & 600 & 427 \\

$m_{\tilde{\chi}^{0}_1}$~(GeV) & $[125,\,320]$ & $[100,\,440]$ & 142 & 132 \\
$m_{\tilde{\chi}^{0}_2}$~(GeV) & $[120,\,550]$ & $[110,\,850]$ & 274 & 275 \\
$m_{\tilde{\chi}^{+}_1}$~(GeV) & $[110,\,550]$ & $[96,\,850]$ & 273 & 275 \\

$\Omega h^2$ & $[0.10,\,0.13]$ & $[0.09,\,0.14]$ & 0.11 & 0.12 \\
$\sigma_{\rm Xe} \times 10^{40}$~(cm$^2$) & $[2,\,3]$ & $[1,\,23]$ & 19 & 4 \\
                                          & $\cup\ [6,\,18]$ & & & \\

${\cal B}(b \rightarrow s\gamma) \times 10^4$ & $[3.2,\,3.6]$ & $[3.0,\,3.8]$ & 3.5 & 3.4 \\
${\cal B}(B_s \rightarrow \mu^+\mu^-) \times 10^9$ & $[2.9,\,3.3]$ & $[2.5,\,4.5]$ & 3.9 & 3.0 \\

$\Delta a_{\mu} \times 10^{10}$ & $[-3,\,18]$ & $[-6,\,36]$ & 25 & 20 \\
\hline
\end{tabular}
\caption{\label{tab:bci-light} 68\% and 95\% BCIs and values of two example points for various parameters and observables in the light sneutrino case with $m_{\tilde{g}} > 1$~TeV.}
\end{center}
\end{table}

\begin{table}[t]
\begin{center}
\begin{tabular}{|c|c|c|c|c|}
\hline
 & 68\% BCI & 95\% BCI & best fit & quasi-mean \\
 &          &          & point    & point \\
\hline
$m_{\tilde{\nu}_{1\tau}}$~(GeV) & $[53,\,56]$ & $[49,\,63]$ & 188 & 193 \\
                                & $\cup\ [90,\,255]$ & $\cup\ [80,\,375]$ & & \\
$m_{\tilde{\nu}_{2\tau}}$~(GeV) & $[1400,\,2600]$ & $[1100,\,3000]$ & 1813 & 1919 \\
$\sin\theta_{\tilde{\nu}_{\tau}}$ & $[0.016,\,0.033]$ & $[0.013,\,0.049]$ & 0.028 & 0.026 \\
$A_{\tilde\nu_\tau}$~(GeV) & $[300,\,750]$ & $[200,\,1100]$ & 516 & 547 \\
$m_{\tilde{\nu}_{1e}}$~(GeV) & $[350,\,1800]$ & $[200,\,2500]$ & 1116 & 863 \\
$m_{\tilde{\nu}_{2e}}$~(GeV) & $[1300,\,2700]$ & $[600,\,2950]$ & 1262 & 1325 \\
$\sin\theta_{\tilde{\nu}_{e}}$ & $[0,\,0.43]$ & $[0,\,0.65]$ & 0.09 & 0.35 \\
                               & $\cup\ [0.96,\,1]$ & $\cup\ [0.78,\,1]$ & & \\
$A_{\tilde\nu_e}$~(GeV) & $[0,\,1750]$ & $[0,\,3600]$ & 180 & 1900 \\
$\tan \beta$ & $[5.7,\,33.0]$ & $[2.9,\,54.0]$ & 63.1 & 37.2 \\
$\mu$~(GeV) & $[-800,\,-420]$ & $[-2620,\,-300]$ & 361 & 512 \\
            & $\cup\ [180,\,2900]$ & $\cup\ [180,\,3000]$ & & \\
$A_t$~(GeV) & $[-4200,\,3000]$ & $[-6700,\,6200]$ & 2005 & $-216$ \\
$M_A$~(GeV) & $[1350,\,2850]$ & $[700,\,3000]$ & 2706 & 588 \\

$m_{h^0}$~(GeV) & $[115,\,123]$ & $[114,\,129]$ & 123.6 & 119.1 \\

$m_{\tilde{g}}$~(GeV) & $[2000,\,2950]$ & $[1350,\,3050]$ & 1417 & 1863 \\
$m_{\tilde{t}_1}$~(GeV) & $[1950,\,2900]$ & $[1150,\,3000]$ & 1623 & 2475 \\

$m_{\tilde{e}_R}$~(GeV) & $[1100,\,2650]$ & $[500,\,2900]$ & 1264 & 1280 \\
$m_{\tilde{\tau}_1}$~(GeV) & $[1400,\,2600]$ & $[1050,\,2950]$ & 1803 & 1911 \\

$m_{\tilde{\chi}^{0}_1}$~(GeV) & $[280,\,480]$ & $[170,\,500]$ & 201 & 270 \\
$m_{\tilde{\chi}^{0}_2}$~(GeV) & $[500,\,950]$ & $[250,\,1000]$ & 341 & 488 \\
$m_{\tilde{\chi}^{+}_1}$~(GeV) & $[500,\,950]$ & $[250,\,1000]$ & 339 & 487 \\

$\Omega h^2$ & $[0.10,\,0.12]$ & $[0.09,\,0.13]$ & 0.11 & 0.11 \\
$\sigma_{\rm Xe} \times 10^{45}$~(cm$^2$) & $[2,\,5]$ & $[1,\,20]$ & 3 & 3 \\

${\cal B}(b \rightarrow s\gamma) \times 10^4$ & $[3.2,\,3.5]$ & $[3.0,\,3.8]$ & 3.7 & 3.8 \\
${\cal B}(B_s \rightarrow \mu^+\mu^-) \times 10^9$ & $[2.9,\,3.3]$ & $[2.4,\,4.7]$ & 2.4 & 2.3 \\

$\Delta a_{\mu} \times 10^{10}$ & $[-1.3,\,2.3]$ & $[-3.7,\,8.6]$ & 9.4 & 4.5 \\
\hline
\end{tabular}
\caption{\label{table:bci-hnd} 68\% and 95\% BCIs and values of two example points for various parameters and observables in the heavy non-democratic sneutrino case.}
\end{center}
\end{table}

\begin{table}[t]
\begin{center}
\begin{tabular}{|c|c|c|c|c|}
\hline
 & 68\% BCI & 95\% BCI & best fit & quasi-mean \\
 &          &          & point    & point \\
\hline
$m_{\tilde{\nu}_{1\tau}}$~(GeV) & $[51,\,61]$ & $[51,\,66]$ & 63 & 230 \\
                                & $\cup\ [115,\,280]$ & $\cup\ [85,\,385]$ & & \\
$m_{\tilde{\nu}_{2\tau}}$~(GeV) & $[1600,\,2700]$ & $[450,\,500]$ & 430 & 1703 \\
                                & & $\cup\ [1000,\,3000]$ & & \\
$\sin\theta_{\tilde{\nu}_{\tau}}$ & $[0.020,\,0.038]$ & $[0.014,\,0.054]$ & 0.014 & 0.022 \\
$A_{\tilde\nu_\tau}$~(GeV) & $[0,\,20]$ & $[0,\,100]$ & 15 & 363 \\
                           & $\cup\ [450,\,1000]$ & $\cup\ [300,\,1300]$ & & \\
$m_{\tilde{\nu}_{1e}}$~(GeV) & $[51,\,61]$ & $[51,\,66]$ & 62 & 239 \\
                             & $\cup\ [115,\,280]$ & $\cup\ [85,\,385]$ & & \\
$m_{\tilde{\nu}_{2e}}$~(GeV) & $[1600,\,2700]$ & $[450,\,500]$ & 420 & 1747 \\
                             & & $\cup\ [1000,\,3000]$ & & \\
$\sin\theta_{\tilde{\nu}_{e}}$ & $[0.020,\,0.038]$ & $[0.014,\,0.054]$ & 0.015 & 0.020 \\
$A_{\tilde\nu_e}$~(GeV) & $[0,\,20]$ & $[0,\,100]$ & 15 & 353 \\
                        & $\cup\ [450,\,1000]$ & $\cup\ [300,\,1300]$ & 15 & 353 \\
$\tan \beta$ & $[4.9,\,32.2]$ & $[3.5,\,55.0]$ & 50.9 & 17.6 \\
$\mu$~(GeV) & $[-1800,\,-1400]$ & $[-2800,\,-350]$ & 151 & 244 \\
            & $\cup\ [-800,\,-550]$ & $\cup\ [100,\,2900]$ & & \\
            & $\cup\ [180,\,2800]$ & & & \\
$A_t$~(GeV) & $[-3800,\,3200]$ & $[-6700,\,6100]$ & $-4878$ & $-218$ \\
$M_A$~(GeV) & $[1400,\,2900]$ & $[650,\,3000]$ & 1082 & 1010 \\

$m_{h^0}$~(GeV) & $[116,\,124]$ & $[114,\,129]$ & 127.6 & 115.4 \\

$m_{\tilde{g}}$~(GeV) & $[2000,\,2950]$ & $[1350,\,3050]$ & 2294 & 2352 \\
$m_{\tilde{t}_1}$~(GeV) & $[1900,\,2900]$ & $[1150,\,3000]$ & 2742 & 1332 \\

$m_{\tilde{e}_R}$~(GeV) & $[1600,\,2700]$ & $[1000,\,3000]$ & 427 & 1748 \\
$m_{\tilde{\tau}_1}$~(GeV) & $[1600,\,2700]$ & $[1000,\,3000]$ & 421 & 1702 \\

$m_{\tilde{\chi}^{0}_1}$~(GeV) & $[300,\,485]$ & $[170,\,500]$ & 145 & 236 \\
$m_{\tilde{\chi}^{0}_2}$~(GeV) & $[500,\,1000]$ & $[250,\,1050]$ & 160 & 252 \\
$m_{\tilde{\chi}^{+}_1}$~(GeV) & $[500,\,1000]$ & $[250,\,1050]$ & 153 & 245 \\

$\Omega h^2$ & $[0.10,\,0.12]$ & $[0.09,\,0.14]$ & 0.11 & 0.11 \\
$\sigma_{\rm Xe} \times 10^{45}$~(cm$^2$) & $[3,\,10]$ & $[0.5,\,40]$ & 1.1 & 1.0 \\

${\cal B}(b \rightarrow s\gamma) \times 10^4$ & $[3.2,\,3.5]$ & $[3.0,\,3.8]$ & 3.4 & 3.3 \\
${\cal B}(B_s \rightarrow \mu^+\mu^-) \times 10^9$ & $[2.9,\,3.3]$ & $[2.4,\,4.8]$ & 3.0 & 3.0 \\

$\Delta a_{\mu} \times 10^{10}$ & $[-1.0,\,1.4]$ & $[-3.1,\,4.7]$ & 2.4 & 1.2 \\
\hline
\end{tabular}
\caption{\label{table:bci-hd} 68\% and 95\% BCIs and values of two example points for various parameters and observables in the heavy democratic sneutrino case.}
\end{center}
\end{table}

\clearpage

%================================================================================
\section{Logarithmic priors for the sneutrino parameters}
%================================================================================

\label{logprior}

An alternative to the uniform prior is to use a logarithmic prior, in which all orders of magnitude are equally likely.
In fact, one may argue that the log prior is the least informative, i.e.~the more
objective, prior associated with a dimensionful quantity. This can be
obtained from the Jeffreys prior based on Fisher information~\cite{Jeffreys46}.

To probe the impact of the log priors on our sampling, we again run 8 chains with $10^6$ iterations each, in the light sneutrino and in the heavy non-democratic sneutrino case, but assuming logarithmic priors on the sneutrino mass parameters $m_{\tilde{\nu}_i}$. For all the other parameters, including the mixing angle $\sin \theta_{\tilde{\nu}_i}$, we assume an uniform prior as before. We stress that the sneutrino $A$-terms, $A_{\tilde{\nu}_i}$, are derived from the sneutrino masses (see eq.~(\ref{eq:mixingA})) and are thus sensitive to the change of prior. 

\begin{figure}[b] 
   \centering
   \includegraphics[width=7cm]{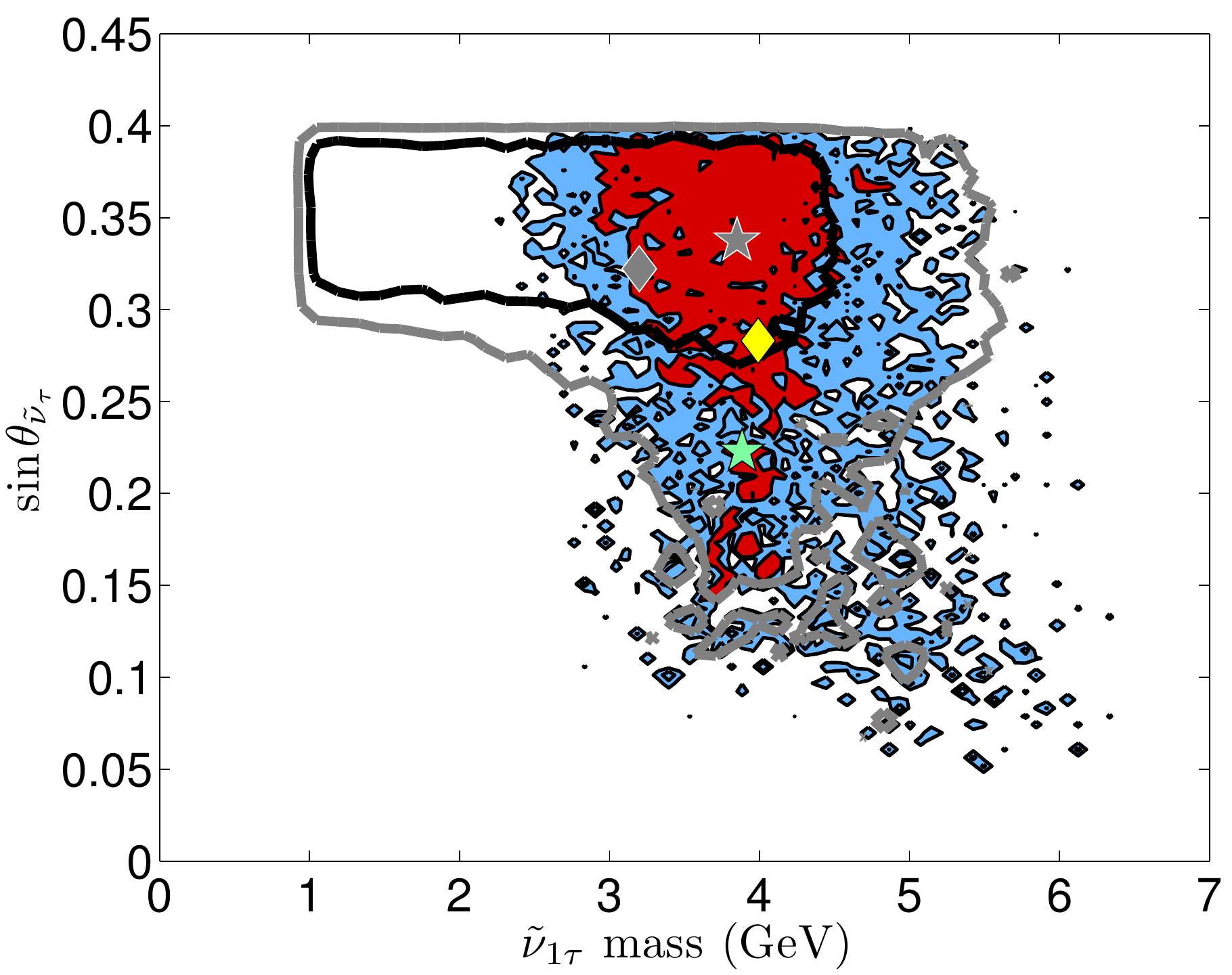} \quad
   \includegraphics[width=7cm]{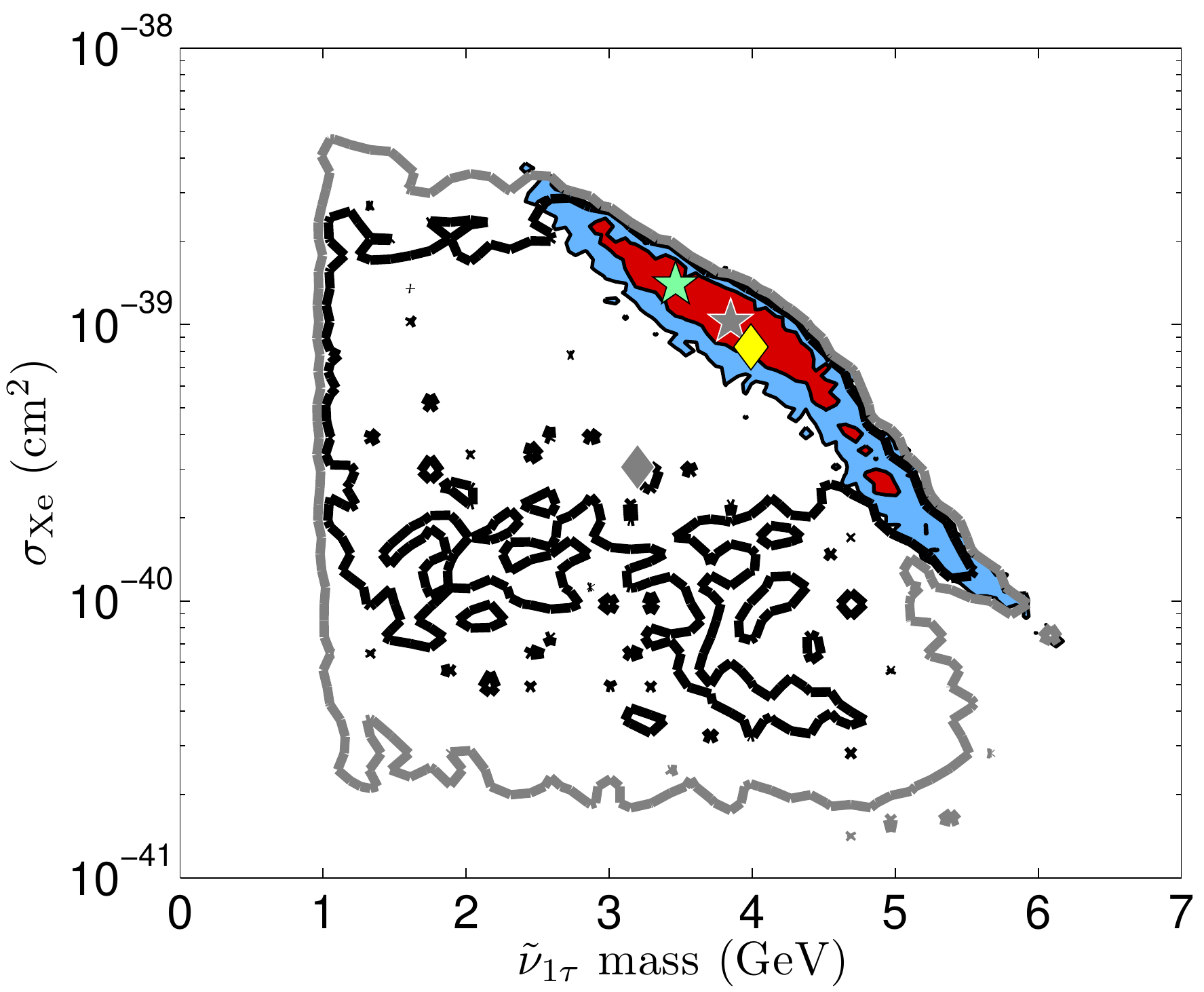}
   \caption{Posterior PDFs in 2D of $\sin\theta_{\tilde{\nu}_{\tau}}$ (left) and $\sigma_{\rm Xe}$ (right) versus $m_{\tilde\nu_{1\tau}}$ for the LD case using logarithmic priors. The red and blue areas are the  68\% and 95\% BCRs, respectively. The green stars mark the highest posterior, while the yellow diamonds mark the mean of the PDF.}
   \label{fig:log-ld}
\end{figure}

In Fig.~\ref{fig:log-ld}, we show the 2-dimensional posterior PDFs of $\sin\theta_{\tilde{\nu}_{\tau}}$ versus $m_{\tilde\nu_{1\tau}}$ and of $\sigma_{\rm Xe}$ versus $m_{\tilde\nu_{1\tau}}$ in the light case. As expected, light sneutrino masses are favored, and as a consequence larger mixing angles are favored (because of $m_{\tilde\nu_{2\tau}}$ being lighter on average, which implies a smaller $A_{\tilde{\nu}_{\tau}}$). With the post-LHC gluino mass limit, however, the PDFs for log prior and uniform prior are quite similar, cf.~Figs.~\ref{fig:light-2d-sinth} and~\ref{fig:light-2d-sigXe}.

\begin{figure}[t] 
   \centering
   \includegraphics[width=7cm]{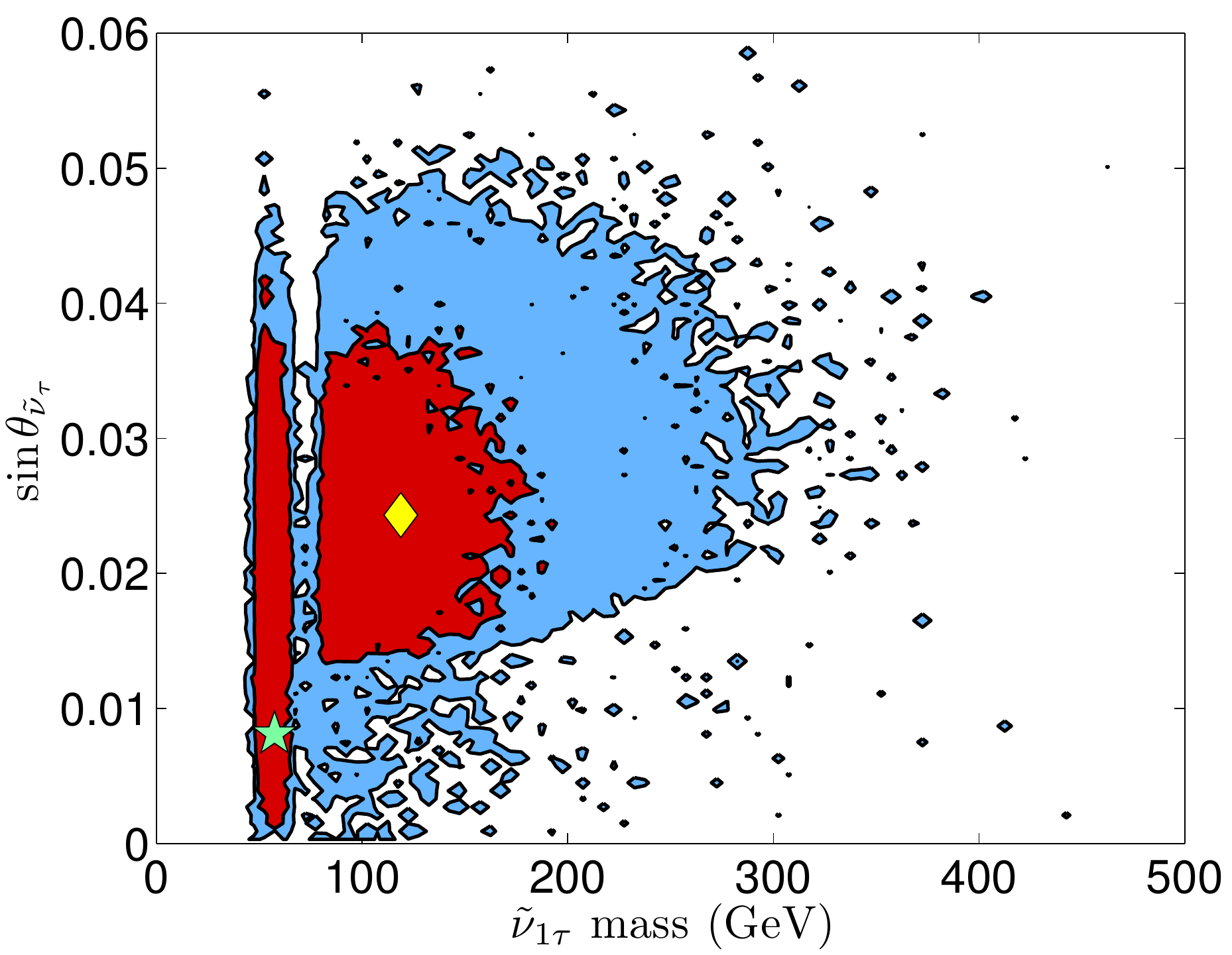} \quad
   \includegraphics[width=7cm]{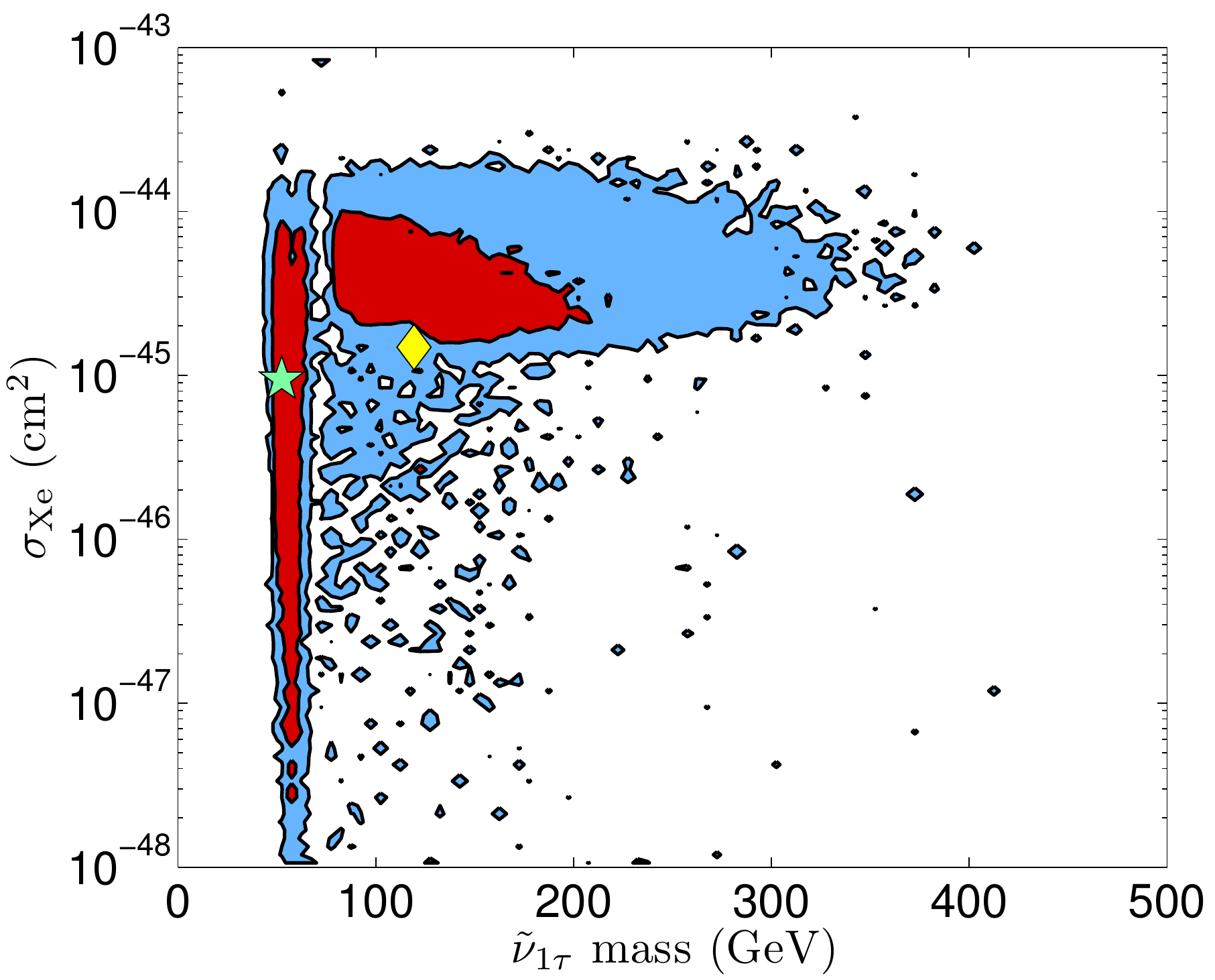}
   \caption{Posterior PDFs in 2D of $\sin\theta_{\tilde{\nu}_{\tau}}$ (left) and $\sigma_{\rm Xe}$ (right) versus $m_{\tilde\nu_{1\tau}}$ for the HND case using logarithmic priors. The red and blue areas are the  68\% and 95\% BCRs, respectively. The green stars mark the highest posterior, while the yellow diamonds mark the mean of the PDF.}
   \label{fig:log-hnd}
\end{figure}

Figure~\ref{fig:log-hnd} is the same as Fig.~\ref{fig:log-ld} but for the HND case. The main change as compared to the uniform prior case, see Fig.~\ref{fig:hnd-2d}, is again the preference for lighter sneutrino masses, roughly  $m_{\tilde\nu_{1\tau}}\lesssim 250$~GeV instead of $m_{\tilde\nu_{1\tau}} \lesssim 375$~GeV at 95\%~BC. Furthermore, the light Higgs resonance region has a probability of 33\% in the log prior case, as compared to 3\% in the case of uniform prior.

We note the extension of the 95\% BCR to low mixing angles and scattering cross sections around 100 GeV: it results from co-annihilation, mainly with the NLSP sneutrinos (also being lighter due to the log prior), but also with a light neutralino or stau. The constraint on the gluino mass from the LHC remains largely irrelevant in the HND case: $p(m_{\tilde{g}}>1~\rm{TeV}) = 94\%$ instead of $99\%$ with uniform priors.

In summary, imposing log priors in the sneutrino sector does not lead to dramatic changes, however it highlights the light Higgs resonance region and the various co-annihilation possibilities in the heavy non-democratic case. These specific cases could lead to a sizable decrease of the scattering and annihilation cross sections thus making it difficult to test the model with future direct and indirect detection experiments.

%\clearpage

%================================================================================
\providecommand{\href}[2]{#2}\begingroup\raggedright\endgroup
%================================================================================

\end{document}